\documentclass[prd,altaffilletter,twocolumn,
superscriptaddress,floatfix,nofootinbib]{revtex4} 

\usepackage{amsfonts,amsmath,graphicx,bm}

\usepackage[pdftex,
       colorlinks=true,
       bookmarksopen=true]{hyperref}
\usepackage{ifpdf}
\usepackage{multirow}
\usepackage[colorlinks=true]{hyperref}
\usepackage{url}

\usepackage{color}

\newcommand{\cambridge}{Department of Applied Mathematics and
Theoretical Physics, University of Cambridge, Cambridge CB3 0WA, UK}

\newcommand{\glasgow}{SUPA, School of Physics and Astronomy,
  University of Glasgow, Glasgow, G12 8QQ, UK}

\newcommand{\subrm}[1]{{\scriptscriptstyle\mathrm{#1}}}

\def\today{\number\day\space\ifcase\month\or
January\or February\or March\or April\or May\or June\or
July\or August\or September\or October\or November\or December\fi
\space\number\year}
\newcount\mins \newcount\hours
\def\now{\hours=\time \mins=\time
	\divide\hours by60 \multiply\hours by60 \advance\mins by-\hours
	\divide\hours by60 
	\number\hours:\ifnum\mins<10 0\fi\number\mins }

\begin{document}

\title{
  Lattice QCD calculation of the $\bm{{B}_{(s)}\to D_{(s)}^{*}\ell{\nu}}$
  form factors at zero recoil and implications for $\bm{|V_{cb}|}$
}

\author{Judd \surname{Harrison}} 
\email[]{jgihh2@cam.ac.uk}
\affiliation{\cambridge}

\author{Christine~T.~H.~\surname{Davies}} 
\affiliation{\glasgow}

\author{Matthew \surname{Wingate}}
\affiliation{\cambridge}

\collaboration{HPQCD Collaboration}
\email[]{http://www.physics.gla.ac.uk/HPQCD}


\begin{abstract}
  We present results of a lattice QCD calculation of $B\to D^*$ and
  $B_s\to D_s^*$ axial vector matrix elements with both states at rest.  
  These zero recoil matrix elements provide the normalization
  necessary to infer a value for the CKM matrix element $|V_{cb}|$ from
  experimental measurements of $\bar{B}^0\to D^{*+}\ell^-\bar{\nu}$ and
  $\bar{B}^0_s\to D_s^{*+}\ell^-\bar{\nu}$ decay.  Results are derived from
  correlation functions computed with highly improved staggered quarks (HISQ)
  for light, strange, and charm quark propagators, and nonrelativistic QCD for
  the bottom quark propagator.  The calculation of correlation functions 
  employs MILC Collaboration ensembles over a range of three lattice spacings.
  These gauge field configurations include sea quark effects of charm, 
  strange, and equal-mass up and down quarks.  We use ensembles with 
  physically light up and down quarks, as well as heavier values.
  Our main results are $\mathcal{F}^{B\to D^*}(1)= 0.895\pm
  0.010_{\mathrm{stat}}\pm{{0.024}_{\mathrm{sys}}}$ and $\mathcal{F}^{B_s\to
    D_s^*}(1)= 0.883\pm 0.012_{\mathrm{stat}}\pm{0.028_{\mathrm{sys}}}$.  We
  discuss the consequences for $|V_{cb}|$ in light of recent investigations
  into the extrapolation of experimental data to zero recoil.
\end{abstract}

\maketitle


\section{Introduction}

Precise measurements of quark flavour-changing interactions offer one way to
uncover physics beyond the Standard Model.  As successful as the Standard
Model appears to be so far, there will continue to be progress reducing
experimental and theoretical uncertainties, as well as making new
measurements. Existing tensions in the global fits to the
Cabibbo-Kobyashi-Maskawa (CKM) parameters may become outright inconsistencies,
or new measurements of rare decays may differ significantly from Standard
Model predictions.

Measurements of the exclusive semileptonic decay $\bar{B}^0 \to
D^{*+}\ell^-\bar{\nu}$ provided the first estimations of the magnitude of CKM
matrix element $V_{cb}$
\cite{Albrecht:1987ej,Bortoletto:1989qb,Fulton:1990bx,Albrecht:1991iz,Barish:1994mu,Buskulic:1995hk,Buskulic:1996yq,Abbiendi:2000hk,Abreu:2001ic,Adam:2002uw,Abdallah:2004rz,Aubert:2007rs,Aubert:2007qs,Aubert:2008yv,Dungel:2010uk,Abdesselam:2017kjf}. This
channel still provides one of three precise methods of determining $|V_{cb}|$.
Measurements for the differential branching fraction are fit to a function of
$q^2$, the lepton invariant mass-squared, and extrapolated to the zero-recoil
point (maximum $q^2$).  Then lattice QCD results for the relevant hadronic
matrix element are used to infer $|V_{cb}|$.  The most recent HFLAV
experimental average \cite{Amhis:2016xyh} combined with the Fermilab/MILC
lattice result \cite{Bailey:2014tva} gives $|V_{cb}| = (38.71 \pm
0.47_{\mathrm{exp}} \pm 0.59_{\mathrm{th}})\times 10^{-3}$.

Measurements of the inclusive $b\to c$ decays $B\to X_c\ell\nu$, combined
with an operator product expansion offer a complementary method.  The
latest estimate is $|V_{cb}| = (42.21 \pm 0.78)\times 10^{-3}$
\cite{Bevan:2014iga,Alberti:2014yda}.  The discrepancy between the inclusive
and exclusive result described above is at the $3\sigma$ level.  

Recently it has been suggested that the inclusive/exclusive difference could
be due to model-dependence implicit in extrapolating experimental data for
$B\to D^*\ell \nu$ to the zero recoil point. The CLN parametrization
\cite{Caprini:1997mu} has been used in recent analyses since it takes
advantage of heavy quark symmetries to improve unitarity constraints in the
form factor shape function. This had several advantages for some time, but
with increased precision in the experimental data, it is possible that
uncertainties arising from these constraints are no longer negligible.  In
fact, recent work \cite{Bernlochner:2017jka,Bigi:2017njr,Grinstein:2017nlq,Bigi:2017jbd,Jaiswal:2017rve,Bernlochner:2017xyx} has
shown that replacing the CLN parametrization by the BGL parametrization
\cite{Boyd:1997kz} yields a determination of $|V_{cb}|$ which is 
as much as 10\% higher, in much better agreement with the $|V_{cb}|$
from inclusive decays.

One can also use the exclusive decay $B \to D\ell\nu$ to estimate $|V_{cb}|$.
Historically this has not given as precise a determination due to 
having to contend with background from $B\to D^*\ell\nu$.  Recent progress
has come from new measurements and joint fits to experimental and lattice
\cite{Bailey:2015rga,Na:2015kha}
data over a range of $q^2$ using so-called $z$-parametrizations
\cite{Boyd:1994tt,Bourrely:2008za}.  The latest result using $B\to D\ell\nu$
results is $|V_{cb}| = (40.85 \pm 0.98)\times 10^{-3}$ \cite{Aoki:2016frl},
in acceptable agreement with either the $B\to D^* \ell\nu$ or $B\to X_c\ell\nu$
determinations.

It is worth noting that a determination of $|V_{cb}|$ is important beyond
semileptonic $b\to c$ decays. Due to insufficient direct knowledge of the 
top-strange coupling, Standard Model predictions which depend on
$V_{ts}$ rely on CKM unitarity, and therefore on $V_{cb}$.  For example the
$K^0$-$\bar{K}^0$ mixing parameter $\epsilon_K$ depends sensitively on
$V_{cb}$; taking $\sin 2\beta$ as an input, then 
$\epsilon_K \propto |V_{cb}|^4$ at leading order \cite{Buras:2008nn}.

In this article we present the details and results of a lattice calculation of
the zero-recoil form factor needed to extract $|V_{cb}|$ from experimental
measurements of the $B\to D^*\ell\nu$ and $B_s \to D_s^*\ell\nu$ decay rates.  
This work differs from the
Fermilab/MILC calculation \cite{Bailey:2014tva} in the following
respects: (1) the gauge field configurations are the next generation MILC
ensembles \cite{ Bazavov:2010ru,Bazavov:2012xda,Bazavov:2015yea} which include effects of $2+1+1$ flavours of sea quarks using the
highly improved staggered quark (HISQ) action \cite{Follana:2006rc}; (2) the fully relativistic HISQ action is used for valence light, strange, {\it and} charm quarks; (3) the nonrelativistic QCD (NRQCD) action \cite{Lepage:1992tx} is used for
the bottom quark.  Therefore, this work represents a statistically 
independent, complementary calculation to Ref.~\cite{Bailey:2014tva}, with
different formulations in many respects.  The two main advantages of using the
HISQ action is that discretization errors are reduced and that the MILC
HISQ ensembles include configurations with physically light $u/d$ quark
effects.  We reported preliminary results 
in recent proceedings \cite{Harrison:2016gup}.

Other groups are applying different methods to calculate $B_{(s)}\to
D^{(*)}_{(s)}$ form factors.  Two-flavour twisted-mass configurations have
been used to estimate the $B_s \to D_s$ form factors near zero recoil
\cite{Atoui:2013zza}; however the uncertainties with this formulation are
quite large.  Work has also recently begun using the domain wall action
for light, strange, and charm quarks \cite{Flynn:2016vej}.  Having results
for the form factors from several groups, each using different approaches,
would be very helpful and could lead to a further reduction in uncertainties
by allowing global fits to uncorrelated numerical and experimental data.


This paper is structured as follows. Sec.~\ref{sec:ff} briefly
introduces the hadronic matrix elements of interest and sets some
notation. In Sec.~\ref{sec:params} we list the inputs to our
computation and summarize the correlation functions calculated.  The
matching between lattice and continuum currents is discussed in
Sec.~\ref{sec:Matching}.  Sec.~\ref{sec:lattresults} is the most
important section for the expert reader; there we discuss the fits to
the correlation functions, the treatment of discretization and quark
mass errors, and estimates of other systematic uncertainties.  We
summarize the result of the lattice calculation in
Sec.~\ref{sec:results}.  In Sec.~\ref{sec:Vcb} we investigate the
implications of the new lattice result in the context of renewed
scrutiny of the extrapolation of experimental data to zero recoil; there we
propose a simplified series expansion as the one least likely to introduce
hidden theoretical uncertainties into a form factor parametrization. We
offer brief conclusions in Sec.~\ref{sec:concl}.  Several appendices are
provided which contain further definitions and details in hopes of making
the manuscript as self-contained as possible.  These are noted at appropriate
places in the body of the paper.

A reader more interested in the results and consequences than the details
of our calculation can safely focus a first reading on Sections \ref{sec:ff},
\ref{sec:results}, \ref{sec:Vcb}, and \ref{sec:concl}, possibly referring
to Appendix \ref{app:exptfits}.

\section{Form factors}
\label{sec:ff}

This section simply summarizes standard notation relating the
differential decay rate, the relevant hadronic matrix elements, and the 
corresponding form factors. Throughout the section we refer to 
$\bar{B}^0\to D^{*+}\ell^-\bar{\nu}$ decay, but the expressions for 
$\bar{B}_s^0\to D_s^{*+}\ell^-\bar{\nu}$ are the same, \textit{mutatis 
mutandis}.

The differential decay rate, integrated over angular variables, is
given in the Standard Model by
\begin{align}
  \frac{d\Gamma}{dw}(\bar{B}^0\rightarrow D^{*+}l^-\bar{\nu}_l) =
  \frac{G_F^2 M_{D^*}^3|\bar{\eta}_{EW} V_{cb}|^2}{4\pi^3} \hspace{5mm}
  \nonumber \\
  \times(M_B-M_{D^*})^2\sqrt{w^2-1}\,\chi(w)|\mathcal{F}(w)|^2
  \label{eq:dGdwF}
\end{align}
where $w = v\cdot v'$ is the scalar product of the $B$ and $D^*$
4-velocities, and $\chi(w)$ is a known kinematic function normalized
so that $\chi(1)=1$ (see App.~\ref{app:exptfits}).  The coefficient
$\bar{\eta}_{EW}$ accounts for electroweak corrections due to box
diagrams in which a photon or $Z$ boson is exchanged in addition to a
$W$ boson as well as the Coulomb attraction of the final-state charged
particles \cite{Sirlin:1981ie,Ginsberg:1968pz,Atwood:1989em}. The form
factor $\mathcal{F}(w)$ is a linear combination of hadronic form
factors parametrizing the matrix elements of the $V-A$ weak current,
i.e.\
\begin{align}
  \langle & D^*(p',\epsilon)|\bar{c}\gamma^\mu  b|B(p)\rangle
 = \frac{2i V(q^2)}{M_B + M_{D^*}}\,
  \epsilon^{\mu\nu\rho\sigma}\epsilon^*_\nu p'_\rho p_\sigma \nonumber \\
\langle & D^*(p',\epsilon)|\bar{c}\gamma^\mu \gamma^5 b|B(p)\rangle = 
 2M_{ D^*}A_0(q^2)\frac{\epsilon^*\cdot q}{q^2} q^\mu \nonumber\\
 & +(M_B+M_{ D^*})A_1(q^2)\Big[ \epsilon^{*\mu} - \frac{\epsilon^*\cdot q}{q^2} q^\mu \Big]\nonumber \\
&-A_2(q^2)\frac{\epsilon^*\cdot q}{M_B+M_{ D^*}}\Big[ p^\mu + p'^\mu - \frac{M_B^2-M_{ D^*}^2}{q^2}q^\mu \Big] \,.
\end{align}
The only contribution to $\mathcal{F}(w)$ at zero recoil, $w=1$, is
from the matrix element of the axial vector current; this reduces to
\begin{align}
  \langle { D^*}(p,\epsilon)|&\bar{c}\gamma^j \gamma^5 b|B(p)\rangle = (M_B+M_{ D^*})A_1(q_{\mathrm{max}}^2)\epsilon^{*j}
  \label{eq:zerorecoilME}
\end{align}
for $j=1,2,3$.  It is sometimes conventional to work with form factors
defined within heavy quark effective theory (HQET).  Of relevance to this work,
we write
\begin{equation}
  h_{A_1}\!(w) = \frac{2\sqrt{M_BM_{ D^*}}}{M_B+M_{ D^*}}
  \frac{A_1(q^2)}{1 - \tfrac{q^2}{(M_B+M_{D^*})^2}} \,.
\end{equation}
At zero recoil, where $w=1$ and $q^2 = q^2_{\mathrm{max}}$,
\begin{align}
  \mathcal{F}(1) = h_{A_1}\!(1) = \frac{M_B+M_{ D^*}}{2\sqrt{M_BM_{ D^*}}}
  \, A_1(q_{\mathrm{max}}^2) \,.
  \label{eq:hA1_A1}
\end{align}
For brevity throughout the paper we will usually use the $h_{A_1}$ notation.
When we wish to specifically refer to the $B_s\to D_s^*$ form factor, we write
$h^s_{A_1}$, so
\begin{equation}
  \mathcal{F}^{B\to D^*}\!(1) = h_{A_1}\!(1) ~~\mbox{and}~~
  \mathcal{F}^{B_s\to D_s^*}(1) = h_{A_1}^s\!(1) \,.
\end{equation}
These are the quantities we calculate here.

\begin{table*}
\centering
\caption{Details of the gauge configurations used in this work. We refer to sets
  1, 2 and 3 as `very coarse', sets 4, 5 and 6 as `coarse' and sets 7 and 8 as
  `fine'. The lattice spacings were determined from the $\Upsilon (2S - 1S)$
  splitting in \cite{latticespacing}.  Sets 3, 6 and 8 use light quarks with
  their physical masses. $u_0$ is the tadpole improvement factor, here we use
  the Landau gauge mean link. The final column specifies the total number of
  configurations multiplied by the number of different start times used for
  sources on each. In order to improve statistical precision we use random 
  wall sources. }
 \begin{tabular}{c c c c c c c c} 
 \hline\hline
Set &  $a(\text{fm})$ & $L/a \times T/a$ & $am_l$ & $am_s$ & $am_c$ & $u_0$ & $n_\text{cfg} \times n_\text{t}$\\ [0.1ex] 
 \hline
1 & 0.1474 & $16 \times 48$ & 0.013 & 0.065 &0.838 & 0.8195  & 960$\times$16 \\ 
2 &  0.1463 & $24 \times 48$  & 0.0064  & 0.064&0.828 & 0.8202 &960$\times$4\\
3 & 0.1450 & $32 \times 48$  & 0.00235  & 0.0647&0.831 & 0.8195  &960$\times$4\\
 \hline
4 & 0.1219 & $24 \times 64$  & 0.0102 & 0.0509 & 0.635 & 0.8341 &960$\times$4\\
5 & 0.1195 & $32 \times 64$  & 0.00507  & 0.0507 &0.628 & 0.8349  &960$\times$4\\
6 & 0.1189 & $48 \times 64$  & 0.00184 & 0.0507&0.628 & 0.8341 &960$\times$4\\

 \hline
7 & 0.0884 & $32 \times 96$ & 0.0074 &0.037&0.440 & 0.8525 &960$\times$4\\ 
8 & 0.08787 & $64 \times 96$ & 0.00120& 0.0363 & 0.432 & 0.8518 &540$\times$4\\ 
 \hline\hline
\end{tabular}
\label{tab:params}
\end{table*}

\begin{table}
\centering
\caption{Valence quark masses and parameters used to calculate
  propagators. The $s$ and $c$ valence masses were tuned using results from \cite{PhysMasses}
  and the $b$ mass was taken from \cite{latticespacing}. $(1+\epsilon_\text{Naik})$ is
  the coefficient of the charm Naik term and $c_{i}$ are the perturbatively
  improved coefficients appearing in the NRQCD action correct through
  $\mathcal{O}(\alpha_s v^4)$  
  \cite{latticespacing}. The last column gives the $T$ values used in three point functions. These have changed from those presented in \cite{Harrison:2016gup} on the very coarse ensembles as it was found that $T=10,11,12,13$ resulted in excessive noise on Set 3, which resulted in poor fit stability and the relatively low value of $\mathcal{F}(1)$ on this ensemble. }
 \begin{tabular}{c c c c c c c c c} 
 \hline\hline
Set &   $am_s^\text{val}$ & $am_c^\text{val}$ & $am_b$ & $\epsilon_\text{Naik}$& $c_1$,$c_6$ & $c_5$&$c_4$ & $T$ \\ [0.1ex] 
 \hline
1 & 0.0641& 0.826 & 3.297& $-0.345$ &1.36 &1.21 &1.22& 6,7,8,9  \\ 
2 & 0.0636& 0.828 & 3.263&$ -0.340$  &1.36 &1.21 &1.22&  6,7,8,9 \\
3 & 0.0628& 0.827 & 3.25&$ -0.345$ & 1.36& 1.21&1.22& 6,7,8,9 \\
 \hline
4 & 0.0522& 0.645& 2.66& $-0.235$ &1.31 &1.16 &1.20 & 10,11,12,13\\
5 & 0.0505& 0.627 &2.62& $-0.224$ &1.31 & 1.16&1.20& 10,11,12,13\\
6 & 0.0507 & 0.631  &2.62 & $-0.226$ &1.31 & 1.16&1.20& 10,11,12,13\\
 \hline
7 & 0.0364&0.434 &1.91&$ -0.117$ & 1.21&1.12 & 1.16& 15,18,21,24\\
8 & 0.0360 & 0.4305 &1.89&$ -0.115$ &1.21 &1.12 & 1.16& 10,13,16,19\\ 
 \hline\hline
\end{tabular}
\label{tab:vqparams}
\end{table}

\section{Lattice parameters and methodology}
\label{sec:params}

Here we give specific details about the lattice calculation.  Once again
many of the expressions will refer to $B\to D^*$ matrix elements, but 
they apply for any spectator quark mass.

The gluon field configurations that we use were generated by the MILC
collaboration and include 2+1+1 flavours of dynamical HISQ quarks in the sea
and include 3 different lattice spacings
\cite{Bazavov:2010ru,Bazavov:2012xda,Bazavov:2015yea}. The $u$ and $d$ quarks
have equal mass, $m_u = m_d \equiv m_l$, and in our calculations we use the
values $m_l/m_s = 0.2$, $0.1$ and the physical value $1/27.4$
\cite{lightmassratio}. The $s$ and $c$ quarks in the sea are also well-tuned
\cite{PhysMasses} and included using the HISQ action.  The gauge action is the
Symanzik improved gluon action with coefficients correct to
$\mathcal{O}(\alpha_s a^2, n_f \alpha_s a^2)$ \cite{GaugeAction}.
Table~\ref{tab:params} gives numerical values for the lattice spacings, quark
masses, and other parameters describing the ensembles we used.

In calculating correlation functions, we slightly tune the valence $s$
and $c$ masses closer to their physical values.  The $d$, $s$, and $c$
quark propagators were computed using the MILC
code~\cite{MILCgithub}. The $b$ quark is simulated using perturbatively improved
non-relativistic QCD \cite{NRQCD,latticespacing}, which takes
advantage of the non-relativistic nature of the $b$ quark dynamics in
$B$ mesons and produces very good control over discretization
uncertainties. Details of the gauge, NRQCD, and HISQ actions used are
given in Appendices~\ref{sec:GAction}, \ref{sec:NRQCDAction}, and
\ref{sec:HISQ}, respectively.  In Table~\ref{tab:vqparams} we record
the parameters used in calculating quark propagators.

In order to extract the form factor from lattice calculations we must compute the set of Euclidean correlation functions
\begin{align}
C_{B2pt}(t)_{ij}&= \langle \mathcal{O}(t)_{Bi} \mathcal{O}^\dagger(0)_{Bj}     \rangle \nonumber\\
C^{\mu\nu}_{D^*2pt}(t)_{ij}&= \langle \mathcal{O}^\mu(t)_{D^*i} \mathcal{O}^{\dagger\nu}(0)_{D^*j}     \rangle \nonumber\\
C^{\mu\kappa}_{3pt}(T,t,0)_{ij}&=\langle \mathcal{O}^\mu(T)_{D^*i} 
{J}^\kappa(t) \mathcal{O}^\dagger(0)_{Bj}     \rangle
\end{align}
where each interpolating operator $\mathcal{O}_i$ is projected onto zero
spatial momentum by summing over spatial lattice points and the current
${J}^\kappa$ is one of several lattice
currents (see Sec.~\ref{sec:Matching}). The indices $i$ and $j$ label
different smearing functions. We use three different smearing operators on
each of the $B$ and $D^*$ interpolating operators.

In implementing $ \mathcal{O}^\mu(t)_{D^*i}$ we use an unsmeared operator and
two gauge covariant Gaussian smearings, implemented by applying
$\left(1-\frac{r_{D^*}^2\nabla^2}{n}\right)^n$ to the field. Here the
derivative is stride-2 in order not to mix the staggered-taste meson
multiplets. $r_{D^*}$ is the radius (in lattice units) chosen to give good
overlap with the ground state, and $n$ is chosen to give a good
approximation to a Gaussian while maintaining numerical stability.  For the
$B$ we use a local operator as well as two Gaussian smearings, implemented as
$\frac{1}{N}e^{-(x-y)^2/r_B^2}$, where again $r_B$ is a radius in lattice
units and $N$ is an overall normalization.  Since
the $B$ smearings are not gauge invariant, the gauge fields are fixed to
Coulomb gauge. We refer to the local operator as $l$ and the Gaussian smearings as $g2$ and $g4$ corresponding to radii of $2a$ and $4a$ respectively. We use the same choices of radii for both $B$ and $D^*$ smearings. The smearing parameters are given in Table~\ref{tab:radii}.

\begin{table}
\caption{Values of $r$, taken to be the same, for the $B_{(s)}$ and $D_{(s)}^*$ Gaussian smearings on each set and the accompanying $n$ values for the $D_{(s)}^*$ smearings. We chose to fix the radii in lattice units rather than physical units as this seemed to result in more consistent numerical stability of the covariant Gaussian smearing operator when moving between lattices. }
 \begin{tabular}{c c c c c} 
 \hline\hline
Set &  $r_{g2}/a$ & $r_{g4}/a$ & $n_{g2}$ & $n_{g4}$ \\[0.1ex] 
\hline
1,2,3 & 2 & 4 & 30  & 30  \\ 
\hline
4,5,6 & 2 & 4 &  30 &  30 \\
\hline
7,8 & 2 & 4 &  30 & 40 \\ 
 \hline\hline
\end{tabular}
\label{tab:radii}
\end{table}

The interpolating operators themselves are
\begin{align}
  \mathcal{O}_B(x) &= \sum_{y}\bar{\psi}_u(x) \gamma^5 \Delta (x,y) \Psi_b(y)
  \nonumber\\
  \mathcal{O}_{D^*}^i (x)&=\sum_{y} \bar{\psi}_u(x) \gamma^i \Delta (x,y)
  \psi_c(y+a\hat{i})
\end{align}
where $\Delta(x,y)$ is the appropriate smearing function discussed above. 
In distinction to the continuum quark fields $b$, $c$, $s$, and $u$
of Sec.~\ref{sec:ff}, here we denote the NRQCD $b$ field by $\Psi_b$ and the
staggered fields, written as 4-component Dirac spinors (see 
App.~\ref{sec:HISQ}), by $\psi$ with the appropriate flavour subscript.

We checked both the point-split and local $D^*$ interpolating operators on the
very coarse, physical point ensemble (Set 3) and found no significant difference in
statistical noise or central value of either the $D^*$ mass or the matrix
element. We primarily used the point-split current as it was simpler to implement
in our framework.  The results quoted below for the $B\to D^*$ fits use the point-split
vector current, except for Set 3 where results are given for the local vector current.
The results below for $B_s \to D_s^*$ form factors were obtained using the local
vector current.

In order to improve statistics we multiply our smeared sources with random walls to produce, on average, 
the effect of multiple sources. Taking the all-to-all 2-point function as an example we have
\begin{align}
C_{2pt}(t,0)_{ij} =& \sum_{xy,\delta} \langle \bar{\psi}_1(x,t)\Gamma\psi_2(x+\delta_{sink},t) \times  \nonumber \\
&\bar{\psi}_2(y,0)\Gamma\psi_1(y+\delta_{src},0) \rangle \Delta_i(\delta_{sink})\Delta_j(\delta_{src}) \nonumber\\
= &\sum_{xy,\delta} \text{tr}\Big[   \Gamma G_2(x,t;y,0)\Gamma \Delta_j(\delta_{src})\times \nonumber \\
& G_1(y+\delta_{src},0;x-\delta_{sink},t) \Delta_i(\delta_{sink}) \Big] \,.
\end{align}
Exact computation requires an inversion for each value of $y$ being summed over. Instead we generate a random vector $\xi$ satisfying
\begin{equation}
  \lim_{N \rightarrow \infty}\sum_l^N \xi_{al}(x) \xi_{bl}(y)^* =
  \delta(x,y)\delta_{ab} \,.
\end{equation}
$N$ here is the number of random vector wall sources. The average over configurations further suppresses violations of this relation; in practice a single random wall per colour, setting $N = N_c = 3$, is sufficient. Inserting the above relation into the 2-point function
\begin{align}
C_{2pt}(t,0)_{ij} =&\sum_{xyz,\delta,l} \text{tr}\Big[   \Gamma G_2(x,t;z,0) \xi(z) \Gamma \times \nonumber \\
\Delta_j(\delta_{src}) & \xi^\dagger(y-\delta_{src})  G_1(y,0;x-\delta_{sink},t) \Delta_i(\delta_{sink}) \Big] \nonumber \\
=&\sum_{xyz,\delta,l} \text{tr}\Big[   \Gamma G_2(x,t;z,0) \xi(z) \Gamma \times \nonumber \\
\gamma^5 \big[  \Delta_i(\delta_{sink}) &  G_1(x-\delta_{sink},t;y,0) \Delta_j(\delta_{src}) \xi(y-\delta_{src}) \big]^\dagger \gamma^5  \Big] 
\end{align}
where we have used $\gamma^5$ hermiticity. The naive propagators $G$ are built from staggered quarks and the full form of the correlation function contractions in terms of NRQCD and staggered propagators is given in Appendix \ref{sec:3pt}.

These correlation functions can be expressed in terms of amplitudes and decaying exponentials by inserting a complete basis of states. Projecting onto zero momentum and setting $q = (M_B-M_{D^*},0,0,0)$ this gives
\begin{align}
C_{B2pt}(t)_{ij} &= \sum_{n,a=0,1} (-1)^{at} B_{ai}^{n} B_{aj}^{n}e^{-M_{B_a^n} t} \nonumber\\
C_{D^*2pt}(t)_{ij}& =  \sum_{n,a=0,1} (-1)^{at} A_{ai}^{n} A_{aj}^{n}e^{-M_{D^{*n}_a} t}\nonumber\\
C_{3pt}(T,t,0)_{ij} &= \sum_{ab = 0,1}\sum_{nm} (-1)^{a(T-t) + bt} A_{ai}^{n} B_{bj}^{m} \nonumber\\
		& \times V ^{nm}_{ab}  e^{-M_{D^{*m}_a}(T-t) - M_{B_b^n}t} 
\label{corrfuncts}
\end{align}
where
\begin{align}
B_{ai}^{n}   				   &= \frac{  \langle \Omega | \mathcal{O}^i_{B} | B_a^n \rangle}{ \sqrt{{2M_{B_a^n}}} }  \nonumber \\
A_{ai}^{n} 				   &= \frac{  \langle \Omega | \mathcal{O}^i_{D^*} | {D^*}_a^n \rangle}{ \sqrt{{2M_{D^{*n}_a}}} }\nonumber \\
V ^{nm}_{ab}               &= \frac{\langle {D^*}_a^n| J |B_b^m
  \rangle}{\sqrt{{2M_{D^{*n}_a}} {2M_{B_b^m}}} } \,.
\label{eq:fitparams}
\end{align}
Note that we have included contributions from opposite-parity states, which
depend on imaginary time like $(-1)^t$ and arise from using staggered quarks
\cite{Follana:2006rc}, by introducing the sum over $a$ and $b$. When either $a$ or $b$
is nonzero the corresponding term in the sum is multiplied by a sign factor
which oscillates between $1$ and $-1$ in time. We are only interested in the
terms with $a=b=0$ here; however in order to extract these, the oscillating
terms must be fit away. For our choice of operators the $A$, $B$ and $V$ parameters
are real \cite{timereversal}.  We discuss our
fits to these correlation functions in Sec.~\ref{ssec:corrfit}.

\section{One-loop Matching }
\label{sec:Matching}

We require a lattice current with the same matrix elements as the continuum
current to a given order. The matching of lattice and continuum currents is done in \cite{Matching2013} through
$\mathcal{O}(\alpha_s,\alpha_s/am_b,\Lambda_{\subrm{QCD}}/m_b)$,where $\Lambda_{\subrm{QCD}}$ is a typical QCD scale of a few hundred MeV, following the
method used in \cite{Matching1998}. Using power counting in powers of $\Lambda_{\subrm{QCD}}/m_b$ a
set of lattice currents is selected. At the order to which we work in this paper
only the following currents contribute
\begin{align}
{J^{(0)i}_\text{latt}}(x) &= \bar{\psi}_c \gamma^i\gamma^5 \Psi_b \nonumber\\
{J^{(1)i}_\text{latt}}(x) &=-\frac{1}{2am_b} \bar{\psi}_c \gamma^i\gamma^5 \bm{\gamma}
\cdot \Delta \Psi_b \,.
\end{align}
It is convenient for us to also compute the matrix elements of operators 
entering at $O(\alpha_s\Lambda_{\subrm{QCD}}/m_b)$
\begin{align}
{J^{(2)i}_\text{latt}}(x) & = -\frac{1}{2am_b}\bar{\psi}_c \bm{\gamma} \cdot \overleftarrow{\Delta} \gamma^0  \gamma^i\gamma^5  \Psi_b \nonumber \\
{J^{(3)i}_\text{latt}}(x) & = -\frac{1}{2am_b}\bar{\psi}_c \gamma^0\gamma^5
{\Delta}^i  \Psi_b \,.
\end{align}
This allows for a configuration-by-configuration check of the code: namely
that at zero recoil, the three-point correlation functions satisfy the relation
$C_{3pt J^{(1)}} + C_{3pt J^{(2)}} - 2C_{3pt J^{(3)}} = 0$.  This identity is derived
using integration by parts and the fact that $\gamma^0\Psi_Q = \Psi_Q$.

The full matching is a double expansion in $\Lambda_{\subrm{QCD}}/m_b$ and
in $\alpha_s$.  The matched current is given by
\begin{align}
\mathcal{J}^i = Z[(1+\alpha_s(\eta-\tau)){J^{(0)i}_\text{latt}} +
{J^{(1)i}_\text{latt}}] + O\left(\frac{\alpha_s \Lambda_{\subrm{QCD}}}{m_b}\right)
\label{matchcurr}
\end{align}
where $ Z$ is a multiplicative factor from the tree-level massive-HISQ wave function renormalization for the HISQ $c$ quark. The one-loop coefficients $\eta$ and $\tau$ respectively account for the renormalization of $J^{(0)i}_\text{latt}$
and for the mixing of $J^{(1)i}_\text{latt}$ into $J^{(0)i}_\text{latt}$.
Numerical values for the perturbative coefficients relevant for
the ensembles used are given in Table~\ref{tab:matching} \cite{Matching2013}.

Matrix elements of currents of order  $\alpha_s^n \Lambda_{\subrm{QCD}}/m_b$ vanish to all orders in $\alpha_s$ according
to Luke's theorem \cite{LUKE}. We will denote by $V$ the
matrix elements of the currents $J_{\mathrm{latt}}$ divided by meson mass factors,
as in (\ref{eq:fitparams}) with $a=b=0$ and $n=m=0$. Luke's theorem implies the combination
\begin{align}
V_{\mathrm{sub}}^{(1)i} = V^{(1)i}- \alpha_s \tau V^{(0)i} \,,
\label{eq:subtractedV}
\end{align}
which represents the the physical, sub-leading matrix element,
should be very small, only different from zero due to systematic
uncertainties.

\begin{table}[t]
  \caption{\label{tab:matching}Tree-level $Z$ factors and one-loop matching coefficients, used in (\ref{matchcurr}), calculated at lattice quark masses appropriate to each of our gauge-field ensembles \cite{Marginalization}. We also give values on each ensemble for the strong coupling constant in the $V$ scheme at a scale of $2/a$ (from results in \cite{Matching2013}). }
\begin{tabular}{ c c c c c } 
\hline \hline
  Set  &  $Z$         & $-\eta$  & $\tau$ & $\alpha_V(2/a)$ \\     
\hline
1 &  0.9930& 0.260(3)& 0.0163(1)& 0.346  \\
2 &  0.9933& 0.260(3)& 0.0165(1)& 0.344  \\
3 &  0.9930&0.260(3)& 0.0165(1)& 0.343   \\
4 &  0.9972& 0.191(3)& 0.0216(1)& 0.311  \\
5 &  0.9974& 0.185(3)& 0.0221(1)& 0.308  \\
6 &  0.9974&0.185(3)& 0.0221(1)& 0.307   \\
7 &  0.9994&0.091(3)& 0.0330(1)& 0.267   \\
8 &  0.9994&0.091(3)& 0.0330(1)& 0.267   \\
 \hline \hline
\end{tabular}
\end{table}


\section{Analysis of numerical data}
\label{sec:lattresults}

In this section we discuss the two main aspects of numerical analysis.
First we present fits to the correlation functions, allowing us to 
determine $h_{A_1}\!(1)$ on each of the 8 ensembles.  Second, we discuss
how we infer a physical value for $h_{A_1}\!(1)$ with an error estimate
for uncertainties associated with current matching, discretization, and
dependence on quark masses.

\subsection{Fits to correlation functions}
\label{ssec:corrfit}

We fit the three correlation functions defined in (\ref{corrfuncts})
simultaneously using the \texttt{corrfitter} package developed by Lepage
\cite{Fitting,corrfitter}. This minimises
\begin{equation}
  \chi^2(p) = \sum_{t,t^\prime} \Delta C(t,p)\sigma^{-2}_{t,t^\prime}\Delta C(t^\prime,p)
  + \sum_i \frac{(p^i-p^i_{\text{prior}})^2}{\sigma_{p^i_\text{prior}}^2}
\end{equation}
with respect to $p$, where $\Delta C(t,p) =\overline{C}(t) - C_\text{TH}(t,p)$ and $p^i$ is the $i$th parameter in the theory, $p^i_\text{prior}$ is its prior value with error $\sigma_{p^i_\text{prior}}$. The correlation matrix $\sigma_{t,t^\prime}$ includes all correlations between data points. Fitting correlators from all smeared sources and sinks simultaneously requires the use of as SVD cut on the eigenvalues when determining the inverse of $\sigma^2$. We also exclude points close in time to the source and sink to suppress excited state contributions and speed up the fit.

\begin{figure}
    \includegraphics[width=0.5\textwidth]{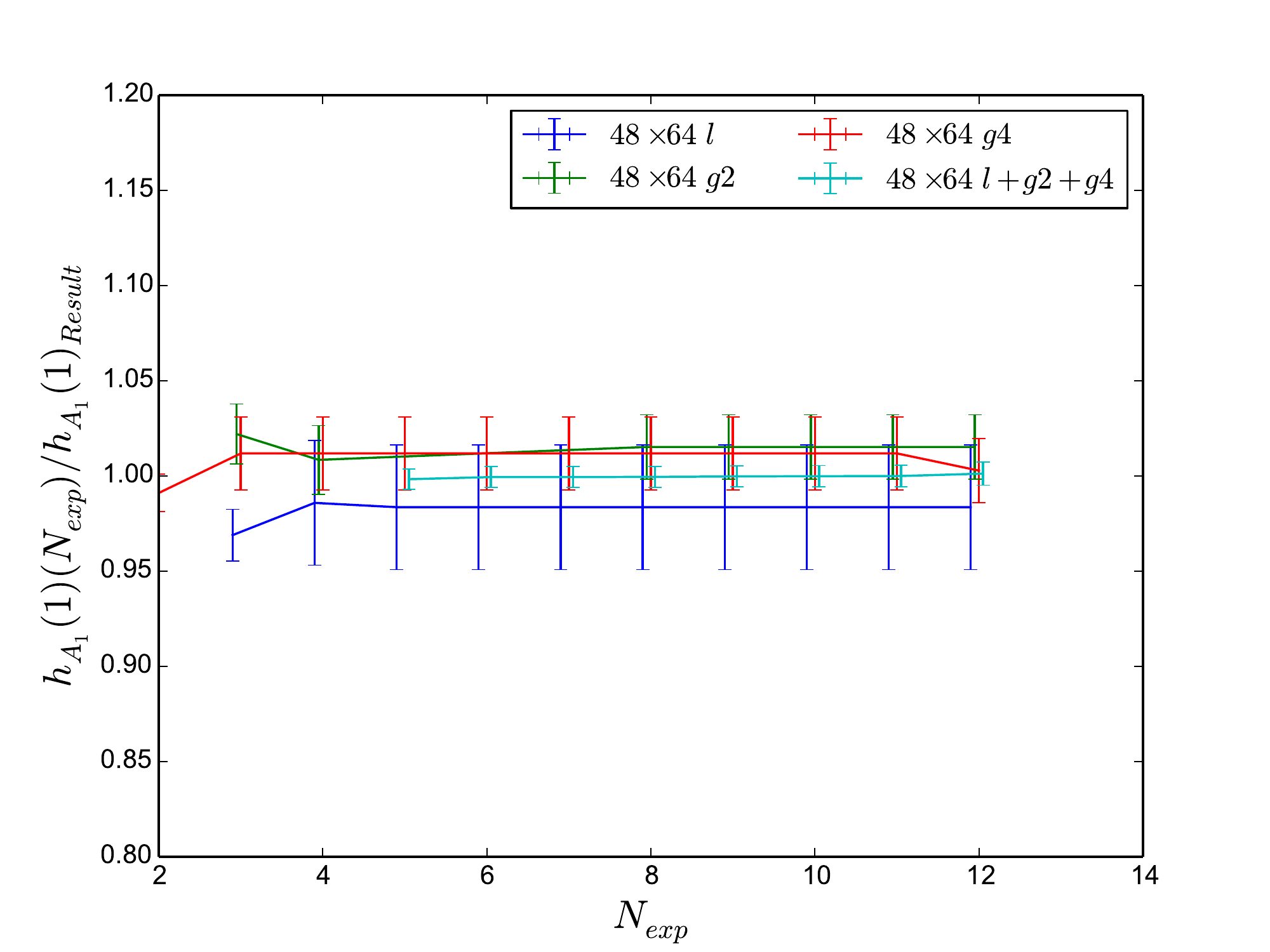}
  \caption{Fractional $h_{A_1}\!(1)$ variation with the number of
    exponentials used in the fit function on Set 6, showing fits done
    using each individual source/sink smearing (local $l$, or Gaussian with 2
    radii, $g2$ and $g4$) as well as the full $3\times 3$ matrix
    fit. }
 \label{4864ratiosmearings}
\end{figure}

We look at the effectiveness of the various smearings by fitting each
smearing diagonal, i.e. equal radii, set of two and three point
correlator functions independently and comparing the result to the
full fit. Figure \ref{4864ratiosmearings} shows an example of this;
plots for the full data set appear in Fig.~\ref{fig:Nexpplots} in
Appendix \ref{corrfitplots}.  In these plots we only include the
results of fits with $\chi^2/\text{dof} < 1.2$. We give the ground
state and oscillating state two point fit parameters for our full
simultaneous fits in Table \ref{corrfitparamstab}. The $C_{B2pt}(t)$
fit amplitudes, the energies and $B_{ai}^n$ parameters of
(\ref{corrfuncts}), are in good agreement with those in
Ref.~\cite{Bmeson}.

\begin{table*}
\centering
\caption{\label{corrfitparamstab}Ground state and oscillating state local amplitudes and masses from our fits. Note that on Set 3 and for all the $D_s^*$ data we use the local vector operator, otherwise we use the point-split operator; therefore, the amplitudes $A$ are not comparable between different operators. Also note that the tabulated $B$ ``masses'' are the NRQCD ``simulation
  energies'' $aE^{\mathrm{sim}}$, representing the nonperturbative contribution
  to the $B$ meson binding energy. The $B$ parameters are in good agreement with those in \cite{Bmeson}.
  }
\begin{tabular}{c c c c c c c c c} \hline\hline
Set & $A_{0l}^0 $  &$A_{1l}^0$ & $aM_{D^{*0}_0}$   & $aM_{D^{*0}_1}$  & $B_{0l}^0$  &$B_{1l}^0$ & $aM_{B^0_0}$   & $aM_{B^0_1}$    \\  
\hline      
1 &$0.1420(12)$ &$0.110(10)$ &$1.5465(19)$ &$1.815(22)$ &$0.2287(17)$ &$0.232(14)$ &$0.5667(14)$ &$0.815(13)$ \\
2 &$0.1338(17)$ &$0.087(12)$ &$1.5304(28)$ &$1.742(26)$ &$0.2171(20)$ &$0.200(24)$ &$0.5534(18)$ &$0.770(18)$ \\
3 &$0.1710(14)$ &$0.092(13)$ &$1.5226(18)$ &$1.675(25)$ &$0.2099(17)$ &$0.214(14)$ &$0.5433(15)$ &$0.761(14)$ \\
4 &$0.1006(23)$ &$0.081(20)$ &$1.2599(31)$ &$1.499(30)$ &$0.1700(23)$ &$0.104(54)$ &$0.4825(21)$ &$0.638(46)$ \\
5 &$0.0951(14)$ &$0.081(10)$ &$1.2289(23)$ &$1.459(18)$ &$0.1611(24)$ &$0.095(54)$ &$0.4745(22)$ &$0.621(42)$ \\
6 &$0.09636(52)$ &$0.0479(87)$ &$1.23244(99)$ &$1.354(22)$ &$0.15739(69)$ &$0.1674(58)$ &$0.46809(80)$ &$0.6523(58)$ \\
7 &$0.06466(40)$ &$0.0520(35)$ &$0.91551(88)$ &$1.0838(82)$ &$0.10762(64)$ &$0.1241(35)$ &$0.37950(76)$ &$0.5437(40)$ \\
8 &$0.05912(40)$ &$0.0502(23)$ &$0.89583(99)$ &$1.0477(71)$ &$0.09884(69)$ &$0.1131(26)$ &$0.36473(98)$ &$0.5042(32)$ \\

\\ \hline
& $A^{s0}_{0l} $  &$A^{s0}_{1l}$ & $aM_{D^{*0}_{s0}}$   & $aM_{D^{*0}_{s1}}$  & $B^{s0}_{0l}$  &$B^{s0}_{1l}$ & $aM_{B^0_{s0}}$   & $aM_{B^0_{s1}}$    \\  
\hline
3 &$0.1987(13)$ &$0.136(14)$ &$1.58655(79)$ &$1.868(14)$ &$0.25554(42)$ &$0.2460(75)$ &$0.60639(28)$ &$0.8862(50)$ \\
6 &$0.13689(81)$ &$0.0918(75)$ &$1.28341(45)$ &$1.5094(94)$ &$0.18822(14)$ &$0.1669(58)$ &$0.51657(11)$ &$0.7277(36)$ \\
8 &$0.08233(40)$ &$0.0618(23)$ &$0.93657(49)$ &$1.1142(50)$ &$0.11867(55)$ &$0.1212(17)$ &$0.40136(48)$ &$0.5698(15)$ \\
 \hline\hline
\end{tabular}
\end{table*}

\begin{table}
  \caption{\label{tab:mixing}Matrix elements, with meson factors
    defined in (\ref{eq:fitparams}), of currents contributing at
    $\mathcal{O}(\alpha_s\Lambda_{\subrm{QCD}}/M_B)$ for $B
    \rightarrow D^*$. Note the approximate cancellation between the
    mixing down term $\alpha_s\tau V^{(0)}$ and $V^{(1)}$ to give
    a small $V^{(1)}_\mathrm{sub}$ as we would
    expect from Luke's theorem. Note $V^{(2)}$ is numerically smaller
    than its parametric estimate $\alpha_s \Lambda_\mathrm{QCD}/m_b \approx 0.03$. }
\begin{tabular}{ c c c } 
\hline \hline
Set& $V^{(1)}_\mathrm{sub}$&$V^{(2)}$\\            
\hline
3 & $-0.0050(8)$ & $0.0138(8) $\\
6 & $-0.0044(5)$ & $0.0101(4)  $\\
8 & $-0.0031(7)$  & $0.0060(8) $\\
 \hline \hline
\end{tabular}
\end{table}

Table~\ref{tab:mixing} gives results for matrix elements corresponding to the currents
$J^{(1)}_\text{latt}$ and $J^{(2)}_\text{latt}$.  One can see that Luke's
theorem holds, in that $V^{(1)}_{\mathrm{sub}}$ is very small. Results are also given for $V^{(2)}$ as well as numerical values for $\alpha_s \Lambda_\mathrm{QCD}/m_b$. While it is important to remember that there are absent mixing down factors from the current $J^{(0)}$  contributing at $\mathcal{O}(\alpha_s \Lambda_\mathrm{QCD}/m_b)$ it is encouraging to see that $V^{(2)}$ is small compared to its expected order.

On each ensemble, we obtain a value for the zero-recoil form factors
$h_{A_1}^{(s)}\!(1)$.  As in the continuum expressions
(\ref{eq:zerorecoilME}) and (\ref{eq:hA1_A1}) we have
\begin{align}
  h_{A_1}\!(1)|_{\mathrm{latt}} = V^{\mathcal{J}} \equiv 
\frac{\langle {D^*}| \mathcal{J} |B \rangle}{\sqrt{{2M_{D^{*}}}{2M_{B}}}}
\,   
\end{align}
and similarly for $h_{A_1}^s\!(1)|_{\mathrm{latt}}$. We write $V^{\mathcal{J}}$ here to make clear that we fit combinations of three point correlators that correspond to the insertion of the current given by (\ref{matchcurr}). Results for $h_{A_1}\!(1)$ 
on each ensemble are presented in Table~\ref{tab:resultsF}. We computed
$h_{A_1}^s\!(1)$ on the physical-point lattices only since chiral 
perturbation theory predicts this quantity to be much less sensitive
to the sea quark mass than the spectator quark mass. (In fact we will
see that the spectator quark mass dependence is also small).

\begin{table}
  \caption{\label{tab:resultsF}Fit results for the zero-recoil form
    factor $  h_{A_1}\!(1)_\text{latt}=V^\mathcal{J}$ for both $B\rightarrow D^*$
    and $B_s\rightarrow D_s^*$. }
\begin{tabular}{ c c c } 
\hline\hline
  Set &$h_{A_1}\!(1)_\text{latt}$ &$h_{A_1}^s(1)_\text{latt}$ \\
\hline
1 & 0.8606(91) &  \\
2 & 0.871(13)  &  \\
3 & 0.8819(96) &  0.8667(42)\\
4 & 0.8498(94) &  \\
5 & 0.8570(84) &  \\
6 & 0.8855(50) &  0.8662(61)\\
7 & 0.8709(75) &  \\
8 & 0.8886(63) &  0.8715(44)\\
 \hline\hline
\end{tabular}
\end{table}


\subsection{Chiral-continuum extrapolation} 
\label{subsec:chiral-cont}

By carrying out the calculation using 8 ensembles, spanning 3 values of
lattice spacing and 3 values of the light quark mass, we can quantify
many of the systematic uncertainties by performing a least-squares fit to
a function which accounts for unphysical parameters or truncation errors.
Below we describe how the fits address each of these sources of
uncertainty then present results of the fits.

There are two types of systematic error for which we must account.
The first type are truncation errors about which the numerical data
contain no information.  In this class are the higher-order (in
$\Lambda_{\subrm{QCD}}/m_b$) current corrections truncated in the
perturbative matching described in Sec.~\ref{sec:Matching}.  The
numerical data contain no information about
$\Lambda_{\subrm{QCD}}^2/m_b^2$ or $\alpha_s\Lambda_{\subrm{QCD}}/m_b$
corrections, so we add to each data point nuisance terms
\begin{align}
\label{eq:HO_HQET}
  &\left.h_{A_1}\!(1)\right|_{\mathrm{fit}} =
  \left.h_{A_1}\!(1)\right|_{\mathrm{latt}} \nonumber \\
  &~~+ e_4\frac{\Lambda_{\subrm{QCD}}^2}{m_b^2} \left[1 +
    {e_5}\Delta_{am_b} + {e_6}\Delta_{am_b}^2
    \right]\\
  &~~+ e_7\frac{\alpha_s\Lambda_{\subrm{QCD}}}{m_b} \left[1 +
    {e_8}\Delta_{am_b} + {e_9}\Delta_{am_b}^2
    \right]\nonumber
\end{align}
where
\begin{equation}
\Delta_{am_b} = (am_b -2.5)/2.5 \nonumber
\end{equation}
and $e_4$, $e_5$, $e_6$, $e_7$, $e_8$, and $e_9$ are Gaussian distributed variables, with mean and standard deviation $\mu (\sigma)$,
with $e_4 = 0(0.5)$, $e_7 = 0(0.3)$ and $e_{5,6,8,9} = 0(1)$, 100\% correlated between
each data point.  The $e_5$, $e_6$, $e_8$ and $e_9$ terms reflect the fact that the
coefficients of the truncated $\Lambda_{\subrm{QCD}}^2/m_b^2$ and $\alpha_s \Lambda_\subrm{QCD}/m_b$ terms will
be slowly varying functions of $am_b$. Our choice of $e_7$ is motivated by the magnitude of $V^{(2)}$ and the expectation that Luke's theorem will hold at this order.

The second type of systematic uncertainties arise from truncation,
discretization, or tuning errors about which we can draw inferences from
our Monte Carlo calculation.  Consider the unknown $\alpha_s^2$ corrections to the current normalization.  In contrast to the truncation of the $\Lambda_{\subrm{QCD}}/m_b$ expansion,
the numerical data is, at least in principle, sensitive to $O(\alpha_s^2)$
corrections through the running of the coupling on the different lattice
spacings.  In addition the results have dependence on the lattice spacing and the light quark mass that can be mapped out using theoretical expectations. For the light quark mass dependence this is based on chiral perturbation theory.  Therefore we fit the data points
to the functional form
\begin{align}
  &\left.h_{A_1}\!(1)\right|_{\mathrm{fit}} = (1 + B) \delta_a^B + C
  \frac{M_\pi^2}{\Lambda_\chi^2} + \delta_a^g \frac{g^2}{48\pi^2 f^2 }
  \times \text{chiral logs} \nonumber\\ & ~+ \gamma_1
  \alpha_s^2\left[1+\frac{\gamma_5}{2} (am_b-2)+\frac{\gamma_6}{4}
    (am_b-2)^2\right] V^{\mathcal{J}} \,.
  \label{eq:fitform}
\end{align}
The first term accounts for the deviation of the physical $h_{A_1}\!(1)$
from the static quark limit value of 1.  The fit parameter $B$ is given
a prior of $0(1)$. We take as priors $\gamma_1 = 0(0.5)$, $\gamma_{5,6} = 0(1)$.  Discretization
and quark mass tuning errors are included in $\delta_a^B$, to be described
further below.

The second and third terms in (\ref{eq:fitform})
give the leading dependence on the light quark mass, parametrized by
$M_\pi^2$ divided by the chiral scale $\Lambda_\chi$, which we set to be
1 GeV.  The coefficient of the chiral logs depends on the
$D^*D\pi$ coupling $g$, which we take as $0.53(8)$ following
\cite{Bailey:2014tva}, and on the pion decay constant in the physical pion mass
limit $f = 130$ MeV. The $D^*-D$ mass splitting,
$\Delta_{m_c}$, appearing in the chiral logs is taken as 142 MeV. The
uncertainties from $f$ and $\Delta_{m_c}$ are negligible compared to
the error on $g$ and are not included.
Further details about the staggered chiral perturbation theory
\cite{SCHIPT} input to (\ref{eq:fitform}) are given in Appendix~\ref{chipt}.
We will return to discuss $\delta_a^g$ shortly.

The fourth term in (\ref{eq:fitform}) is present in the fit since
the current matching has truncation errors of $O(\alpha_s^2)$.
The truncated term would have some mild dependence on $am_b$, which
is reflected in the ansatz for this term.

The $\delta_a^B$ and $\delta_a^g$ in (\ref{eq:fitform}) parametrize
how discretization and quark mass tuning errors could enter the fit
form.  These originate from the gauge action, the NRQCD action and the
HISQ action. In all three actions discretization errors appear as even
powers of $a$, hence we include multiplicative factors
\begin{equation}
\delta_{a} = \big( b_0+ b_1 (a\Lambda_{\subrm{QCD}} )^2 + b_2 (a\Lambda_{\subrm{QCD}} )^4 +  b_3 (a\Lambda_{\subrm{QCD}})^6 \big) \,.
\end{equation}
Each factor $b_i$ contains a distinct sea quark tuning error dependence
\begin{equation}
b_i = \kappa_i \delta_b^i \delta_c^i \delta_\text{sea}^i
\end{equation}
where the $\kappa_i$ are given a Gaussian prior $0(0.5)$.  Note that we do not include a $\kappa_0$ term for the $\mathcal{O}(a^0)$ piece as such a term would not represent a mistuning error or discretization effect. The product on the
right-hand side allows for effects of small mistunings in the sea quark masses
and the valence charm and bottom quark masses.
 For the sea $u/d$ and $s$ quarks we include a multiplicative factor 
\begin{equation}
\delta_{sea} = 1+ c_1 (\delta x_{\text{sea}} /m_\text{sea}^\text{phys} )+ c_2 (\delta x_{\text{sea}}/m_\text{sea}^\text{phys} )^2
\end{equation}
where $m_\text{sea} = 2m_l+m_s$ and $\delta x_{\text{sea}} = m_\text{sea} - m_\text{sea}^\text{phys}$. The physical masses are taken from \cite{Bmeson2}  and are computed using the $\eta_s$ mass. We take $m_l^\text{phys}/m_s^\text{phys} = 27.4$ \cite{lightmassratio}. We also include the multiplicative factor
\begin{equation}
\delta_{c} = 1+ d_1 (\delta m_c /m_c^\text{phys} )+ d_2(\delta m_c /m_c^\text{phys} )^2
\end{equation}
where $\delta m_c = m_c - m_c^\text{phys}$, with physical mass taken from \cite{PhysMasses}, and the factor
\begin{equation}
\delta_{b} = 1+ f_1 (\delta m_b /m_b^\text{phys} )+ f_2(\delta m_b /m_b^\text{phys} )^2
\end{equation}
with $\delta m_b = m_b - m_b^\text{phys}$ where $m_b^\text{phys}$ is
determined from the spin-averaged kinetic mass of the $\Upsilon$ and
$\eta_b$\cite{latticespacing}. $c_i$, $d_i$, and $f_i$ are given prior values
of $0(0.5)$. We neglect the effects of the very small mistuning of the light quark masses from their physical value which we expect to be small.

Finite volume corrections to the staggered chiral perturbation theory are given in \cite{SCHIPT}. Evaluating these expressions on our lattices, we have found that finite volume effects are at least an order of magnitude smaller than the leading $\mathcal{O}(\alpha_s^2)$ error on the unphysical lattices. On sets 3, 6 and 8 the finite volume effects are larger, around half a percent in size. This is significant at the order to which we work. To account for these effects we subtract the finite volume correction to $h_{A_1}(1)$ from our data for these ensembles. We further discuss finite volume effects in Appendix \ref{chipt}.

\begin{figure}
  \centering
  \includegraphics[width=0.48\textwidth]{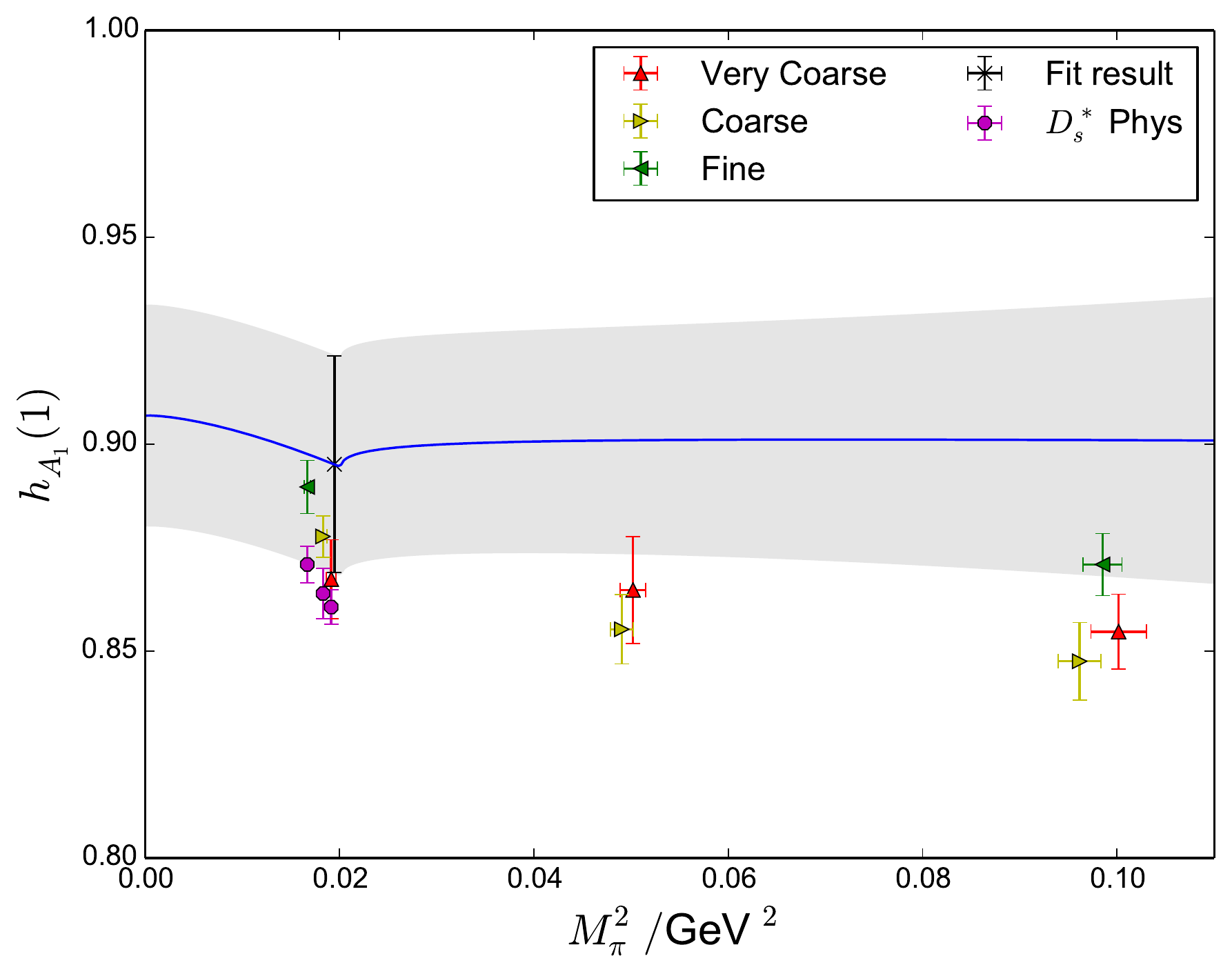}
  \caption{\label{fig:BDstar_mpisq}Fit to our data using staggered chiral perturbation theory. Finite volume corrections are included in the data points, visible only for the physical pion mass points. The blue line and grey band are the continuum chiral perturbation theory result and error extrapolated from our lattice data. The error band includes systematic errors coming from matching uncertainties and hence has a much larger error than any of the data points, which are only shown with their statistical error.}
\end{figure}

\begin{figure}
    \centering
    \includegraphics[width=0.5\textwidth]{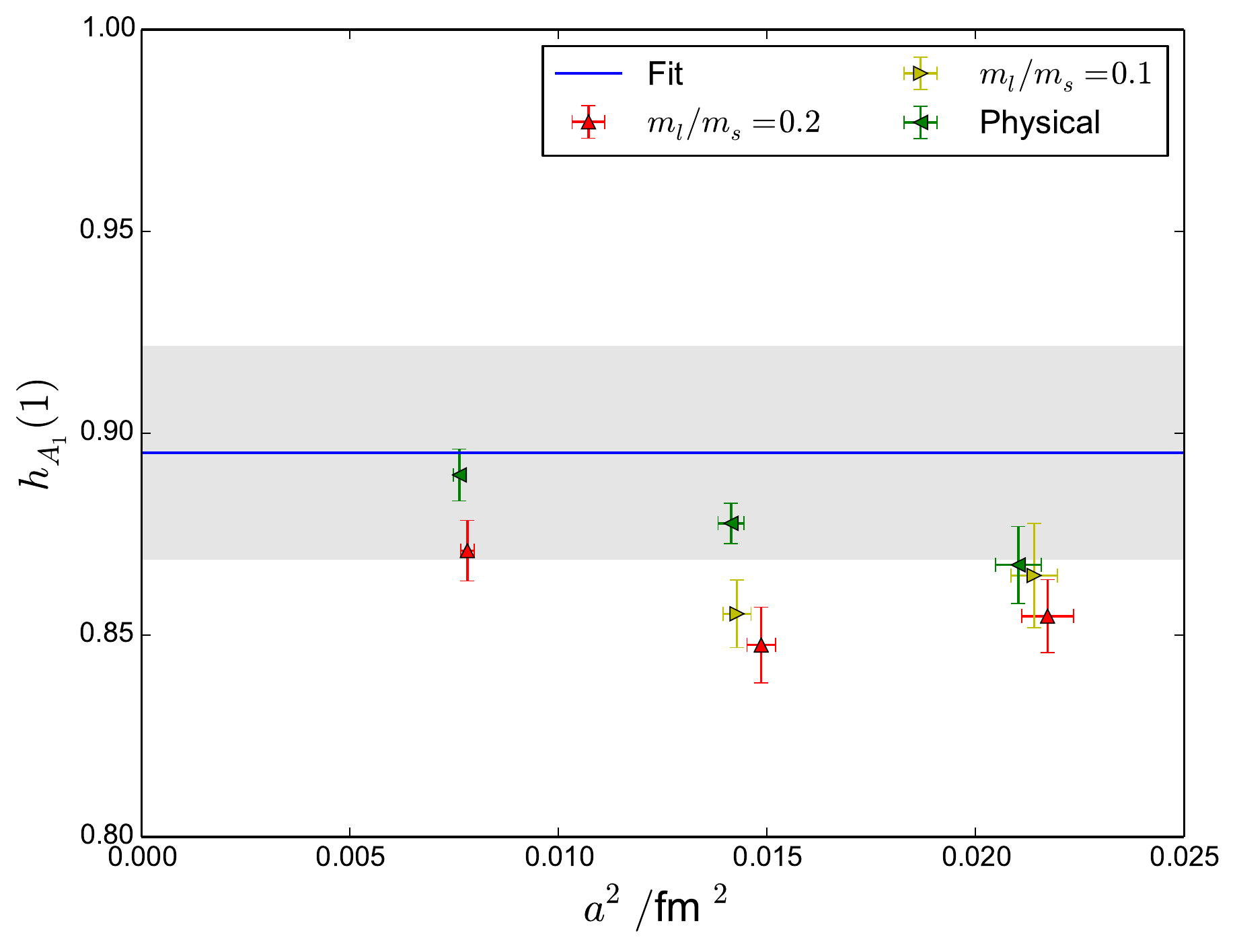}
  \caption{\label{fig:BDstar_asq}Plot showing the $a^2$ dependence of our $B\to D^*$ data. Finite volume corrections are included in the data points, visible only for the physical pion mass points. The blue line with grey error band shows the physical result for the form factor determined by the fit described in the text.}
\end{figure}

\begin{figure}
    \centering
    \includegraphics[width=0.514\textwidth]{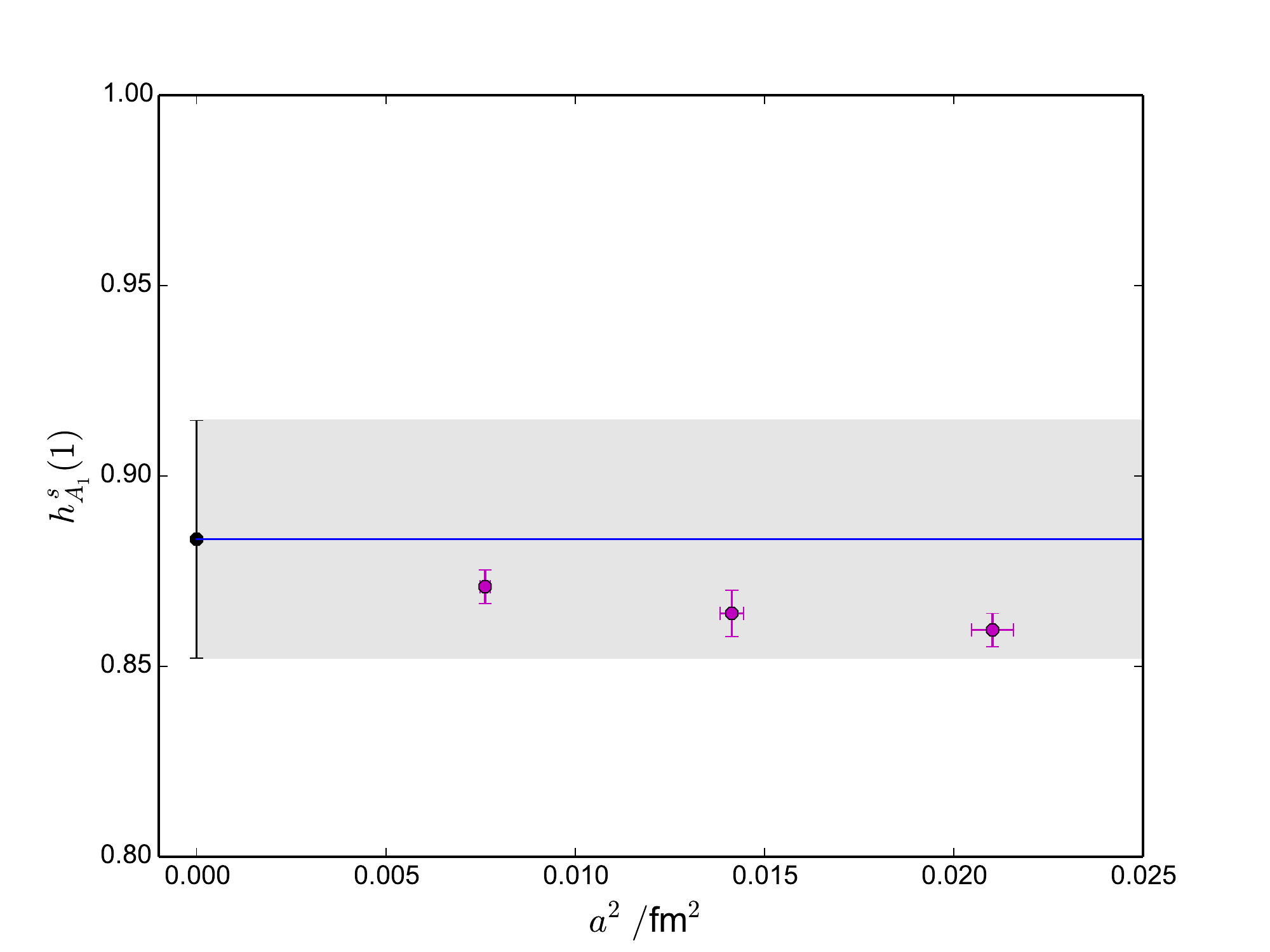}
  \caption{\label{fig:BsDsstar_asq}Lattice spacing dependence of our results for the $B_s\rightarrow D_s^*$ zero recoil form factor. The blue line with grey error band shows the physical result for the form factor determined by the fit described in the text.}
\end{figure}

The calculation on each ensemble of the form factor for $B_s \rightarrow
D_s^*$ decay is equivalent to the $B\rightarrow D^*$ calculation, with the
light quark propagator replaced with a strange quark propagator. The analysis
is substantially more straightforward, both because the data is less noisy and
because no chiral extrapolation is required.  Before fitting the lattice data,
we include a term to account for the absence of
$O(\Lambda_{\subrm{QCD}}^2/m_b^2)$ and $O(\alpha_s \Lambda_{\subrm{QCD}}/m_b)$ effects, as in
(\ref{eq:HO_HQET}), using the same Gaussian variables $e_4$, $e_5$, $e_6$,
$e_7$, $e_8$, and $e_9$.

For the continuum-chiral fit to the $h_{A_1}^s\!(1)$ we take the
functional form to be the following, where $\delta^{sB}_a B^s$ has the same form and priors as the term included for the $B\rightarrow D^*$:
\begin{align}
&\left.h^s_{A_1}\!(1)\right|_{\mathrm{fit}} = (1 + B^s) \delta_a^{sB} \nonumber \\
&~~+\gamma_1 \alpha_s^2\Big[1+\frac{\gamma_5}{2}(am_b-2)
    +\frac{\gamma_6}{4}(am_b-2)^2\Big] V^{(0)}
  \label{eq:fitform_s}
\end{align}
where $\gamma_1$, $\gamma_5$ and $\gamma_6$ are the same as in (\ref{eq:fitform}) because these terms represent the same higher order 
matching corrections. We run the $B_s \rightarrow D^*_s$ fit simultaneously with the $B \rightarrow D^*$ fit.

The NRQCD and HISQ systematics are the same as before, and we expect negligible isospin breaking and finite volume effects.  In Figure \ref{fig:BDstar_mpisq} we show the $M_\pi ^2$ dependence of our $B\rightarrow D^*$ data and the extrapolated continuum chiral form.

\begin{table*}
\centering
\caption{Results for parameters in the chiral-continuum fits, Eq.~(\ref{eq:fitform}) and
  (\ref{eq:fitform_s}). Higher order terms retain their prior values and are not shown while $\kappa_2^B = -0.17(25)$ and $\kappa_2^B = -0.05(42)$ for $h_{A_1}\!(1)$ and $h^s_{A_1}\!(1)$ respectively.}
\begin{tabular}{ccccccccc}\hline\hline 
 & & $c_1$ & $c_2$ & $d_1$ &$d_2$ &$f_1$& $f_2$ \\
 \hline 
 $h_{A_1}\!(1) $&$\delta^B_{a^0}$ &$-0.15(12)$&0.27(29)&0.24(40)&0.0(5)&0.24(40)&0.0(5)\\
 \hline 
 $h^s_{A_1}\!(1) $&$\delta^B_{a^0}$ &$-0.03(22)$&0.05(35)&0.0(5)&0.0(5)&0.0(5)&0.0(5)\\
\\ \hline
&&$B$&$C$&$g$&$\gamma_1$&$\gamma_5$&$\gamma_6$\\
 \hline 
$h_{A_1}\!(1)$ &&$-0.091(27)$&$-0.02(24)$&0.521(78)&$-0.14(44)$&0(1)&-0.15(97)\\
$h^s_{A_1}\!(1)$ &&$-0.117(31)$&-- &-- &$-0.14(44)$&0(1)&-0.15(97)\\
\hline\hline
\label{chiralcontfitparams}
\end{tabular}
\end{table*}

We present results for the $h_{A_1}\!(1)$ and $h^s_{A_1}\!(1)$ fit parameters $B$,
$\gamma_i$, $\kappa_i$, $c_i$, $d_i$, $f_i$ in Table \ref{chiralcontfitparams}. Plots showing the $a^2$ dependence of our $B\rightarrow D^*$ and $B_s \rightarrow D_s^*$ data are shown in Figures \ref{fig:BDstar_asq} and \ref{fig:BsDsstar_asq} respectively, together with the result of our fit. The $\mathcal{O}(a^4)$ and $\mathcal{O}(a^6)$ parameters default to their prior values, while the $\mathcal{O}(a^2)$ parameters are consistent with zero. We tried various modifications to our fit, the results of which we present in Appendix~\ref{chipt}.
Table~\ref{tab:partial_errors} presents a summary and combination of the
uncertainties in our results for $h_{A_1}\!(1)$ and $h^s_{A_1}\!(1)$.

\begin{table}
 \caption{\label{tab:partial_errors}Partial errors (in percentages) for
   $h_{A_1}^{(s)}(1)$.  A full accounting of the breakdown of systematic errors is made difficult by the fact that smaller priors not well constrained by the data are mixed in a correlated way by the fitter; these are reflected in the total systematic uncertainty. Note that the uncertainty from missing
   $\alpha_s^2$ terms in the matching for $h_{A_1}\!(1)$ and $h^s_{A_1}\!(1)$ is constrained somewhat by the fit; a naive estimate would give $3.5\%$ on the
   fine lattices.}
 \centering
 \begin{tabular}{c c c c} 
\\ \hline\hline
Uncertainty & $h_{A_1}\!(1)$ & $h^s_{A_1}\!(1)$ & $h_{A_1}\!(1)/h^s_{A_1}\!(1)$
    \\ [0.5ex]  \hline
$\alpha_s^2$& 2.1 & 2.5 &0.4\\
$\alpha_s \Lambda_{\subrm{QCD}}/m_b$&      0.9 & 0.9&0.0 \\
$( \Lambda_{\subrm{QCD}}/m_b)^2$&      0.8 & 0.8&0.0 \\
 $a^2$&      0.7 & 1.4 &1.4\\
$g_{D^*D\pi}$ & 0.2 & 0.03 &0.2\\
\hline
Total systematic & 2.7 & 3.2 & 1.7\\
\hline
Data &     1.1  & 1.4 &1.4\\
\hline
Total & 2.9 & 3.5 & 2.2\\
\hline\hline
\end{tabular}
\end{table}

\subsection{Isospin breaking effects}
\label{subsec:isospin}

The effects of electromagnetic interactions and $m_u\ne m_d$ on $h_{A_1}\!(1)$ are negligible compared 
to the dominant uncertainties quoted in Table~\ref{tab:partial_errors}.  
We find only a variation of $0.25\%$ in the chiral-continuum fits to $h_{A_1}\!(1)$ whether the  
$\pi^0$ or $\pi^+$ mass is used as the input value for the physical limit.  Electroweak and Coulomb
effects in the decay rate (\ref{eq:dGdwF}) are presently accounted for at leading order by a single multiplicative factor $\bar\eta_{EW}$ to be discussed below in Sec.~\ref{sec:Vcb}.  As lattice QCD uncertainties
are reduced in the future, it will be desirable to more directly calculate the effects of electromagnetism in a 
lattice QCD+QED calculation, where $m_u \ne m_d$ can also be implemented.

\section{Results and Discussion}
\label{sec:results}

We have calculated the zero recoil form factor for $B \rightarrow D^* \ell \nu $ decay using the most physically realistic gluon field configurations currently available along with quark discretizations that are highly improved. Our final result for the form factor, including all sources of uncertainty, is
\begin{equation}
\mathcal{F}^{B\to D^*}\!(1) = h_{A_1}\!(1) = 0.895(10)_\text{stat}(24)_\text{sys}\,.
\end{equation}
It is clear from this treatment that the dominant source of
uncertainty is the $\mathcal{O}(\alpha_s^2)$ uncertainty coming from the
perturbative matching calculation. In principle this could be reduced
by a two-loop matching calculation; however, such calculations in lattice
NRQCD have not been done before. It is worth noting that for our calculation this uncertainty is somewhat constrained by the fit, as is reflected in Table \ref{tab:partial_errors}. It has also been suggested \cite{ANDREWlatt2016} that it
could be estimated using heavy-HISQ $b$ quarks on `ultrafine' lattices
with $a=0.045~\text{fm}$ and $m_ba<1$. There we can use the nonperturbative
PCAC relation and the absolute normalization of the pseudoscalar
current to normalise $J^{(0)}$, using $(m_b+m_c) \hat{P} = Z
\partial_\mu \hat{A}^\mu$ to find the matching coefficient $Z$ and
then comparing matrix elements of this normalized current to the
result using perturbation theory.

Within errors, our result agrees with the result from the Fermilab
Lattice and MILC Collaborations \cite{Bailey:2014tva},
$h_{A_1}\!(1) = 0.906(4)(12)$.  The higher precision achieved in this work is
due to the use of the same lattice discretization for the $b$ and $c$ quarks.
This enabled them to avoid the larger current-matching uncertainties
present in our NRQCD-$b$, HISQ-$c$ work.  Nevertheless, the value of
providing a completely independent lattice QCD result using different formalisms is self-evident.

After combining the statistical and systematic errors in quadrature, a
weighted average of the two lattice results yields $h_{A_1}\!(1) = 0.904(12)$.
We use this value in our discussion in Sec.~\ref{sec:Vcb}.

Our result for the $B_s\rightarrow D_s^*$ zero-recoil form factor is
\begin{align}
  \mathcal{F}^{B_s\to D_s^*}\!(1) = h_{A_1}^s\!(1)
  = 0.883(12)_\text{stat}(28)_\text{sys} \,.
\end{align}
This is the first lattice QCD calculations of this quantity. We see no significant difference between the result for $B\rightarrow D^*$ and $B_s\rightarrow D_s^*$ showing that spectator quark mass effects are very small.  Correlated systematic uncertainties
cancel in the ratio, which we find to be
\begin{align}
  \frac{\mathcal{F}^{B\to D^*}\!(1)}{\mathcal{F}^{B_s\to D_s^*}\!(1)} =
\frac{h_{A_1}\!(1)}{h_{A_1}^s\!(1)} = 1.013(14)_\text{stat}(17)_\text{sys} \,.
\end{align}
We find there to be no significant $U$-spin
($d\leftrightarrow s$) breaking effect at the few percent level.

\section{Implications for $|V_{cb}|$}
\label{sec:Vcb}

Until recently, one would simply combine a world average of lattice
data for $h_{A_1}\!(1)$ with the latest HFLAV result
for the $\bar{B}^0\to D^{*+}\ell^-\nu$ differential branching fraction
extrapolated to zero recoil:
$\bar{\eta}_{EW} \mathcal{F}(1) |V_{cb}|= 35.61(11)(44)\times 10^{-3}$
\cite{Amhis:2016xyh}.  Doing so with the
weighted average of the Fermilab/MILC result and ours yields 
\begin{equation}
  |V_{cb}|_{\subrm{HFLAV}} = (38.9 \pm 0.7)\times 10^{-3} \,,
\end{equation}
where we have used the estimated charge-averaged value of
$\bar{\eta}_{EW} = 1.015(5)$ \cite{Bailey:2014tva}.  The uncertainty
in $|V_{cb}|_{\subrm{HFLAV}}$ is due in equal parts to lattice and
experimental error.

Recent work analyzing unfolded Belle data \cite{Abdesselam:2017kjf}
has called into question the accuracy of what has become the standard
method of extrapolating experimental data to zero recoil
\cite{Bernlochner:2017jka,Bigi:2017njr,Grinstein:2017nlq,Bigi:2017jbd,Jaiswal:2017rve,Bernlochner:2017xyx}. In
order to understand our new result for $h_{A_1}\!(1)$, as well as to
prepare for future lattice calculations and experimental measurements,
we carry out a similar analysis here.  We generally agree with
conclusions already in the literature, but we present a few of our own
suggestions for how one could proceed in the future.

The method used by experiments to date is due to Caprini, Lellouch, and
Neubert (CLN) \cite{Caprini:1997mu}.  Their paramatrization of the form
factors entering the differential decay rate and angular observables
is an expansion about zero-recoil, i.e.\ about $w=1$. (See Appendix
\ref{app:exptfits} for expressions relating experimental observables to form factors.)  In the case of the $h_{A_1}\!(w)$ form factor it was found that
the kinematic variable $z$ gives a more convergent series. Given a
specific choice of $t_0$, 
$z$ depends on the $t=q^2$ as
\begin{align}
  z(t,t_0) = \frac{\sqrt{t_+ - t} - \sqrt{t_+-t_0}}{\sqrt{t_+ - t} +
    \sqrt{t_+-t_0}}
\end{align}
with $t_\pm = (M_B \pm M_{D^*})^2$. Usually one takes $t_0 = t_-$, and
this is the choice assumed throughout this paper.\footnote{One can
  express $z(t,t_-)$ as a function of $w$ as
\[
  z(w) = \frac{\sqrt{w+1} - \sqrt{2}}{\sqrt{w+1} + \sqrt{2}} \,.
\]}

The CLN form factors are given as follows
\begin{align}
  h_{A_1}\!(w) & = h_{A_1}\!(1)[1 - 8 \rho^2 z + (r_{h2r} \rho^2 + r_{h2})z^2
    \nonumber \\
   &\hspace{15mm} + (r_{h3r}\rho^2 + r_{h3})z^3] \nonumber \\    
  R_1(w) & = R_1(1) + r_{11}(w-1) + r_{12}(w-1)^2 \nonumber\\
  R_2(w) & = R_2(1) + r_{21}(w-1) + r_{22}(w-1)^2
  \label{eq:CLNseries}
\end{align}
with the coefficients computed to be \cite{Caprini:1997mu}
\begin{alignat}{4}
  r_{h2r} & = 53\,,  ~&~ r_{h2} & = -15\,, \nonumber \\
  r_{h3r} &=  -231\,, ~&~ r_{h3} &= 91\,, \nonumber \\
  r_{11} & = -0.12\,, ~&~ r_{12}  &= 0.05\,,  \nonumber \\
  r_{21} & = 0.11\,, ~&~ r_{22}  &= -0.06\,.
\label{eq:rcoeff}
\end{alignat}
These numbers are the result of a calculation in HQET, using QCD
sum rules and neglecting
contributions of $\alpha_s\Lambda_{\subrm{QCD}}/m_c$ and 
$(\Lambda_{\subrm{QCD}}/m_c)^2$, as well as smaller effects.
Until recently effects of neglecting these terms have not been included in
fitting the experimental data.  

Ref.~\cite{Caprini:1997mu} claims an accuracy of $2\%$; however this
is based on comparing an expansions in $z$ against some full expressions.
While this tests the convergence of the expansions, it does not test the
accuracy of numerical factors computed in truncated HQET.  In fact
the data do not require any higher order terms in $z$ or $w-1$.  We
found no effect when including a $z^4$ term or $(w-1)^3$ terms in
(\ref{eq:CLNseries}) with Gaussian priors allowing the coefficient $r_{h4}$
to be up to $O(10^3)$ and $r_{13}$, $r_{23}$ to be up to $O(1)$.

\begin{table}
\caption{\label{tab:CLNfits}Fits to the unfolded Belle data using the
  CLN parametrization.  The first fit does not account for any
  uncertainties in the $r$ coefficients (\ref{eq:rcoeff}).  The next
  three include the $r$ coefficients as Gaussian priors with widths of
  10\%, 20\% or 100\% uncertainties, respectively. The final two fits
  assign $10\%$ or $20\%$ uncertainty to the coefficients in
  $h_{A_1}\!(w)$ and allow the coefficients of $R_1(w)$ and $R_2(w)$ to
  be $O(1)$.}
\begin{tabular}{ccccc}\hline\hline
fit & $I$ & $\rho^2$ & $R_1(1)$ & $R_2(1)$ \\ \hline 
0\% & 0.0348(12) & 1.17(15) & 1.386(88) & 0.912(76)    \\
10\% & 0.0349(13) & 1.19(16) & 1.387(88) & 0.914(76)   \\ 
20\% & 0.0352(13) & 1.24(19) & 1.390(88) & 0.922(78)  \\
100\% & 0.0367(16) & 1.64(31) & 1.397(94) & 0.941(96)   \\
$h$:10\%, $R$:$0(1)$ & 0.0359(14) & 1.29(17) & 1.19(22) & 1.05(18)    \\ 
$h$:20\%, $R$:$0(1)$ & 0.0359(14) & 1.31(19) & 1.19(22) & 1.04(19)  \\
\hline\hline
\end{tabular}
\end{table}

Nevertheless none of this accounts for higher order terms in the HQET.
We can get some idea of how the fit is affected by allowing the $r$
coefficients (\ref{eq:rcoeff}) to be fit parameters with Gaussian
priors, with means equal to the CLN values but with widths which we
vary. Table~\ref{tab:CLNfits} shows the results of fitting to the CLN
parametrization.  We present six variations, which we describe below.
In order to infer $|V_{cb}|$ from the lattice $h_{A_1}\!(1)$ and the
fit to data, the main output is the combination
\begin{align}
  I & = |\bar{\eta}_{EW} V_{cb}| \,h_{A_1}\!(1) \,.
  \label{eq:intercept}
\end{align}

In the first fit, we treat the $r$-coefficients (\ref{eq:rcoeff})
as pure numbers; this has been the standard treatment until recently.
The value of $I$ we obtain agrees with the unfolded fit result
of Belle \cite{Abdesselam:2017kjf}, $I = 34.9(1.5)$.

It would be better to include estimates of HQET truncation errors in
the fits.  We implement this by treating the $r$-coefficients as fit
parameters, adding Gaussian priors with central values as in
(\ref{eq:rcoeff}) and with widths equal to our uncertainty.
Unfortunately it is not clear how accurately these are known at this
order in HQET.  We note both $\alpha_s \Lambda_{\subrm{QCD}}/m_c$ 
and $(\Lambda_{\subrm{QCD}}/m_c)^2$ 
are roughly 0.1, so one approach
is to suggest truncated terms could vary each of the $r$'s by 10\%.
However, some linear algebra has been done after truncating HQET
expressions to arrive at the form factors (\ref{eq:CLNseries}).  This
could enhance (or suppress) the truncation error in some terms, and
the opposite in others.  The fit does not change much if the
uncertainties are $20\%$, but 100\% uncertainties in (\ref{eq:rcoeff})
do affect the fit result.  Most notably, the value of $I$ increases by
5\%, i.e.\ one standard deviation.

The smallness of the coefficients in the expansions of $R_1$ and $R_2$
is likely due to cancellations in the expansions when ratios are taken.
Therefore, assuming a relative error on the $r_{ij}$  ($i,j=1,2$) is
probably not correct.  We present two fits where these coefficients are
given Gaussian priors equal to $0 \pm 1$, while the coefficients in the
expansion of $h_{A_1}\!(w)$ are given 10\% or 20\% uncertainties.  The resulting
values for $I$ lie in between the tightly constrained fits and the $100\%$
uncertainty fit.

Note that the HQET predicts $R_1(1)= 1.27$ and $R_2(1) = 0.80$,
but in most fits in the literature (as here) these are treated as free fit
parameters.  In fact the world average fit values differ from the HQET
estimates: Belle's world averages are $R_1(1)=1.40(3)$ and
$R_2(1)=0.85(2)$ \cite{Abdesselam:2017kjf}.

The fact that the tightly constrained CLN fits describe the data well,
with good $\chi^2$ for example, is a success for HQET.  It shows that
the important physics has been captured within the accuracy of the
theory.  However, now that we are in the high precision era of flavour
physics, we ought to be wary about the accuracy of the assumptions
which go into fitting the data.  The observation that $I$ increases
under a relaxation of assumptions about the $r$-coefficients agrees
with other authors' findings
\cite{Bernlochner:2017jka,Bigi:2017njr,Grinstein:2017nlq,Bigi:2017jbd,Jaiswal:2017rve,Bernlochner:2017xyx}.

An alternative parametrization for the hadronic form factors is the
one proposed by Boyd, Grinstein, and Lebed (BGL) \cite{Boyd:1997kz}.
In their conventions the three form factors entering (assuming the
lepton mass can be neglected) are $f(q^2)$, $F_1(q^2)$, and $g(q^2)$.  
Two of the form factors are kinematically constrained at $q^2 = 0$:
$F_1(0) = (M_B-M_{D^*})f(0)$.  Each
of these is expanded in a Taylor series about $z=0$ after factoring
out a function intended to account for nearby resonances. Abbreviating
$t = q^2$, form factors are parametrized by
\begin{align}
  F(t) & = Q_F(t) \sum_{k=0}^{K_F-1} a_k^{(F)} z^k(t,t_0) \,.
  \label{eq:zexpansion}
\end{align}
Throughout this paper we take $t_0 = t_-$. With appropriately chosen $Q_F$, 
\begin{align}
  Q_F(t) = \frac{1}{B_n(z) \phi_F(z)} \,,
 \label{eq:BGLprefactor}
\end{align}
the magnitudes of the coefficients $a_n^{(F)}$ are bounded by unitarity
constraints.
\begin{align}
  S_{fF} &= \sum_{k=0}^{K_f-1} [(a_k^{(f)})^2 + (a_k^{(F_1)})^2] \le 1\nonumber \\
  S_g & = \sum_{k=0}^{K_g-1} (a_k^{(g)})^2 \le 1\,.
  \label{eq:unitaryconstraint}
\end{align}
Even stronger bounds can be imposed if one is able to include all
the $B^{(*)} \to D^{(*)}$ matrix elements, with (pseudo)scalar and
(axial)vector initial and final states \cite{Bigi:2017jbd}, but this
is outside the scope of our analysis here.

\begin{table}
  \caption{\label{tab:BCspectrum}$B_c$ vector and axial vector masses
    below $BD^*$ threshold ($7.290$ GeV) used in the Blaschke factors.
  Mass differences \cite{Precisehlmesonmasses} are combined with 
  $M_{B_c} = 6.2749(8)$ \cite{Patrignani:2016xqp}. We adopt the model
  estimates of Ref.~\cite{Bigi:2017njr}, up to 3 digits.}
  \centering
  \begin{tabular}{ccc|ccc} \hline\hline
    $M_{1^-}/$GeV & method & Ref. & $M_{1^+}$/GeV & method & Ref.
    \\ \hline
    6.335(6) & lattice & \cite{Precisehlmesonmasses} &
    6.745(14) & lattice & \cite{Precisehlmesonmasses} \\
    6.926(19) & lattice & \cite{Precisehlmesonmasses} &
    6.75 & model & \cite{Devlani:2014nda,Godfrey:2004ya} \\
    7.02 & model & \cite{Devlani:2014nda} & 
    7.15 & model & \cite{Devlani:2014nda,Godfrey:2004ya}\\
    7.28 & model & \cite{Eichten:1994gt} &
    7.15 & model & \cite{Devlani:2014nda,Godfrey:2004ya}\\
    \hline\hline
  \end{tabular}
\end{table}

\begin{figure}
  \centering
\includegraphics[width=0.48\textwidth]{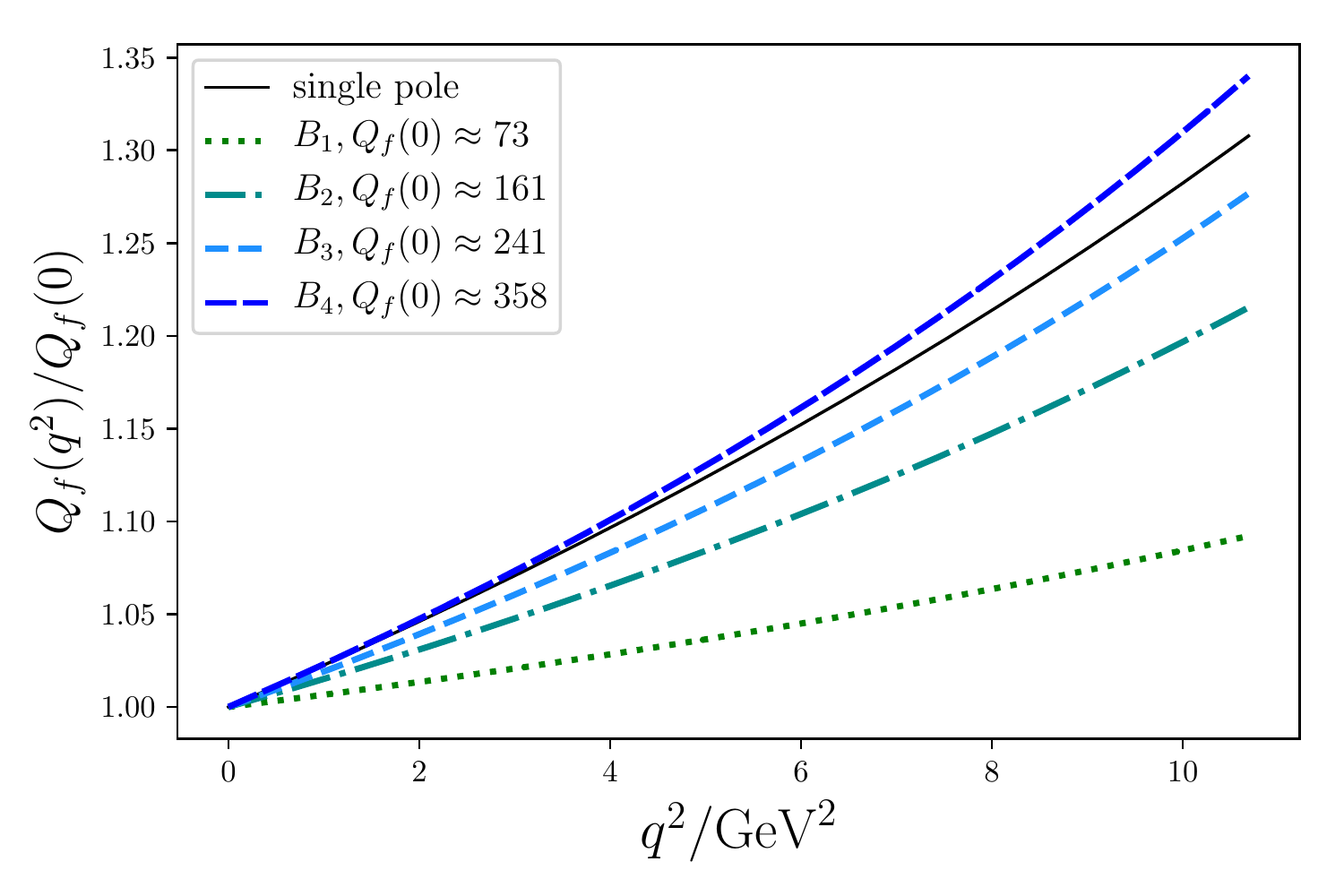}
\includegraphics[width=0.48\textwidth]{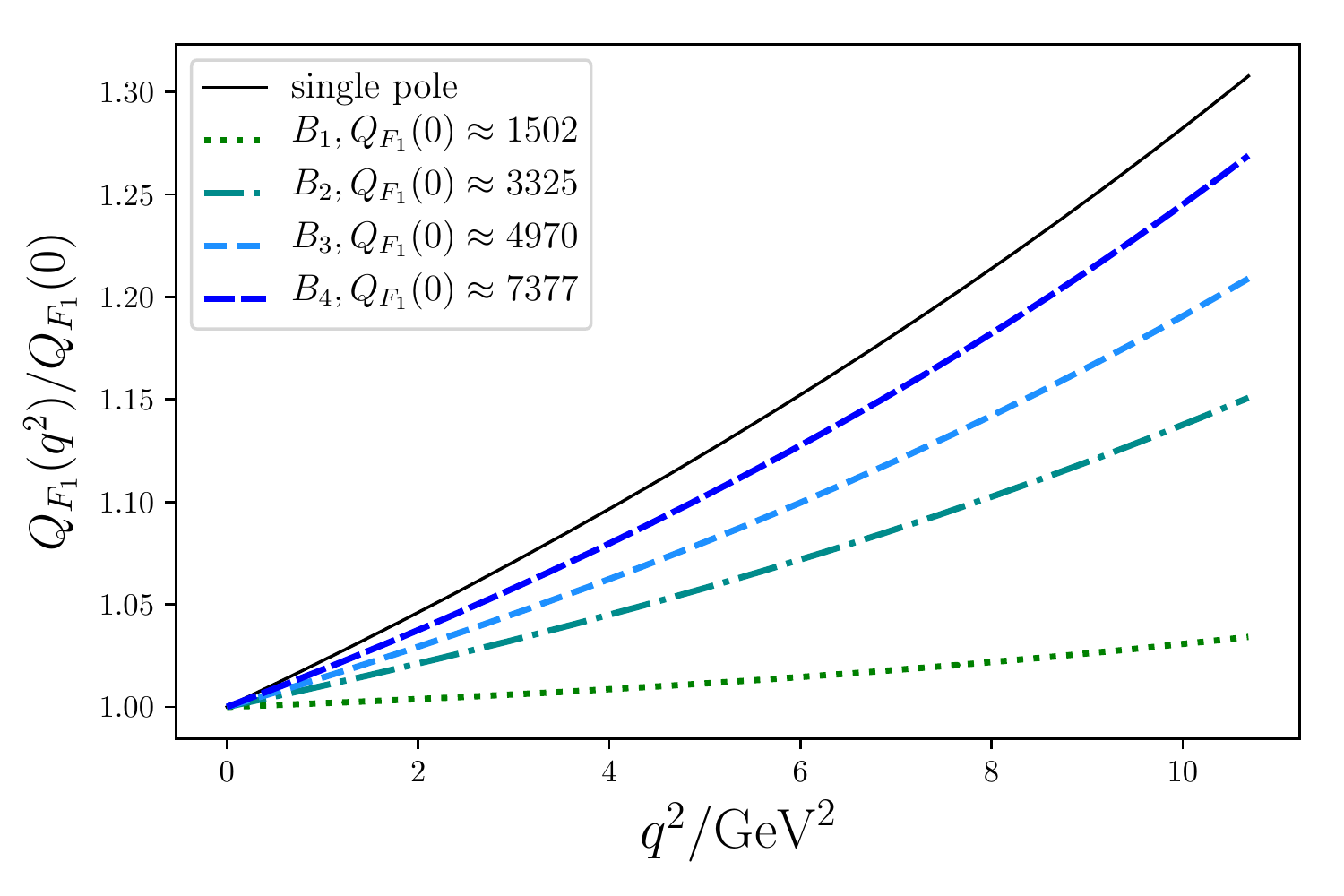}
\includegraphics[width=0.48\textwidth]{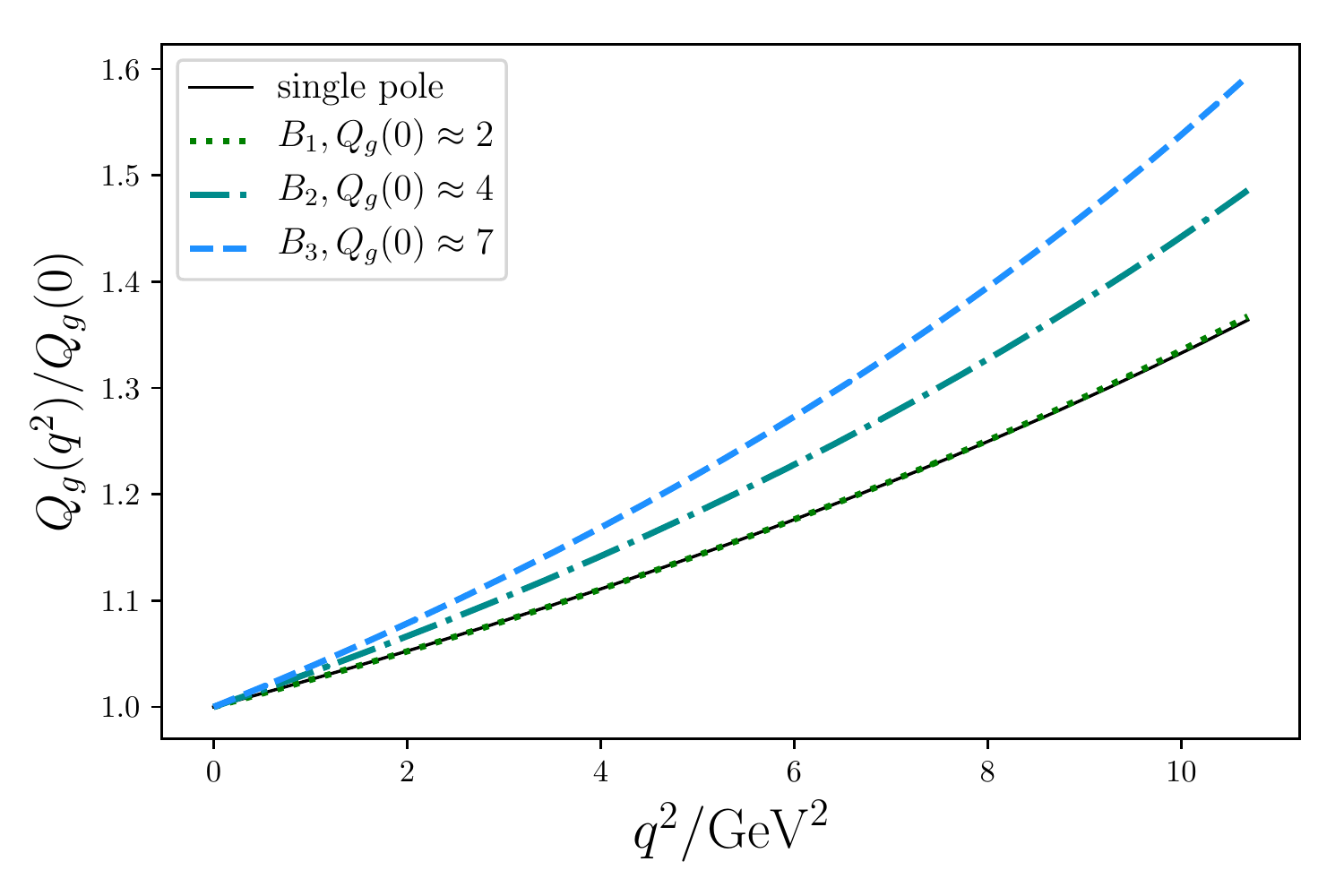}
\caption{\label{fig:poles}Comparison of the prefactor $Q_{F}(q^2)$ for the
BGL and BCL series expansions of form factor $F= f$, $F_1$, and $g$, from
top to bottom. Curves are normalized by $Q_F(0)$, which is given in the legend.}
\end{figure}

The two functions in (\ref{eq:BGLprefactor}) are the
outer functions $\phi_F(z)$, which can be found in the literature
e.g.\ in Refs.~\cite{Boyd:1997kz,Bigi:2017njr,Grinstein:2017nlq},
and the Blaschke factor
\begin{align}
  B_n(z) = \prod_{i=1}^{n} \frac{z - z_{P_i}}{1-z z_{P_i}}\,, ~~
  z_{P_i} = z(M_{P_i}^2, t_-) \,.
\end{align}
The product is over a set of applicable resonances, the vector
$B_c^*$ states for $g$ and the axial vector $B_c$ states for $f$ and
$F_1$. The resonances included in the Blaschke factor should be those with
the appropriate quantum numbers and below scattering threshold.  There
are 4 $B_c$ vector and 4 axial-vector states conjectured to be below
$BD^*$ threshold.  Table~\ref{tab:BCspectrum} lists calculations of
the vector and axial vector $B_c$ resonances.  The
model estimate for the mass of the heaviest vector state is very close to
threshold, so has been left out of several analyses, including here.
The magnitude of the Blaschke factors can be very sensitive to $n$, so
leaving states out reduces the strength of unitarity constraints.
This is illustrated in Fig.~\ref{fig:poles} for the $Q_F(q^2)$ for
$F = f$, $F_1$, and $g$.

\begin{table*}
  \caption{\label{tab:SEfits} Results of $z$-expansion fits
    (\ref{eq:zexpansion}), either using the BGL
    (\ref{eq:BGLprefactor}) or BCL (\ref{eq:BCLprefactor})
    parametrization.  Unitarity constraints are not enforced in the
    fit, but the sums $S_g$ and $S_{fF}$ (\ref{eq:unitaryconstraint})
    are given for reference (see text). The number of $1^+/1^-$ resonances
    included in the Blaschke factor is $n_B^+/n_B^-$.
    Terms up to $O(z^{K-1})$ are
    included in the fits.  Coefficients of higher order terms are
    consistent with zero.}
\begin{tabular}{ccccccccccccc}\hline\hline
fit & $n_B^+$ & $n_B^-$ & $K$ & $I$ & $a^{(f)}_0$ & $a^{(f)}_1$ & $a^{(F_1)}_0$ & $a^{(F_1)}_1$ & $a^{(g)}_0$ & $a^{(g)}_1$ & $S_{fF}$ & $S_g$ \\ \hline 
BGL & 2 & 2 & 2 & 0.0366(14) & $0.03005(39)$ & $-0.120(51)$ & $0.005031(65)$ & $-0.0146(40)$ & $0.032(15)$ & $0.88(50)$ & 0.015(12) & 0.78(89) \\ 
BGL & 2 & 2 & 3 & 0.0376(16) & $0.03004(39)$ & $-0.148(62)$ & $0.005031(65)$ & $-0.030(13)$ & $0.029(14)$ & $0.99(50)$ & 0.13(32) & 0.98(98) \\ 
BGL & 2 & 2 & 4 & 0.0376(16) & $0.03004(39)$ & $-0.148(62)$ & $0.005031(65)$ & $-0.030(13)$ & $0.029(14)$ & $0.99(50)$ & 0.13(33) & 0.98(98) \\ 
BGL & 3 & 3 & 2 & 0.0368(15) & $0.01913(25)$ & $-0.069(36)$ & $0.003204(41)$ & $-0.0073(27)$ & $0.0138(85)$ & $0.63(30)$ & 0.0052(49) & 0.40(38) \\ 
BGL & 3 & 3 & 3 & 0.0379(17) & $0.01913(25)$ & $-0.088(47)$ & $0.003204(41)$ & $-0.0181(86)$ & $0.0125(82)$ & $0.68(31)$ & 0.06(21) & 0.46(41) \\ 
BGL & 3 & 3 & 4 & 0.0379(17) & $0.01913(25)$ & $-0.088(47)$ & $0.003204(41)$ & $-0.0181(87)$ & $0.0125(82)$ & $0.68(31)$ & 0.06(22) & 0.46(42) \\ 
BGL & 4 & 3 & 2 & 0.0369(15) & $0.01228(16)$ & $-0.035(24)$ & $0.002057(26)$ & $-0.0032(18)$ & $0.0138(84)$ & $0.63(30)$ & 0.0014(17) & 0.39(38) \\ 
BGL & 4 & 3 & 3 & 0.0380(17) & $0.01228(16)$ & $-0.049(36)$ & $0.002057(26)$ & $-0.0102(57)$ & $0.0129(86)$ & $0.66(33)$ & 0.04(25) & 0.44(43) \\ 
BGL & 4 & 3 & 4 & 0.0380(17) & $0.01228(16)$ & $-0.049(36)$ & $0.002057(26)$ & $-0.0102(59)$ & $0.0129(85)$ & $0.66(33)$ & 0.04(25) & 0.44(42) \\ 
BCL & -- & -- & 2 & 0.0367(15) & $0.01502(19)$ & $-0.047(27)$ & $0.002946(38)$ & $-0.0029(27)$ & $0.028(13)$ & $0.78(44)$ & 0.0025(26) & 0.60(69) \\ 
BCL & -- & -- & 3 & 0.0378(17) & $0.01502(19)$ & $-0.066(40)$ & $0.002946(38)$ & $-0.0136(82)$ & $0.026(13)$ & $0.82(46)$ & 0.08(38) & 0.67(75) \\ 
BCL & -- & -- & 4 & 0.0382(18) & $0.01502(19)$ & $-0.311(42)$ & $0.002946(38)$ & $-0.0152(83)$ & $0.109(16)$ & $-0.29(38)$ & 0.144(67) & 0.10(22) \\ 
BCL & -- & -- & 5 & 0.0382(18) & $0.01502(19)$ & $-0.311(42)$ & $0.002946(38)$ & $-0.0152(83)$ & $0.109(16)$ & $-0.29(38)$ & 0.144(67) & 0.10(22) \\ 
\hline \hline
\end{tabular}
\end{table*}

Table~\ref{tab:SEfits} shows the results of BGL fits to the unfolded
Belle data \cite{Abdesselam:2017kjf}, varying the number of states
included in the Blaschke factor and the number of terms kept in the
$z$-expansion.  The fits enforce the $q^2=0$ constraint on $F_1(0)$ and
$f(0)$ at the $1\%$ level.
Priors for the coefficients $a_k^{(F)}$ are Gaussians
with mean 0 and standard deviation 1.  Only the $k=0$ and 1 coefficients
are tabulated; the others are not constrained by the data and remain
statistically consistent with 0. As discussed above, the magnitude
of these coefficients depends on the number of states in the Blaschke
factor. Nevertheless, the results for $I$ are insensitive to this.
On the other hand, $I$ does increase by about 0.001, or approximately
$0.7\sigma$, when switching from $K=2$ to higher order polynomials
in $z$. (Results remain the same for $K>4$.)

The fits presented in Table~\ref{tab:SEfits} do not enforce the
unitarity bounds (\ref{eq:unitaryconstraint}), but these bounds are
not close to being saturated unless only two resonances are included
in the Blaschke factors.  Performing the fits with the bounds enforced
did not significantly affect results.  Considering the $n_B > 2$, $K=4$ 
fits, increasing the standard deviation of
the Gaussian priors for the series coefficients $a_k^{(F)}$ by a factor of
2 or 4 had very little effect on the parameters well-determined by the
data, i.e.\ $I$, $a_0^{(f)}$, and $a_0^{(F_1)}$, while the uncertainties
in $a_1^{(f)}$, $a_1^{(F_1)}$, $a_0^{(g)}$, $a_1^{(g)}$, $S_{fF}$, $S_g$,
 increased by factors of 1.5-2.  For most of the $a_k^{(F)}$ with $k \ge 2$,
 the posterior distribution is the same as the prior, the exception being
 that $a_2^{(F_1)} \lesssim 0.5$ seems preferred by the fit, even with wide 
 prior widths.

To the extent that unitarity constraints do not affect the BGL
fits, then a simpler approach would be to represent $Q_{F}(t)$
by a simple pole, as in the simplified series expansion (BCL)
\cite{Bourrely:2008za}.  That is, one can parametrize the form factors
by (\ref{eq:zexpansion}) with 
\begin{equation}
  Q_F(t) = \frac{N_F}{1 - t/M_P^2} \,,
  \label{eq:BCLprefactor}
\end{equation}
with $M_P$ being the mass of the lightest resonance with the appropriate
quantum numbers.
The normalization $N_F$ can be chosen so that the series coefficients
are of the same order of magnitude as in a particular BGL expansion.
With Fig.~\ref{fig:poles} as a guide,
we take $N_f = 300$, $N_{F_1}=7000$, and $N_g=5$.  Once again we fit
with priors for $a_k^{(F)}$ equal to $0\pm 1$.  The results for $I$
show the same behaviour for the BCL fits as for the BGL fits.

The virtue of the BCL fit is in its simplicity.  The BGL fit requires
theory input for the outer functions $\phi_F$: perturbatively calculated
derivatives of two-point functions at $q^2=0$ and $n_I$, the number
of spectator quarks adjusted to account for $SU(3)_F$ breaking.
(In the BGL fits here we take the values given in Table 2 of
Ref.~\cite{Bigi:2017njr}.)  The Blaschke factor requires as input
model estimates for the excited $B_c$ resonances to include in the Blaschke
factor.  If unitarity bounds become tight enough to have an effect on
the fits to data, then the effects of theoretical assumptions needs to be
carefully included in the error analysis.  On the other hand, the BCL fits
only take as additional input the mass of a single resonance, available
to very good precision from lattice QCD.  In the future, fits to the
BCL simplified $z$-expansion could provide a clean, benchmark fit.

\begin{figure}
  \centering
  \includegraphics[width=0.48\textwidth]{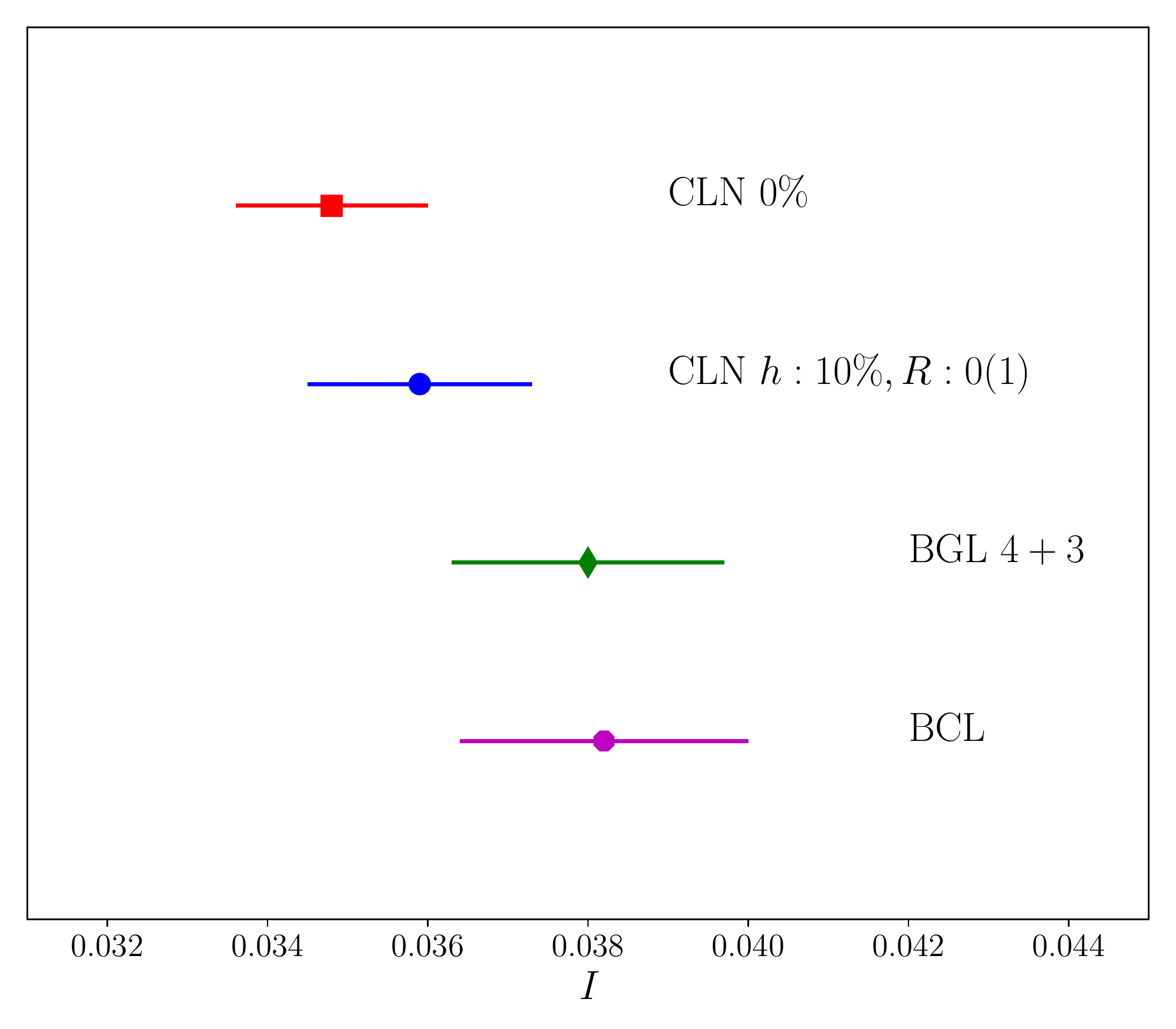}
  \caption{\label{fig:compare_I} Values of
    $I=|\bar{\eta}_{EW}V_{cb}|h_{A_1}\!(1)$ obtained from different fit
    ans\"atze (see text).}
\end{figure}

Fig.~\ref{fig:compare_I} summarizes the consequences to $I$ of different
fitting choices selected from Tables~\ref{tab:CLNfits} and \ref{tab:SEfits}.
The top two points show results from CLN fits including no uncertainties
on the coefficients (\ref{eq:rcoeff}), or 10\% errors on the $r_h$
coefficients and allowing the coefficients in the expansions of $R_{1,2}(w)$
to be $0\pm 1$. The bottom two points are respectively BGL and BCL fits with
$K=4$, and $n_B^+ = 4$, $n_B^-=3$ for the BGL fit.

\begin{figure*}
  \centering
  \includegraphics[width=0.49\textwidth]{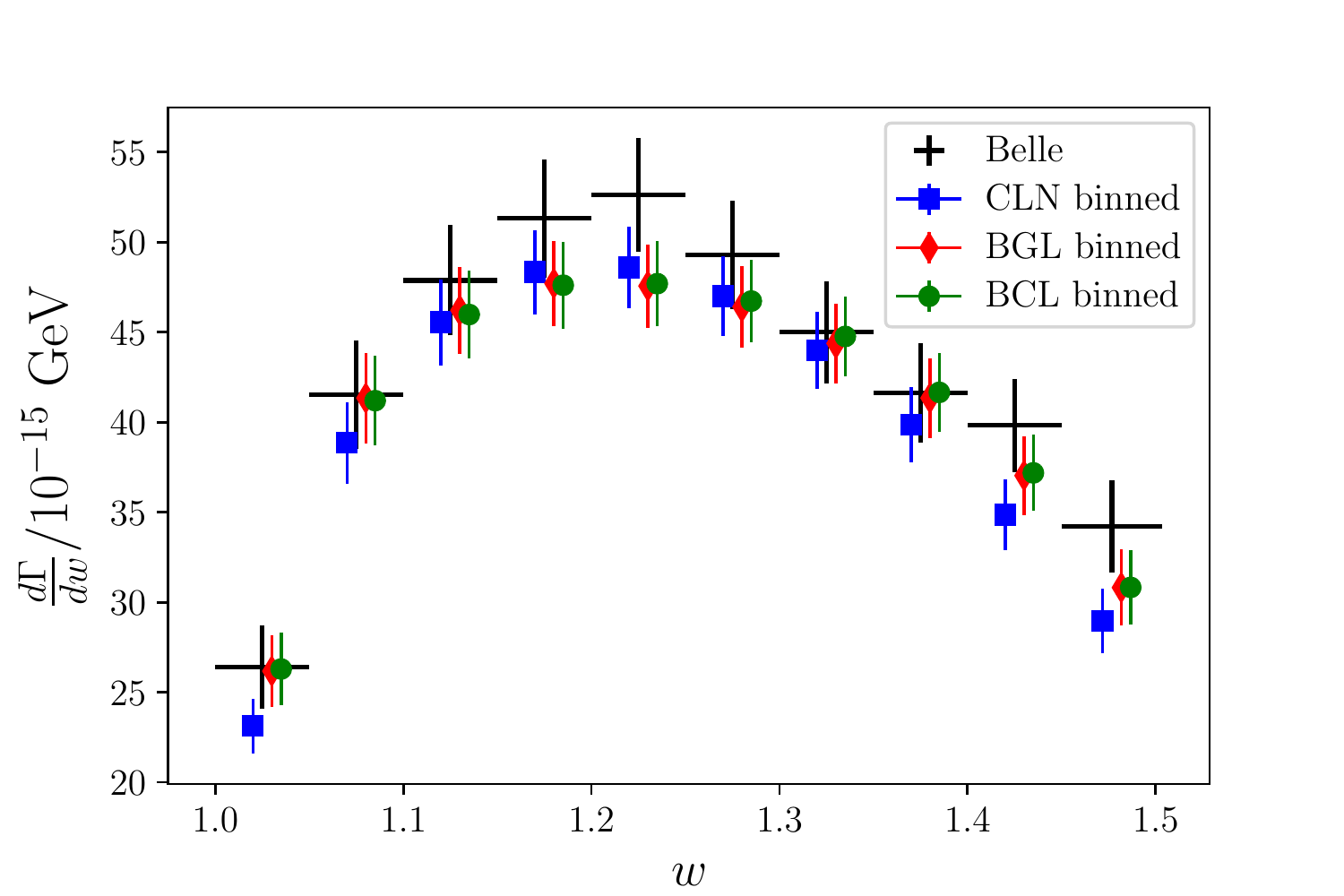}
  \includegraphics[width=0.49\textwidth]{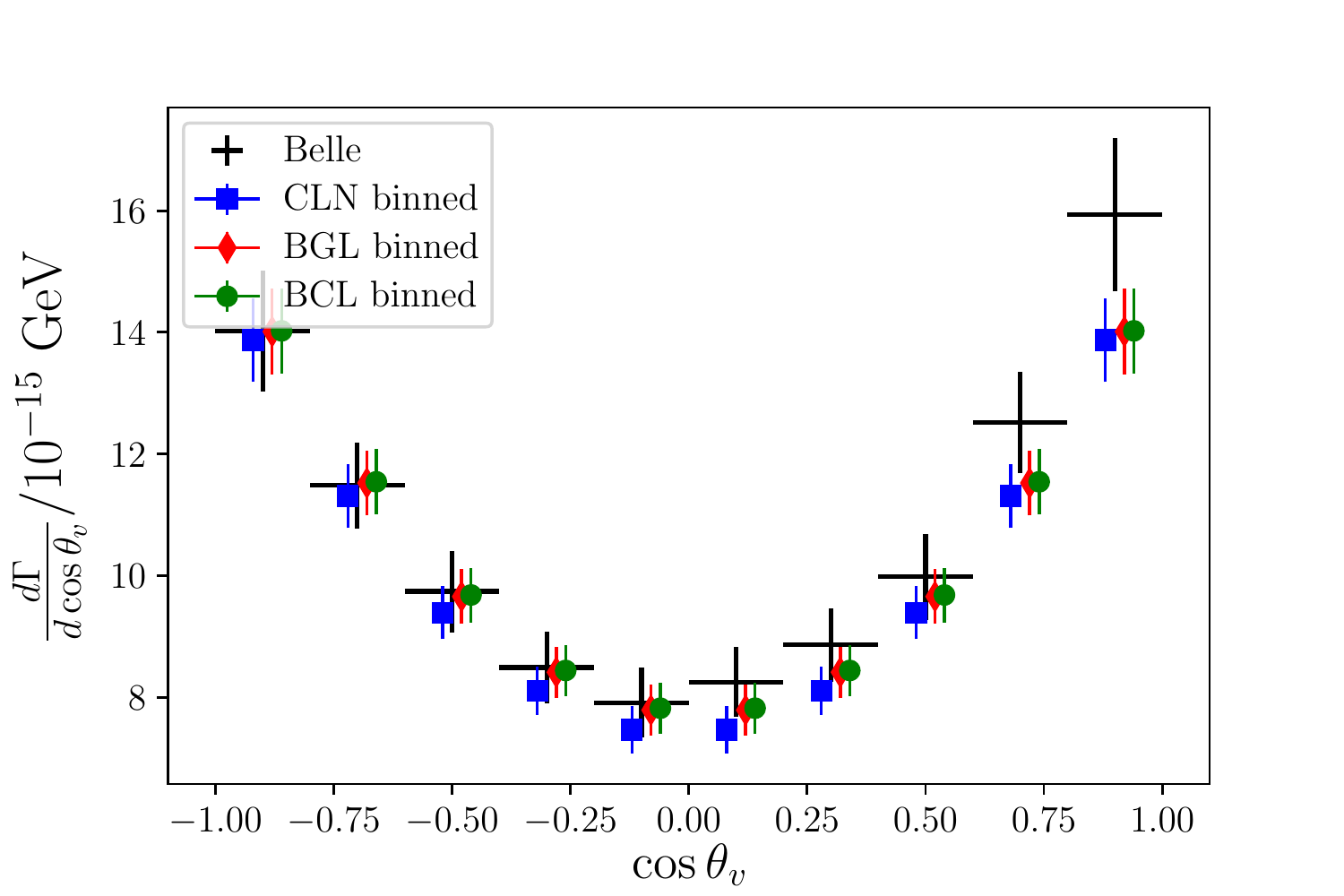}
  \includegraphics[width=0.49\textwidth]{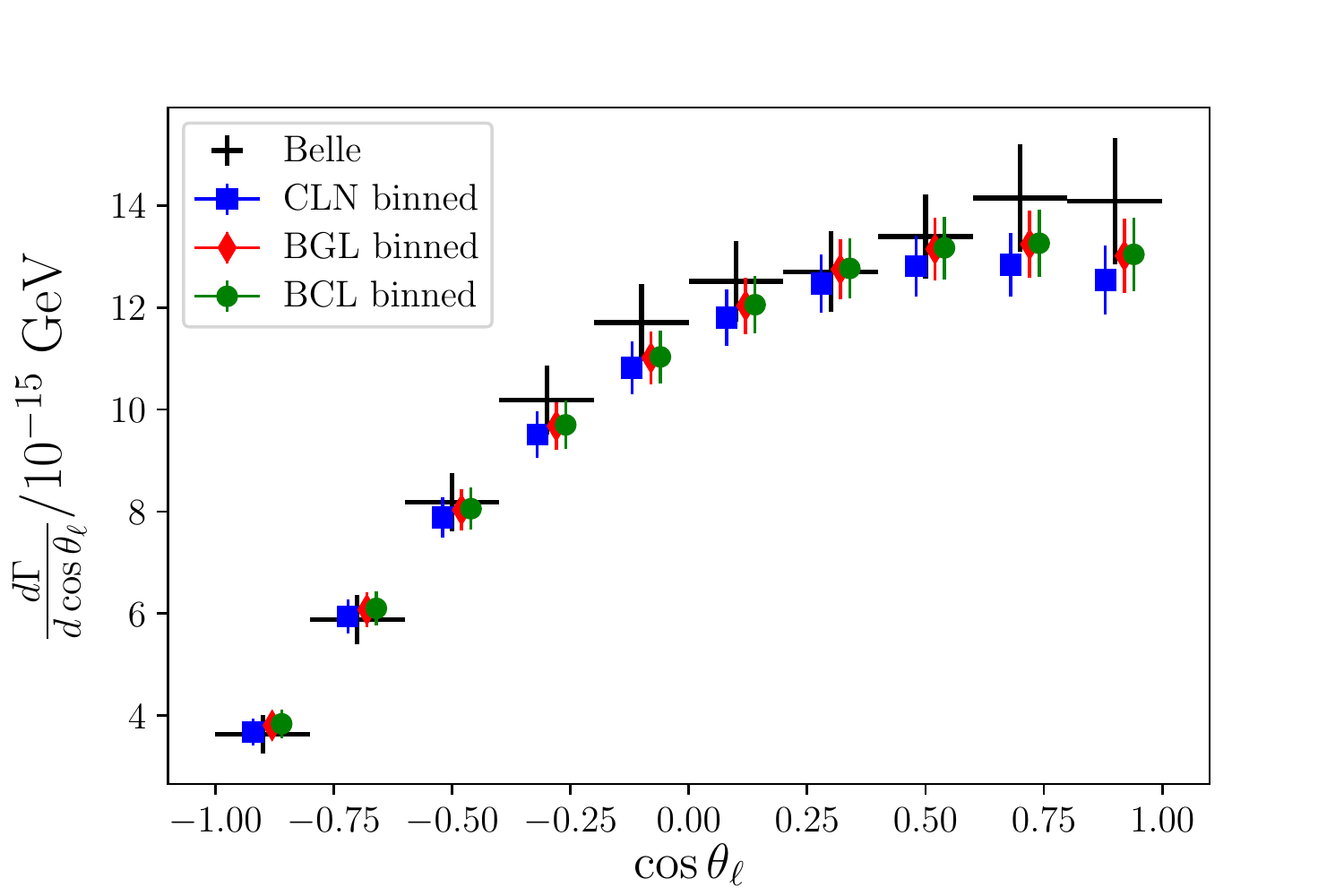}
  \includegraphics[width=0.49\textwidth]{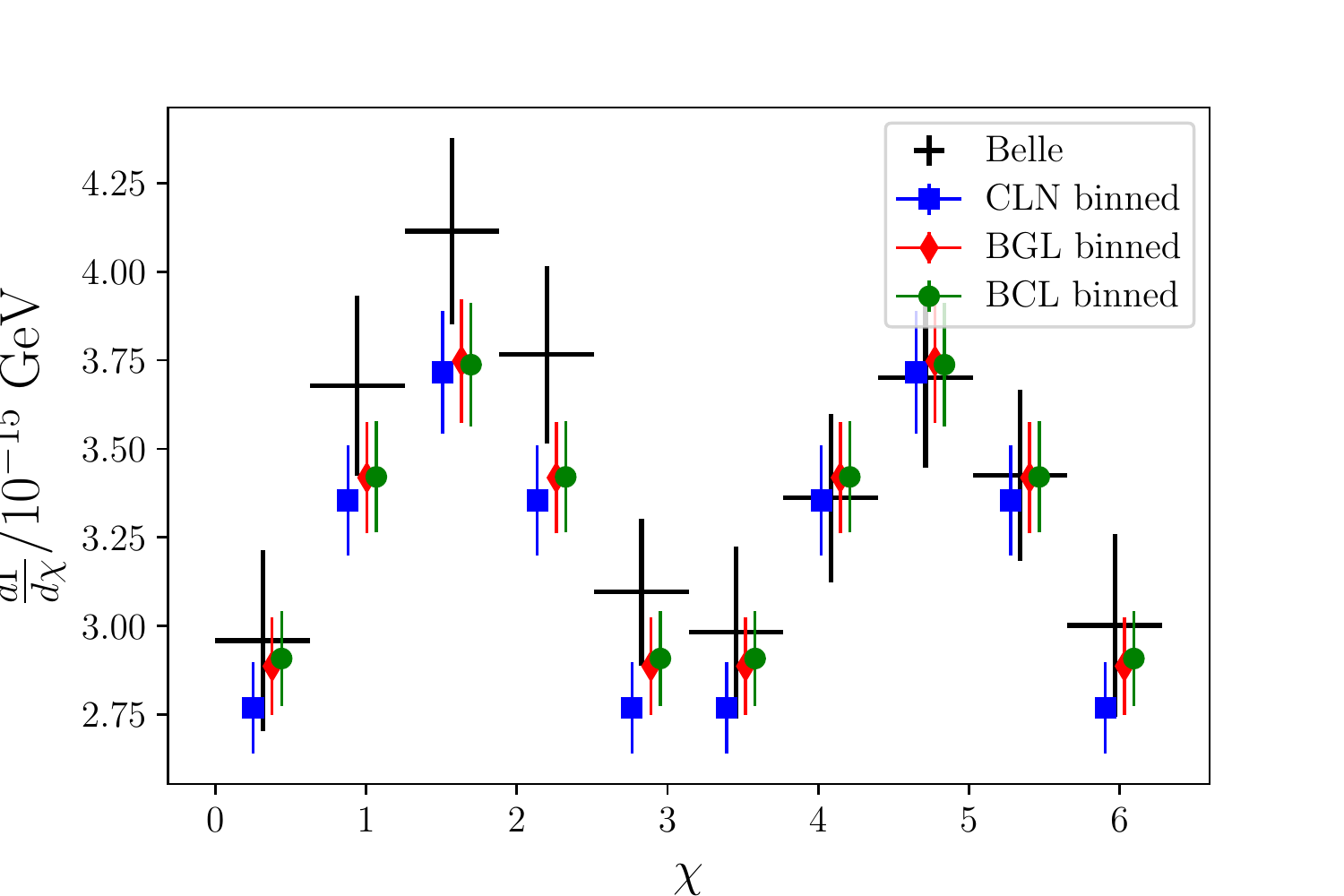}
  \caption{\label{fig:exptdata} Comparison of fit results to
    experimental data \cite{Abdesselam:2017kjf}. The binned fit
    results are slightly offset from the bin midpoints for clarity.
    See Appendix~\ref{app:exptfits} and Ref.~\cite{Abdesselam:2017kjf}
    for definitions.}
\end{figure*}

 In Fig.~\ref{fig:exptdata} we compare the fit results,
integrated over the experimental bins, of the tightly constrained CLN fit and
the BGL and BCL fits (with $K=4$) to the Belle data \cite{Abdesselam:2017kjf}.
The agreement is generally good, with the notable exception of the $d\Gamma/dw$
in the smallest $w$ bin, where the CLN result is in greater tension with the data than
the BGL and BCL results.

\begin{figure}
  \centering
  \includegraphics[width=0.47\textwidth]{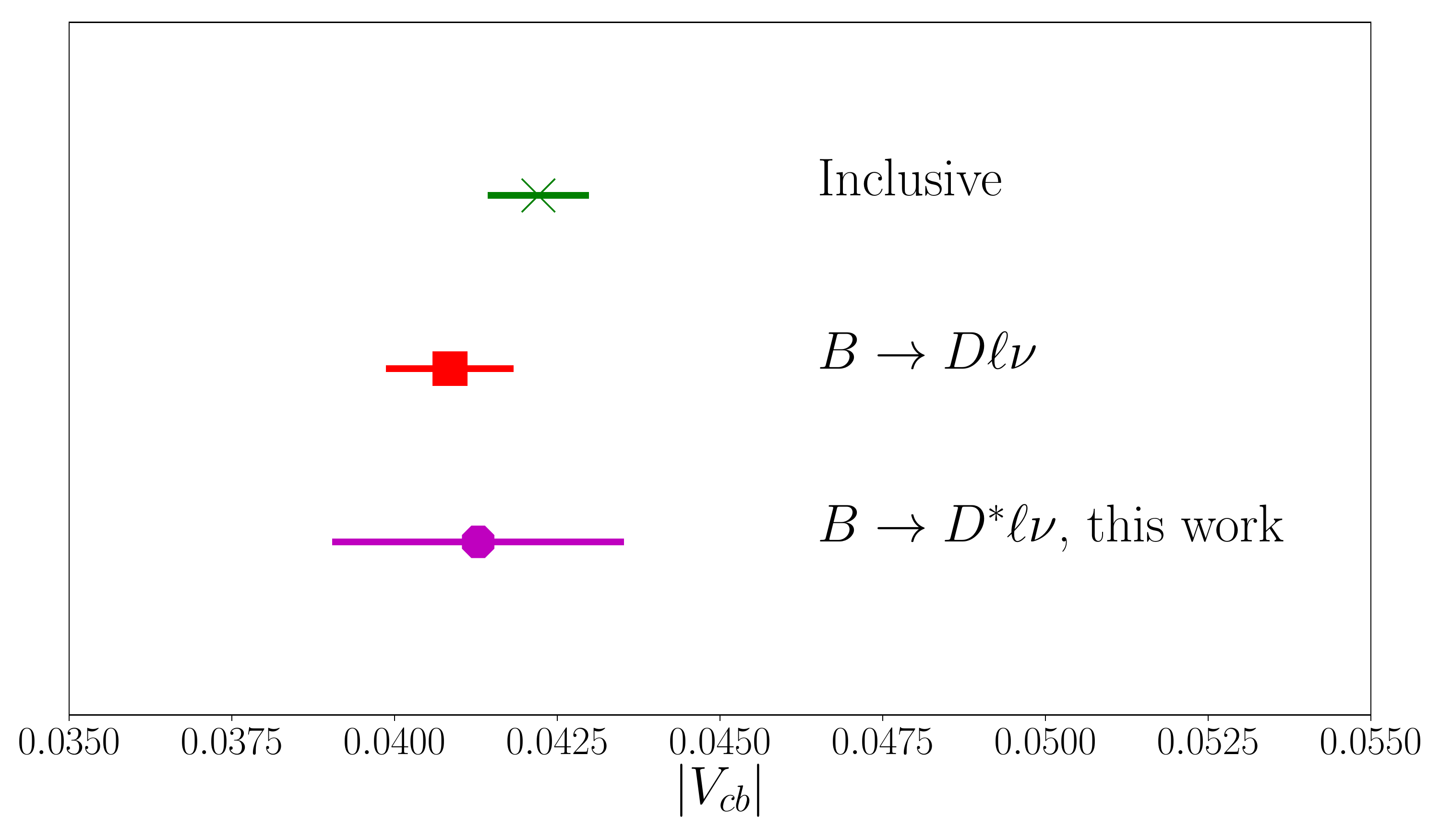}
  \caption{\label{fig:compare_Vcb} Comparison of the $|V_{cb}|$ from (\ref{eq:Vcb}) with
  the latest determinations from $B\to X_c\ell\nu$ \cite{Bevan:2014iga,Alberti:2014yda} 
  and $B\to D\ell\nu$ \cite{Aoki:2016frl}.}
\end{figure}

For the time being, with only one experimental data set available to carry out
these investigations, determinations of $|V_{cb}|$ from $B\to D^*\ell\nu$ are
less certain than has been thought.  The BGL and BCL fits to Belle data
indicate $I = 0.038(2)$.  Ref.~\cite{Bailey:2014tva} cites a private
communication with C.\ Schwanda giving
$\bar\eta_{EW}=\eta_{EW}\eta_{\mathrm{Coulomb}} = 1.0182(16)$ as the product
of the electroweak factor $\eta_{EW}=1.0066(16)$ and a term accounting for
electromagnetic interactions between the charged $D^*$ and lepton in the final
state.  Combining this with the weighted average for $h_{A_1}\!(1)$ from
Fermilab/MILC \cite{Bailey:2014tva} and this work, we arrive at
\begin{equation}
  |V_{cb}| = (41.3 \pm 2.2) \times 10^{-3}
  \label{eq:Vcb}
\end{equation}
where the error is dominated by the experimental and related fitting
uncertainty.  This determination agrees well with both those from inclusive
and exclusive $B\to D\ell\nu$ decays as shown in Fig.~\ref{fig:compare_Vcb}.

One may ultimately obtain a more precise determination of $|V_{cb}|$
by including all relevant information, from HQET, by imposing stronger
unitarity bounds \cite{Bigi:2017jbd}, and including light cone
sum rule calculations of form factors at large recoil \cite{Faller:2008tr}.
Comparison of the different approaches would be helpful to highlight the
impact of including different ingredients.

\section{Conclusions}
\label{sec:concl}

We present new unquenched lattice QCD determinations of the zero-recoil
form factors $h_{A_1}\!(1)$ and $h_{A_1}^s\!(1)$, sometimes denoted
$\mathcal{F}^{B\to D^*}(1)$ and $\mathcal{F}^{B_s\to D_s^*}(1)$, respectively.
We have used 8 ensembles  spanning 3 lattice spacings and 3 values of light-to-strange
quark mass ratios, including the physical point.  Our results are
\begin{align}
\mathcal{F}^{B\to D^*}\!(1) & = h_{A_!}\!(1) =  0.895(10)_\text{stat}(24)_\text{sys}\nonumber  \\
\mathcal{F}^{B_s\to D_s^*}(1)& = h^s_{A_!}\!(1) = 0.883(12)_\text{stat}(28)_\text{sys} \nonumber \\
\frac{\mathcal{F}^{B\to D^*}\!(1)}{\mathcal{F}^{B_s\to D_s^*}(1)}& =
\frac{h_{A_1}\!(1)}{h_{A_1}^s\!(1)} = 1.013(14)_\text{stat}(17)_\text{sys} \,.
\end{align}

This result for $h_{A_1}\!(1)$ provides a valuable, independent check
of the Fermilab/MILC result \cite{Bailey:2014tva}.  We have used
completely independent sets of gauge field configurations and different
formulations for the charm and bottom quarks.  The two results are in 
good agreement.

While the determination of $|V_{cb}|$ using these results is complicated by the
need to investigate assumptions used in extrapolating experimental data to zero
recoil, series expansion fits to the unfolded Belle data yield
\begin{equation}
  |V_{cb}| = (41.3 \pm 2.2) \times 10^{-3} \,.
\end{equation}
This is consistent with recent determinations using exclusive $B\to
D\ell\nu$ and inclusive decays (Fig~\ref{fig:compare_Vcb}).

A reanalysis of BaBar data for the differential decay rate would
complement the unfolded Belle data used here.  We can also look
forward to new data from Belle II, after which the the precision of
$|V_{cb}|$ from $B\to D^*\ell\nu$ is likely to be much improved.
Lattice QCD data away from zero recoil will also help reduce the
uncertainties.  Preliminary results from the Fermilab/MILC
collaboration were presented at the Lattice 2017 conference
\cite{Aviles-Casco:2017nge}.

Our result for the $B_s\to D_s^*$ form factor is the first complete
calculation of $h^s_{A_1}\!(1)$. In the future, measurements of the
exclusive decays with a strange spectator, $B_s \to D_s^{(*)}\ell\nu$,
could also provide a constraint on $|V_{cb}|$.  LHCb has reconstructed
$B_s^0 \to D_s^{*-}\mu^+\nu_\mu$ decays \cite{Aaij:2017vqj}.
Eventually, with properly normalized branching fractions, these will
also provide a method of constraining $|V_{cb}|$.

Spectator quark mass effects are bounded by our calculation of 
the ratio $h^s_{A_1}\!(1)/h_{A_1}\!(1)$ and its consistency with unity.
We find deviations from $d\leftrightarrow s$ symmetry in the
zero recoil $B_{(s)}\to D_{(s)}^*$ form factors to be no more than $2$-$3\%$.


\section*{ACKNOWLEDGMENTS}

We thank R.~R.\ Horgan for useful discussions and P.\ Gambino for
correspondence regarding fits to experimental data.
We are grateful to the MILC collaboration for making publicly available
their gauge configurations and their code MILC-7.7.11 \cite{MILCgithub}.
This work was funded in part by
STFC grants ST/L000385/1 and ST/L000466/1. The
results described here were obtained using the Darwin Supercomputer of the
University of Cambridge High Performance Computing Service as part of the
DiRAC facility jointly funded by STFC, the Large Facilities Capital Fund of
BIS and the Universities of Cambridge and Glasgow. This equipment was funded
by BIS National E-infrastructure capital grant (ST/K001590/1), STFC capital
grants ST/H008861/1 and ST/H00887X/1, and STFC DiRAC Operations grants
ST/K00333X/1, ST/M007073/1, and ST/P002315/1.  MW is grateful for an IPPP
Associateship held while this work was undertaken.

 \bibliography{Master}{}

\begin{thebibliography}{10}

\bibitem{Albrecht:1987ej}
H.~Albrecht {\em et~al.} (ARGUS Collaboration),
\newblock Phys. Lett. B {\bf 197}, 452 (1987).

\bibitem{Bortoletto:1989qb}
D.~Bortoletto {\em et~al.} (CLEO Collaboration),
\newblock Phys. Rev. Lett. {\bf 63}, 1667 (1989).

\bibitem{Fulton:1990bx}
R.~Fulton {\em et~al.} (CLEO Collaboration),
\newblock Phys. Rev. D {\bf 43}, 651 (1991).

\bibitem{Albrecht:1991iz}
H.~Albrecht {\em et~al.} (ARGUS Collaboration),
\newblock Phys. Lett. B {\bf 275}, 195 (1992).

\bibitem{Barish:1994mu}
B.~Barish {\em et~al.} (CLEO Collaboration),
\newblock Phys. Rev. D {\bf 51}, 1014 (1995), arXiv:hep-ex/9406005.

\bibitem{Buskulic:1995hk}
D.~Buskulic {\em et~al.} (ALEPH Collaboration),
\newblock Phys. Lett. B {\bf 359}, 236 (1995).

\bibitem{Buskulic:1996yq}
D.~Buskulic {\em et~al.} (ALEPH Collaboration),
\newblock Phys. Lett. B {\bf 395}, 373 (1997).

\bibitem{Abbiendi:2000hk}
G.~Abbiendi {\em et~al.} (OPAL Collaboration),
\newblock Phys. Lett. B {\bf 482}, 15 (2000), arXiv:hep-ex/0003013.

\bibitem{Abreu:2001ic}
P.~Abreu {\em et~al.} (DELPHI Collaboration),
\newblock Phys. Lett. B {\bf 510}, 55 (2001), arXiv:hep-ex/0104026.

\bibitem{Adam:2002uw}
N.~E. Adam {\em et~al.} (CLEO Collaboration),
\newblock Phys. Rev. D {\bf 67}, 032001 (2003), arXiv:hep-ex/0210040.

\bibitem{Abdallah:2004rz}
J.~Abdallah {\em et~al.} (DELPHI Collaboration),
\newblock Eur. Phys. J. C {\bf 33}, 213 (2004), arXiv:hep-ex/0401023.

\bibitem{Aubert:2007rs}
B.~Aubert {\em et~al.} (BaBar Collaboration),
\newblock Phys. Rev. D {\bf 77}, 032002 (2008), arXiv:0705.4008.

\bibitem{Aubert:2007qs}
B.~Aubert {\em et~al.} (BaBar Collaboration),
\newblock Phys. Rev. Lett. {\bf 100}, 231803 (2008), arXiv:0712.3493.

\bibitem{Aubert:2008yv}
B.~Aubert {\em et~al.} (BaBar Collaboration),
\newblock Phys. Rev. D {\bf 79}, 012002 (2009), arXiv:0809.0828.

\bibitem{Dungel:2010uk}
W.~Dungel {\em et~al.} (Belle Collaboration),
\newblock Phys. Rev. D {\bf 82}, 112007 (2010), arXiv:1010.5620.

\bibitem{Abdesselam:2017kjf}
A.~Abdesselam {\em et~al.} (Belle Collaboration),
\newblock (2017), arXiv:1702.01521.

\bibitem{Amhis:2016xyh}
Y.~Amhis {\em et~al.} (HFLAV Collaboration),
\newblock (2016), arXiv:1612.07233.

\bibitem{Bailey:2014tva}
J.~A. Bailey {\em et~al.},
\newblock Phys. Rev. D {\bf 89}, 114504 (2014), arXiv:1403.0635.

\bibitem{Bevan:2014iga}
A.~J. Bevan {\em et~al.},
\newblock Eur. Phys. J. C {\bf 74}, 3026 (2014), arXiv:1406.6311.

\bibitem{Alberti:2014yda}
A.~Alberti, P.~Gambino, K.~J. Healey, and S.~Nandi,
\newblock Phys. Rev. Lett. {\bf 114}, 061802 (2015), arXiv:1411.6560.

\bibitem{Caprini:1997mu}
I.~Caprini, L.~Lellouch, and M.~Neubert,
\newblock Nucl. Phys. B {\bf 530}, 153 (1998), arXiv:hep-ph/9712417.

\bibitem{Bernlochner:2017jka}
F.~U. Bernlochner, Z.~Ligeti, M.~Papucci, and D.~J. Robinson,
\newblock Phys. Rev. D {\bf 95}, 115008 (2017), arXiv:1703.05330.

\bibitem{Bigi:2017njr}
D.~Bigi, P.~Gambino, and S.~Schacht,
\newblock Phys. Lett. B {\bf 769}, 441 (2017), arXiv:1703.06124.

\bibitem{Grinstein:2017nlq}
B.~Grinstein and A.~Kobach,
\newblock Phys. Lett. B {\bf 771}, 359 (2017), arXiv:1703.08170.

\bibitem{Bigi:2017jbd}
D.~Bigi, P.~Gambino, and S.~Schacht,
\newblock (2017), arXiv:1707.09509.

\bibitem{Jaiswal:2017rve}
S.~Jaiswal, S.~Nandi, and S.~K. Patra,
\newblock (2017), arXiv:1707.09977.

\bibitem{Bernlochner:2017xyx}
F.~U. Bernlochner, Z.~Ligeti, M.~Papucci, and D.~J. Robinson,
\newblock (2017), arXiv:1708.07134.

\bibitem{Boyd:1997kz}
C.~G. Boyd, B.~Grinstein, and R.~F. Lebed,
\newblock Phys. Rev. D {\bf 56}, 6895 (1997), arXiv:hep-ph/9705252.

\bibitem{Bailey:2015rga}
J.~A. Bailey {\em et~al.},
\newblock Phys. Rev. D {\bf 92}, 034506 (2015), arXiv:1503.07237.

\bibitem{Na:2015kha}
H.~Na {\em et~al.},
\newblock Phys. Rev. D {\bf 92}, 054510 (2015), arXiv:1505.03925,
\newblock [Erratum: Phys. Rev. D \textbf{93}, 119906 (2016)].

\bibitem{Boyd:1994tt}
C.~G. Boyd, B.~Grinstein, and R.~F. Lebed,
\newblock Phys. Rev. Lett. {\bf 74}, 4603 (1995), arXiv:hep-ph/9412324.

\bibitem{Bourrely:2008za}
C.~Bourrely, L.~Lellouch, and I.~Caprini,
\newblock Phys. Rev. D {\bf 79}, 013008 (2009), arXiv:0807.2722.

\bibitem{Aoki:2016frl}
S.~Aoki {\em et~al.} (FLAG Collaboration),
\newblock Eur. Phys. J. C {\bf 77}, 112 (2017), arXiv:1607.00299.

\bibitem{Buras:2008nn}
A.~J. Buras and D.~Guadagnoli,
\newblock Phys. Rev. D {\bf 78}, 033005 (2008), arXiv:0805.3887.

\bibitem{Bazavov:2010ru}
A.~Bazavov {\em et~al.} (MILC Collaboration),
\newblock Phys. Rev. D {\bf 82}, 074501 (2010), arXiv:1004.0342.

\bibitem{Bazavov:2012xda}
A.~Bazavov {\em et~al.} (MILC Collaboration),
\newblock Phys. Rev. D {\bf 87}, 054505 (2013), arXiv:1212.4768.

\bibitem{Bazavov:2015yea}
A.~Bazavov {\em et~al.} (MILC Collaboration),
\newblock Phys. Rev. D {\bf 93}, 094510 (2016), arXiv:1503.02769.

\bibitem{Follana:2006rc}
E.~Follana {\em et~al.} (HPQCD Collaboration),
\newblock Phys. Rev. D {\bf 75}, 054502 (2007), arXiv:hep-lat/0610092.

\bibitem{Lepage:1992tx}
G.~P. Lepage, L.~Magnea, C.~Nakhleh, U.~Magnea, and K.~Hornbostel,
\newblock Phys. Rev. D {\bf 46}, 4052 (1992), hep-lat/9205007.

\bibitem{Harrison:2016gup}
J.~Harrison, C.~Davies, and M.~Wingate,
\newblock Proc. Sci. {\bf LATTICE2016}, 287 (2016), arXiv:1612.06716.

\bibitem{Atoui:2013zza}
M.~Atoui, V.~Mor\'enas, D.~Be{\v{c}}irevi\'c, and F.~Sanfilippo,
\newblock Eur. Phys. J. C {\bf 74}, 2861 (2014), arXiv:1310.5238.

\bibitem{Flynn:2016vej}
J.~Flynn {\em et~al.},
\newblock Proc. Sci. {\bf LATTICE2016}, 296 (2016), arXiv:1612.05112.

\bibitem{Sirlin:1981ie}
A.~Sirlin,
\newblock Nucl. Phys. B {\bf 196}, 83 (1982).

\bibitem{Ginsberg:1968pz}
E.~S. Ginsberg,
\newblock Phys. Rev. {\bf 171}, 1675 (1968),
\newblock [Erratum: Phys. Rev. \textbf{174}, 2169 (1968)].

\bibitem{Atwood:1989em}
D.~Atwood and W.~J. Marciano,
\newblock Phys. Rev. D {\bf 41}, 1736 (1990).

\bibitem{latticespacing}
R.~J. Dowdall {\em et~al.} (HPQCD Collaboration),
\newblock Phys. Rev. D {\bf 85}, 054509 (2012), arXiv:1110.6887.

\bibitem{PhysMasses}
B.~Chakraborty {\em et~al.} (HPQCD Collaboration),
\newblock Phys. Rev. D {\bf 91}, 054508 (2015), arXiv:1408.4169.

\bibitem{lightmassratio}
A.~Bazavov {\em et~al.} (Fermilab Lattice and MILC Collaborations),
\newblock Phys. Rev. D {\bf 90}, 074509 (2014), arXiv:1407.3772.

\bibitem{GaugeAction}
A.~Hart, G.~M. von Hippel, and R.~R. Horgan,
\newblock Phys. Rev. D {\bf 79}, 074008 (2009), arXiv:0812.0503.

\bibitem{MILCgithub}
MILC Code Repository, https://github.com/milc-qcd.

\bibitem{NRQCD}
G.~P. Lepage, L.~Magnea, C.~Nakhleh, U.~Magnea, and K.~Hornbostel,
\newblock Phys. Rev. D {\bf 46}, 4052 (1992), arXiv:hep-lat/9205007.

\bibitem{timereversal}
J.~J. Dudek, R.~G. Edwards, and D.~G. Richards,
\newblock Phys. Rev. D {\bf 73}, 074507 (2006), arXiv:hep-ph/0601137.

\bibitem{Matching2013}
C.~Monahan, J.~Shigemitsu, and R.~Horgan,
\newblock Phys. Rev. D {\bf 87}, 034017 (2013), arXiv:1211.6966.

\bibitem{Matching1998}
C.~J. Morningstar and J.~Shigemitsu,
\newblock Phys. Rev. D {\bf 59}, 094504 (1999), arXiv:hep-lat/9810047.

\bibitem{LUKE}
M.~E. Luke,
\newblock Phys. Lett. B {\bf 252}, 447 (1990).

\bibitem{Marginalization}
C.~McNeile, C.~T.~H. Davies, E.~Follana, K.~Hornbostel, and G.~P. Lepage (HPQCD
  Collaboration),
\newblock Phys. Rev. D {\bf 82}, 034512 (2010), arXiv:1004.4285.

\bibitem{Fitting}
G.~P. Lepage {\em et~al.},
\newblock Nucl. Phys. Proc. Suppl. {\bf 106}, 12 (2002), hep-lat/0110175.

\bibitem{corrfitter}
{G.P. Lepage},
\newblock {Corrfitter Version 4.1},
\newblock github.com/gplepage/corrfitter.git.

\bibitem{Bmeson}
B.~Colquhoun {\em et~al.},
\newblock Phys. Rev. D {\bf 91}, 114509 (2015), arXiv:1503.05762.

\bibitem{SCHIPT}
{J. Laiho, R.S. Van de Water},
\newblock Phys. Rev. D {\bf 73}, 054501 (2006), arXiv:hep-lat/0512007.

\bibitem{Bmeson2}
R.~J. Dowdall, C.~T.~H. Davies, R.~R. Horgan, C.~J. Monahan, and J.~Shigemitsu,
\newblock Phys. Rev. Lett. {\bf 110}, 222003 (2013), arXiv:1302.2644.

\bibitem{ANDREWlatt2016}
B.~Colquhoun, C.~Davies, J.~Koponen, A.~Lytle, and C.~McNeile (HPQCD
  Collaboration),
\newblock PoS {\bf LATTICE2016}, 281 (2016), arXiv:1611.01987.

\bibitem{Precisehlmesonmasses}
R.~Dowdall, C.~Davies, T.~Hammant, and R.~Horgan,
\newblock Phys. Rev. D {\bf 86}, 094510 (2012), arXiv:1207.5149.

\bibitem{Patrignani:2016xqp}
C.~Patrignani {\em et~al.} (PDG Collaboration),
\newblock Chin. Phys. C {\bf 40}, 100001 (2016).

\bibitem{Devlani:2014nda}
N.~Devlani, V.~Kher, and A.~K. Rai,
\newblock Eur. Phys. J. A {\bf 50}, 154 (2014).

\bibitem{Godfrey:2004ya}
S.~Godfrey,
\newblock Phys. Rev. D {\bf 70}, 054017 (2004), arXiv:hep-ph/0406228.

\bibitem{Eichten:1994gt}
E.~J. Eichten and C.~Quigg,
\newblock Phys. Rev. D {\bf 49}, 5845 (1994), arXiv:hep-ph/9402210.

\bibitem{Faller:2008tr}
S.~Faller, A.~Khodjamirian, C.~Klein, and T.~Mannel,
\newblock Eur. Phys. J. C {\bf 60}, 603 (2009), arXiv:0809.0222.

\bibitem{Aviles-Casco:2017nge}
A.~Vaquero Avil\'es-Casco {\em et~al.},
\newblock (2017), arXiv:1710.09817.

\bibitem{Aaij:2017vqj}
R.~Aaij {\em et~al.} (LHCb Collaboration),
\newblock (2017), arXiv:1705.03475.

\bibitem{OnLatticePT}
G.~P. Lepage and P.~B. Mackenzie,
\newblock Phys.\ Rev. D {\bf 48}, 2250 (1993), arXiv:hep-lat/9209022.

\bibitem{radcoefficients}
T.~Hammant, A.~Hart, G.~von Hippel, R.~Horgan, and C.~Monahan,
\newblock Phys. Rev. Lett. {\bf 107}, 112002 (2011), arXiv:1105.5309.

\bibitem{PrecDsupdate}
C.~T.~H. Davies {\em et~al.},
\newblock Phys. Rev. D {\bf 82}, 114504 (2010), arXiv:1008.4018.

\bibitem{Follana:2007uv}
E.~Follana, C.~T.~H. Davies, G.~P. Lepage, and J.~Shigemitsu,
\newblock Phys. Rev. Lett. {\bf 100}, 062002 (2008), arXiv:0706.1726.

\bibitem{fat7}
D.~Toussaint and K.~Orginos,
\newblock Nucl. Phys. Proc. Suppl. {\bf 73}, 909 (1999), hep-lat/9809148.

\bibitem{ASQTAD}
G.~P. Lepage,
\newblock Phys. Rev. D {\bf 59}, 074502 (1999), arXiv:hep-lat/9809157.

\bibitem{staggeredmasses}
C.~Aubin and C.~Bernard,
\newblock Phys. Rev. D {\bf 68}, 034014 (2003), arXiv:hep-lat/0304014.

\bibitem{PIMASSES}
R.~J. Dowdall, C.~T.~H. Davies, G.~P. Lepage, and C.~McNeile,
\newblock Phys. Rev. D {\bf 88}, 074504 (2013), arXiv:1303.1670.

\bibitem{Arndt:2004bg}
D.~Arndt and C.~J.~D. Lin,
\newblock Phys. Rev. D {\bf 70}, 014503 (2004), arXiv:hep-lat/0403012.

\bibitem{Richman:1995wm}
J.~D. Richman and P.~R. Burchat,
\newblock Rev. Mod. Phys. {\bf 67}, 893 (1995), arXiv:hep-ph/9508250.

\bibitem{Neubert:1993mb}
M.~Neubert,
\newblock Phys. Rept. {\bf 245}, 259 (1994), arXiv:hep-ph/9306320.

\end{thebibliography}
 \bibliographystyle{h-physrev5_collab}

\appendix

\section{Gauge Action}
\label{sec:GAction}

The gauge action used to generate the configurations is the Symanzik and tadpole improved action of \cite{GaugeAction}, which contains additional rectangle and parallelogram loops to cancel radiative $\mathcal{O}(a^2)$ errors:
\begin{align}
S = &\sum_x \Big[ a_0 P_0(x) + a_1 P_1(x) + a_2 P_2(x) \Big]\nonumber\\
P_0(x) = &\sum_{\mu<\nu} U_{\mu \nu}(x)\nonumber\\
P_1(x) = &\sum_{\mu<\nu} U_{\mu \mu \nu}(x) +U_{\mu \nu \nu}(x)\nonumber\\
P_2(x) = &\sum_{\mu<\nu<\rho} U_{\mu \nu \rho}(x) +U_{\mu \rho \nu}(x) + U_{\rho \mu \nu}(x) + U_{\rho -\mu \nu}(x)\nonumber\\
\end{align}
Where $-\mu$ indicates a Hermitian conjugated gauge link. $a_1$ and $a_2$ are calculated in terms of $a_0$ using lattice perturbation theory. The perturbative coefficients are specified in Ref.~\cite{GaugeAction}.


\section{$b$-Quarks Using NRQCD}
\label{sec:NRQCDAction}

In order to efficiently simulate the bottom quark we employ non-relativistic QCD (NRQCD) \cite{NRQCD}. This formulation has been used for many calculations done by the HPQCD collaboration \cite{latticespacing,PhysMasses,Bmeson2,Bmeson,Na:2015kha}. The action is given in \cite{NRQCD}, which we repeat here for clarity:
\begin{align}
S = a^3 \sum_x \Big[ &\psi^\dagger(x)\psi(x) -\psi(x+a\hat{t})\Big( 1-\frac{aH_0}{2n}\Big)^n\Big(1-\frac{a\delta H}{2}\Big)\nonumber \\
&\times U^\dagger_t(x)\Big(1-\frac{a\delta H}{2}\Big) \Big( 1-\frac{aH_0}{2n}\Big)^n \psi(x) \Big] \,.
\end{align}
The heavy quark propagator then satisfies the simple evolution equation
\begin{align}
G(x+a\hat{t},z) =& \delta(x+a\hat{t},z) + \Big( 1-\frac{aH_0}{2n}\Big)^n \Big(1-\frac{a\delta H}{2}\Big)\nonumber\\
\times& U^\dagger_t(x) \Big(1-\frac{a\delta H}{2}\Big)\Big( 1-\frac{aH_0}{2n}\Big)^nG(x,z)
\end{align}
with $G(x,y) = 0$ for $x_t< y_t$, since the quark part of the action is first order in $D_0$ the propagator has no pole at $-E(p)$ and so is only the retarded part of the full propagator. This allows the bottom quark propagator to be computed by applying the evolution equation iteratively, allowing for faster, less memory intensive calculations and greater statistics.

The NRQCD quark action is tadpole improved \cite{OnLatticePT} and Symanzik improved, with 
\begin{align}
aH_0 =& - \Delta^{(2)}/2am\nonumber\\
a\delta H =& -c_1\frac{(\Delta^{(2)})^2}{8(am)^3} + c_2 \frac{ig}{8(am)^2}\big(\Delta \cdot \tilde{E} - \tilde{E} \cdot \Delta \big)\nonumber\\
&-c_3 \frac{g}{8(am)^2} \sigma \cdot \big( \Delta \times \tilde{E} - \tilde{E} \times \Delta \big)\nonumber\\
&-c_4 \frac{g}{2am}\sigma \cdot \tilde{B} + c_5 \frac{a^2 \Delta^{(4)}}{24am} - c_6 \frac{a(\Delta^{(2)})^2}{16n(am)^2}
\end{align}
where the tilded quantities are the tadpole improved versions. In our simulations here we take the stability parameter $n=4$. The coefficients $c_1$-$c_6$ were computed perturbatively in \cite{latticespacing,radcoefficients} and are given in table II of \cite{Bmeson}.


\section{HISQ Quarks}
\label{sec:HISQ}

For the $u/d$ and $c$ valence quarks in our calculation we use the same HISQ action as for the sea quarks \cite{Follana:2006rc}. The advantage of using HISQ is that $am_q$ discretization errors are under sufficient control that it can be used both for light and for $c$ quarks \cite{PrecDsupdate,Follana:2006rc,Follana:2007uv}. The HISQ action is also numerically inexpensive as a result of the staggering which means we are able to attain better statistics. The valence $u/d$ masses are the same as those in the sea. The masses are given in Table \ref{tab:params}. Below we
summarize a few relevant facts.

The naive Dirac action has a discrete, space time dependent symmetry
\begin{align}
\psi(x) \rightarrow \mathcal{B}_\xi(x)\psi(x) \nonumber \\
\bar{\psi}(x) \rightarrow \bar{\psi}(x)\mathcal{B}^\dagger_\xi(x)
\end{align}
where
\begin{equation}
 \mathcal{B}_\xi(x) = \gamma^{\bar{\xi}}(-1)^{\xi\cdot x}
\end{equation}
and following \cite{Follana:2006rc}
\begin{align}
&\gamma^m = \prod_{i=0}^3 (\gamma^i)^{m_i}\nonumber \\
&\overline{m}_\mu = \sum_{\eta \neq \mu} m_\eta \text{  mod 2}  \,.
\end{align}
The conventions for $\gamma^{\bar{\xi}}$ are specified in the appendices. In momentum space this then gives the relation for the naive quark propagator:
\begin{equation}
S_F(p,q) = \mathcal{B}_\xi(0) S_F(p+\xi\pi,q+\xi\pi)\mathcal{B}_\xi(0)\,.
\end{equation}
One can diagonalise the naive action in spin indices using a position dependent transformation of the fields. There are several choices for such a transformation, here we use:
\begin{align}
\psi(x) \rightarrow \Omega(x)\chi(x) \nonumber \\
\bar{\psi}(x) \rightarrow \bar{\chi}(x)\Omega^\dagger(x)
\end{align}
with $\Omega(x) = \gamma^x$
this yields the action
\begin{equation}
\mathcal{S}  = \sum_{x,i} \bar{\chi}_i(x)(\alpha(x)\cdot \Delta(U) + m_0)\chi_i(x)
\end{equation}
with propagator
\begin{equation}
  \langle\chi_\kappa(x) \bar{\chi}_\delta(y)\rangle =  s(x,y)
  \delta_{\kappa\delta}\,.
\end{equation}
We then need only do the inversion for a single component of $\chi$ and the full naive propagator can be reconstructed trivially by inserting $\Omega$ matrices:
\begin{align}
S_F(x,y)_{\alpha\beta} =& \langle \psi_\alpha(x) \bar{\psi}_\beta(y)\rangle  =\Omega_{\alpha\kappa}(x) \langle\chi_\kappa(x) \bar{\chi}_\delta(y)\rangle \Omega^\dagger_{\delta\beta}(y)  \nonumber\\
 =& \Omega_{\alpha\delta}(x)\Omega^\dagger_{\delta\beta}(y) s(x,y) \,.
\end{align}
In order to remove discretization errors and taste exchange violations the operator $\Delta_\mu (U)$ used in simulations is more elaborate. It retains the feature that $\Delta_\mu(U) \psi(x)$ only contains fields $\psi(x')$ located an odd number of lattice sites away from $x$ in the $\mu$ direction, ensuring that the spin-diagonalization $(12)$ still works. The full, highly improved staggered $SU(3)$-covariant derivative operator is~\cite{Follana:2006rc}:
\begin{equation}
\mathcal{D}^{\text{HISQ}}_\mu = \Delta_\mu(W) - \frac{a^2}{6}(1+\epsilon)\Delta^3_\mu(X)
\end{equation}
with
\begin{align}
W_\mu &= \mathcal{F}^{\text{HISQ}}_\mu U_\mu \nonumber \\
X_\mu &= \mathcal{U}\mathcal{F}_\mu U_\mu \nonumber \\
\mathcal{F}^{\text{HISQ}}_\mu &= \left(   \mathcal{F}_\mu - \sum_{\rho\neq \mu} \frac{a^2 \delta_\rho^2}{2} \right)  \mathcal{U} \mathcal{F}_\mu \nonumber \\
\mathcal{F}_\mu &= \prod_{\rho\neq\mu}   \left(  1 + \frac{a^2 \delta_\rho^{(2)}}{4} \right)_{\text{symm}}  \,.
\end{align}
Where `symm' indicates that the product ordering is symmetrised in $\rho$, $\delta_\rho$ approximates a covariant first derivative on the gauge links and  $\delta^{(2)}_\rho$ approximates a second covariant derivative.
\begin{align}
\delta_\rho U_\mu(x)=& \frac{1}{a}\Big(    U_\rho(x)U_\mu(x+a\hat{\rho})U^\dagger_\rho(x+a\hat{\mu}) \nonumber \\
&- U^\dagger_\rho(x-a\hat{\rho})U_\mu(x-a\hat{\rho})U_\rho(x-a\hat{\rho}+a\hat{\mu})  \Big) \nonumber\\
\delta^{(2)}_\rho U_\mu(x)&= \frac{1}{a^2}\Big(    U_\rho(x)U_\mu(x+a\hat{\rho})U^\dagger_\rho(x+a\hat{\mu})  \nonumber\\
&+  U^\dagger_\rho(x-a\hat{\rho})U_\mu(x-a\hat{\rho})U_\rho(x-a\hat{\rho}+a\hat{\mu}) \nonumber \\
&- 2U_\mu(x)  \Big) \,.
\end{align}
The third covariant derivative term removes order $a^3$ discretization errors coming from the approximation of the derivative. Without the epsilon term, tree level discretization errors appear going as $(ap_\mu)^4$. For the mesons we are interested in quarks are typically nonrelativistic, and so the error is dominated by the energy, and ultimately the mass contribution going as $(am)^4$. For light quarks this is negligible, but for charm physics this must be included since current lattice spacings have $am_c\approx 0.5$. $\epsilon$ can be calculated straightforwardly as an expansion in $(am)^2$ by requiring the tree level dispersion relation $\lim_{p\rightarrow0} {(E^2(p)-m^2)}/{p^2}$ to have its correct value, $1$, to a given order $\mathcal{O}(am)$. The expansion is~\cite{Follana:2006rc}: 
\begin{equation}
\epsilon = - \frac{27}{40}(am)^2 +  \frac{327}{1120}(am)^4 -  \frac{5843}{53760}(am)^6 + \mathcal{O}((am)^8) \,.
\end{equation}
The smearings $\mathcal{F}_\mu$ remove taste changing interactions, since $\delta_\rho^{(2)} \approx -4/a^2$ when applied to a link carrying momentum $q_\rho \approx \pi/a$ The $\mu$ direction needn't be smeared as the original interaction vanishes in this case anyway. The smearing $\mathcal{F}_\mu$, known as ``Fat7" smearing \cite{fat7}, introduces new $\mathcal{O}(a^2)$ errors. These are removed by replacing $\mathcal{F}_\mu$ with ~\cite{ASQTAD}
\begin{equation}
\mathcal{F}^\text{ASQTAD}_\mu = \mathcal{F}_\mu - \sum_{\rho\neq\mu} \frac{a^2\delta_\rho^2}{4} \,.
\end{equation}
Where $\mathcal{F}^\text{ASQTAD}_\mu$ is the gauge link smearing employed in the widely used $a$-squared tadpole improved action. Similar errors originating from the smearing on the third derivative term needn't be corrected as they go as $\mathcal{O}(a^4)$. A single smearing introduces perpendicular gauge links which are themselves unsmeared. To further suppress taste exchange we use multiple smearings. Once such smearing is:
\begin{equation}
\mathcal{F}^\text{ASQTAD}_\mu \mathcal{U} \mathcal{F}^\text{ASQTAD}_\mu 
\end{equation}
where $\mathcal{U}$ is a reunitarization. This combination ensures that each smearing does not introduce any additional $\mathcal{O}(a^2)$ errors, and ensures no growth in the size of two gluon vertices, since the unitarization ensures it is bounded by unity. In the HISQ operator defined in $(17)$ we have moved the entirety of the $\mathcal{O}(a^2)$ corrections to the outermost smearing.

In order to check the taste exchange violations in HISQ one can check for taste-splittings of the pion masses. However since there are more allowed effective taste exchange vertices that there are degenerate pion multiplets this does not guarantee the theory is free of taste exchange. A better check is the explicit calculation of the couplings to taste exchange currents required to remove taste exchange. These are given in ~\cite{Follana:2006rc} in which it is clear that the HISQ action is a significant improvement over the older ASQTAD action.


\section{3-point function}
\label{sec:3pt}
For real, symmetric, stride-2 smearings $\Delta $, suppressing Dirac indices for the moment, and summing over repeated indices and spatial coordinates for zero recoil:
\begin{widetext}
\begin{align}
C_{3pt}(x_0,y_0,z_0) = &\langle \bar{u}_a(x) M_{st}c_a(x+\sigma_1+\delta_{st}) \bar{c}_b(y)\Gamma b_b(y) \bar{b}_c(z+\sigma_2)\gamma u_c(z)\rangle \Delta_1(\sigma_1) \Delta_2(\sigma_2)\nonumber\\
 = & \text{tr} \left[ \Omega^\dagger (x)  M_{st} \Omega(x+\delta_{st}) S_{ab}^c(x+\sigma_1+\delta_{st},y) \Omega^\dagger (y) \Gamma\right]\nonumber \\
&\times \left[ G_{bc}^b (y,z+\sigma_2) \gamma \Omega(z) S_{ca}^l(z,x) \right] \Delta_1(\sigma_1) \Delta_2(\sigma_2)\nonumber\\
 = & \text{tr} \left[ \xi^*_{ea}(x) \Omega^\dagger (x)  M_{st} \Omega(x+\delta_{st}) S_{eb}^c(x+\sigma_1+\delta_{st},y) \Omega^\dagger (y) \Gamma \right]\nonumber \\
&\times \left[ G_{bc}^b (y,z+\sigma_2) \gamma \Omega(z) S_{cd}^l(z,x') \xi_{da}(x') \right] \Delta_1(\sigma_1) \Delta_2(\sigma_2)
\end{align}
where it is understood that when we add $\delta_{st}$ it is modulo the hypercube. We have used the noise condition: 
\begin{equation}
\begin{split}
\xi^*_{ab}(z)\xi_{cb}(y) = \delta_{ac}\delta_{xy}
\end{split}
\end{equation}
to insert the random walls. Setting
\begin{equation}
\text{Ext}_{ba}(y) = G_{bc}^b (y,z+\sigma_2) \gamma \Omega(z) S_{cd}^l(z,x') \xi_{da}(x') \Delta_2(\sigma_2)
\end{equation}
this becomes
\begin{align}
C_{3pt}(x_0,y_0,z_0) = & \text{tr} \left[ \xi^*_{ea}(x) \Omega^\dagger (x)  M_{st}\Omega(x+\delta_{st}) S_{eb}^c(x+\sigma_1+\delta_{st},y) \Omega^\dagger (y) \Gamma \text{Ext}_{ba}(y) \right] \Delta_1(\sigma_1)\nonumber\\
= & \text{tr} \left[ \xi^*_{ea}(x-\sigma_1) \Omega^\dagger (x)  M_{st} \Omega(x+\delta_{st}) S_{eb}^c(x+\delta_{st},y) \Omega^\dagger (y) \Gamma \text{Ext}_{ba}(y) \right] \Delta_1(\sigma_1) \,.
\end{align}
Now, we do not have $S_{ab}^c(x,y)$, we have $S_{ab}^c(y,x)$ so we can use:
\begin{equation}
S^*_{ba}(x,y) = (-1)^y S_{ab}(y,x) (-1)^{x}
\end{equation}
where $(-1)^x$ is shorthand for $ (-1)^{x_0+x_1+x_2+x_3}$. Now
\begin{align}
C_{3pt}(x_0,y_0,z_0) = & \text{tr} \left[ \Omega^\dagger (y)   (-1)^y S_{bc}^{c*}(y,x+\delta_{st})(-1)^{x+\delta_{st}}  \beta_M(x)  \xi^*_{ca}(x-\sigma_1) \Gamma \text{Ext}_{ba}(y) \right] \Delta_1(\sigma_1)
\end{align}
where $\beta_M(x) = \Omega^\dagger (x)  M_{st} \Omega(x+\delta_{st})$ is the local spin-taste phase. Inserting Dirac indices:
\begin{align}
C_{3pt}(x_0,y_0,z_0) = &\Omega^\dagger _{\alpha \beta} (y)   (-1)^y S_{bc}^{c*}(y,x)(-1)^{x}  \beta_M(x+\delta_{st})  \xi^*_{ca}(x-\sigma_1+\delta_{st}) \Delta_1(\sigma_1) \Gamma_{\beta\kappa} \text{Ext}_{ba,\kappa\alpha}(y) \nonumber\\
= & \pm \left[ \Omega_{ \beta \alpha} (y)   (-1)^y S_{bc}^{c}(y,x)(-1)^{x}  \beta_M(x)  \xi_{ca}(x-\sigma_1+\delta_{st})   \Delta_1(\sigma_1) \right]^*\Gamma_{\beta\kappa} \text{Ext}_{ba,\kappa\alpha}(y)  \,.
\end{align}
We recognise $S_{bc}^{c}(y,x)(-1)^{x}  \beta_M(x)  \xi_{ca}(x-\sigma_1+\delta_{st})   \Delta_1(\sigma_1)$ as the MILC KS propagator. The naive active quark that gets made in NRQCD is then:
\begin{align}
\text{Active}_{ab,\alpha\beta}(y) = \Omega_{ \alpha \beta} (y)   (-1)^y S_{ac}^{c}(y,x)(-1)^{x}  \beta_M(x)  \xi_{cb}(x-\sigma_1+\delta_{st})   \Delta_1(\sigma_1) 
\end{align}
and the contractions to do are
\begin{align}
&\text{Current}_{ab,\alpha\beta}(y) = \text{Active}^*_{ba,\kappa \alpha}(y) \Gamma_{\kappa\beta} \nonumber \\
&C_{3pt} = \text{Current}_{ab,\alpha\beta}(y) \text{Ext}_{ba,\beta\alpha}(y) \,.
\end{align}

\section{Correlator Fits}
\label{corrfitplots}
\begin{figure}
    \includegraphics[width=0.4\textwidth]{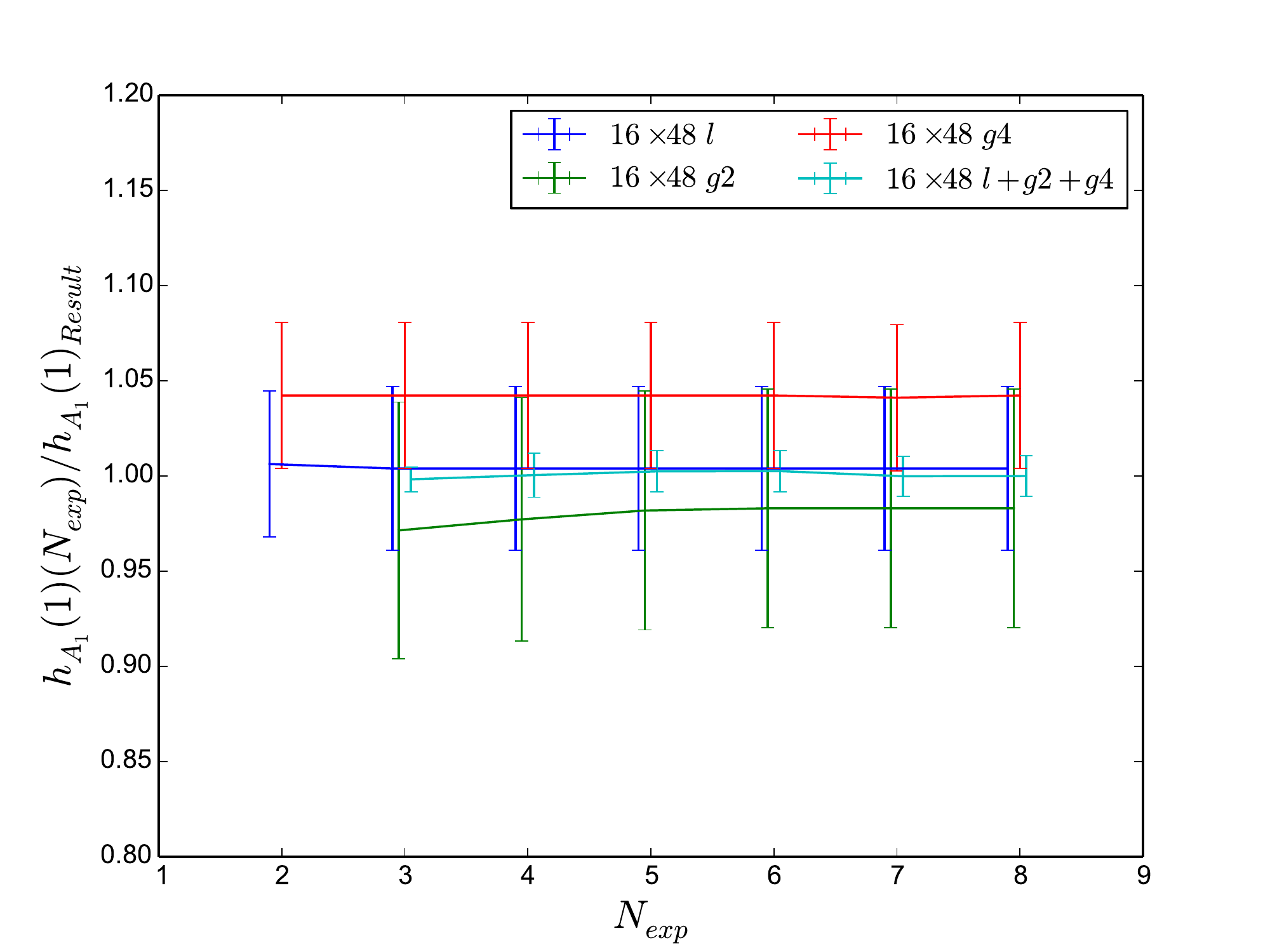}\includegraphics[width=0.4\textwidth]{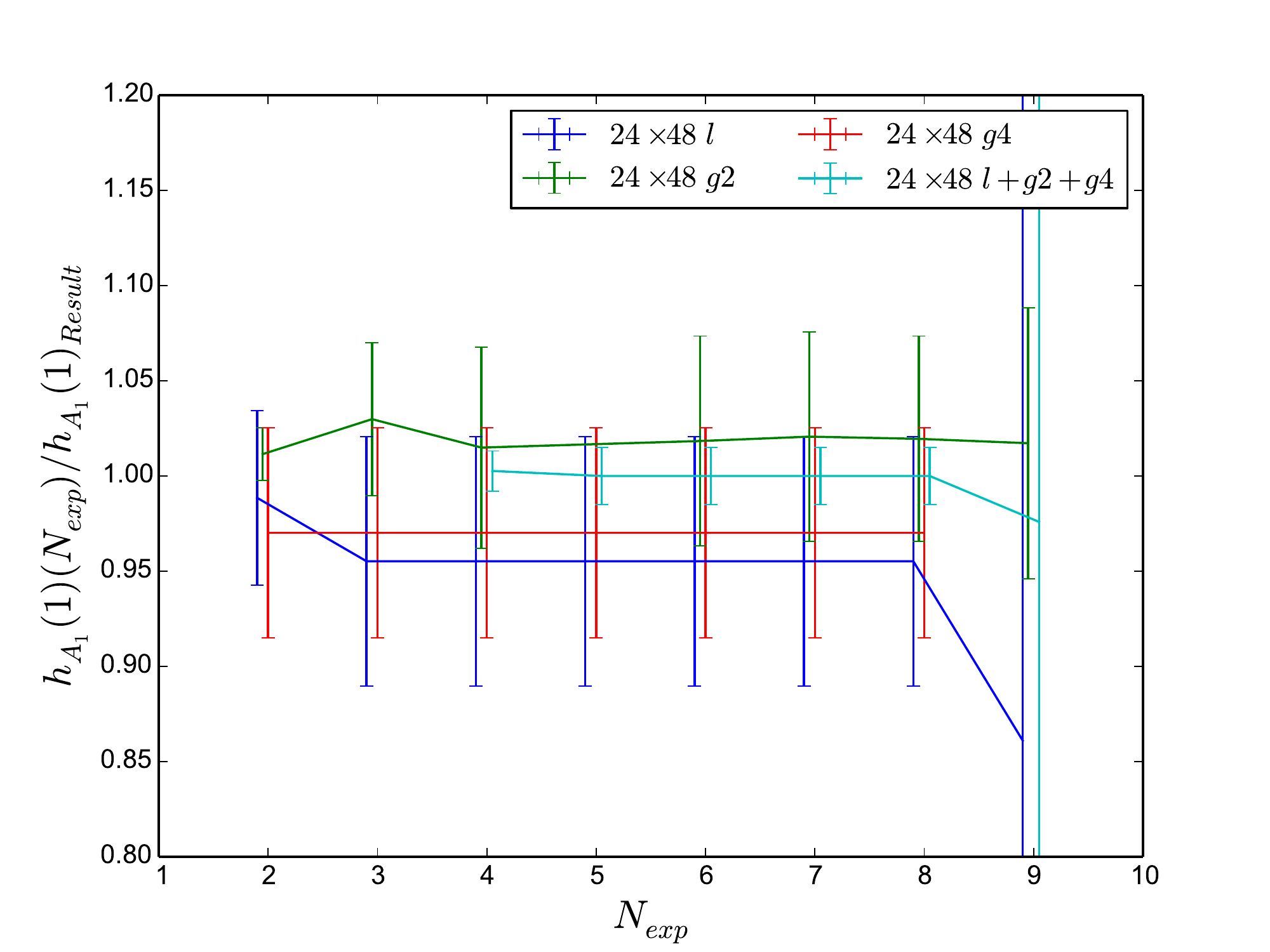}\\[-1mm]
    \includegraphics[width=0.4\textwidth]{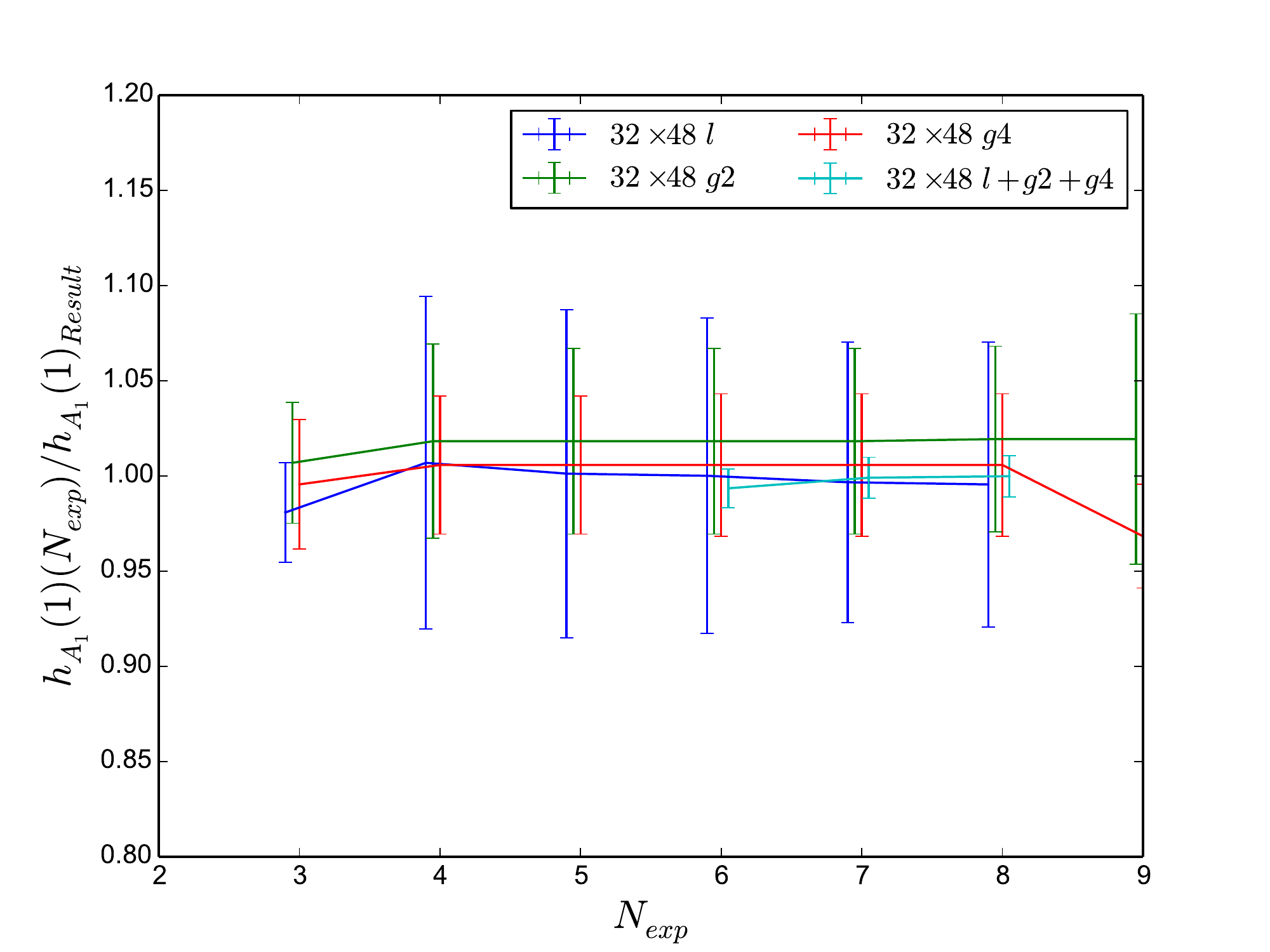}\includegraphics[width=0.4\textwidth]{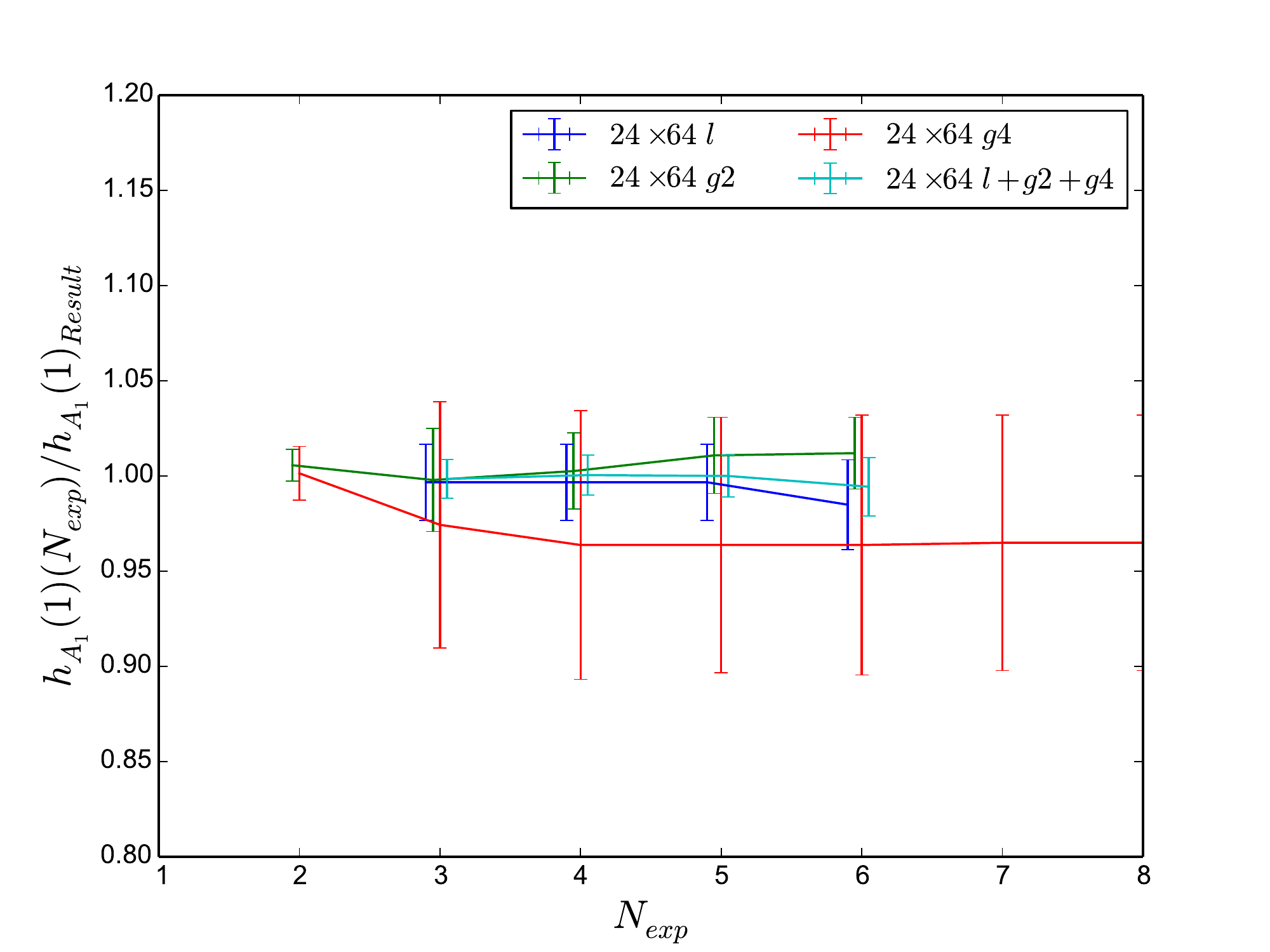}\\[-1mm]
    \includegraphics[width=0.4\textwidth]{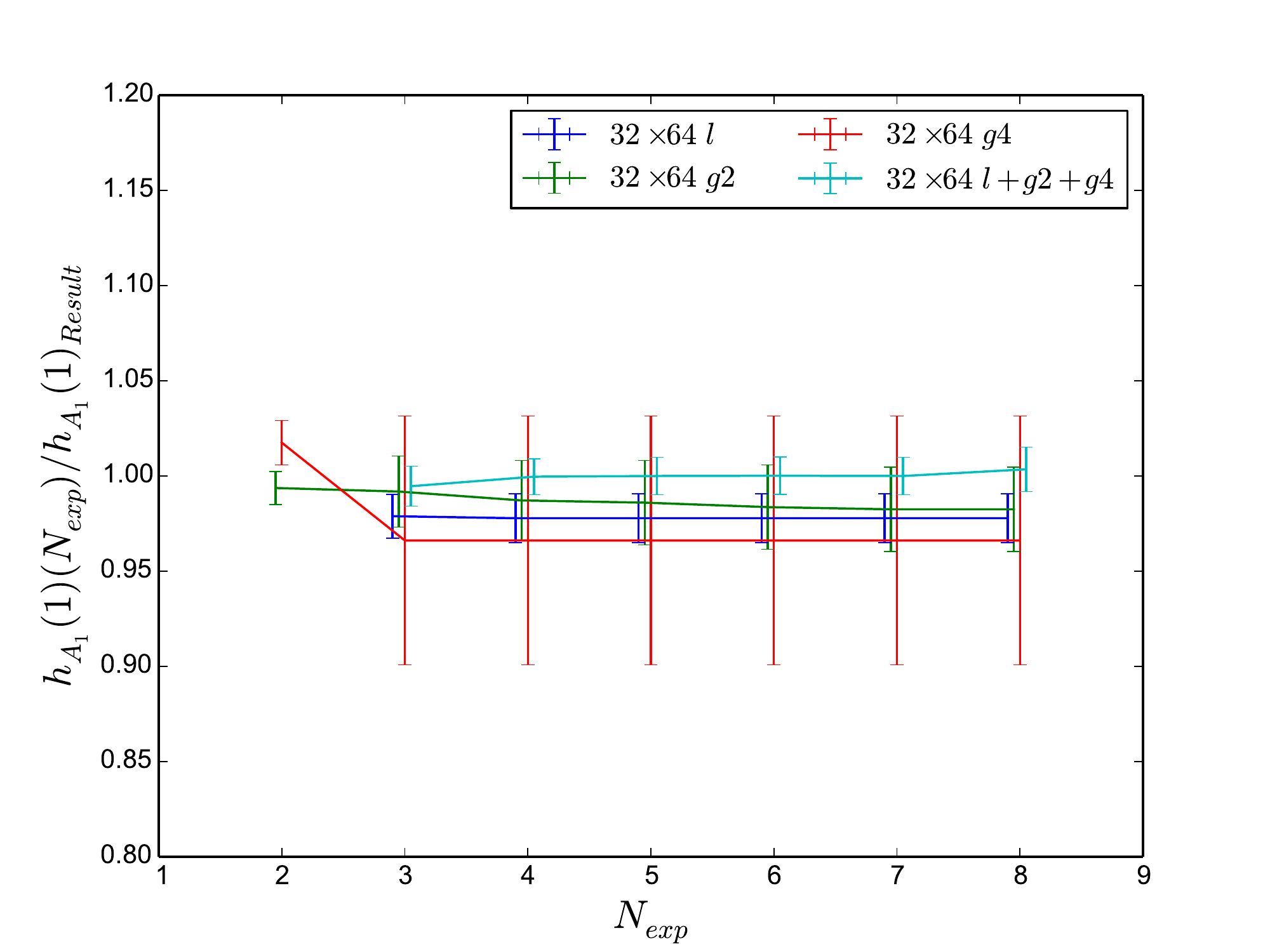}\includegraphics[width=0.4\textwidth]{Nexp_and_smearing_dependence_48_64.pdf}\\[-1mm]
    \includegraphics[width=0.4\textwidth]{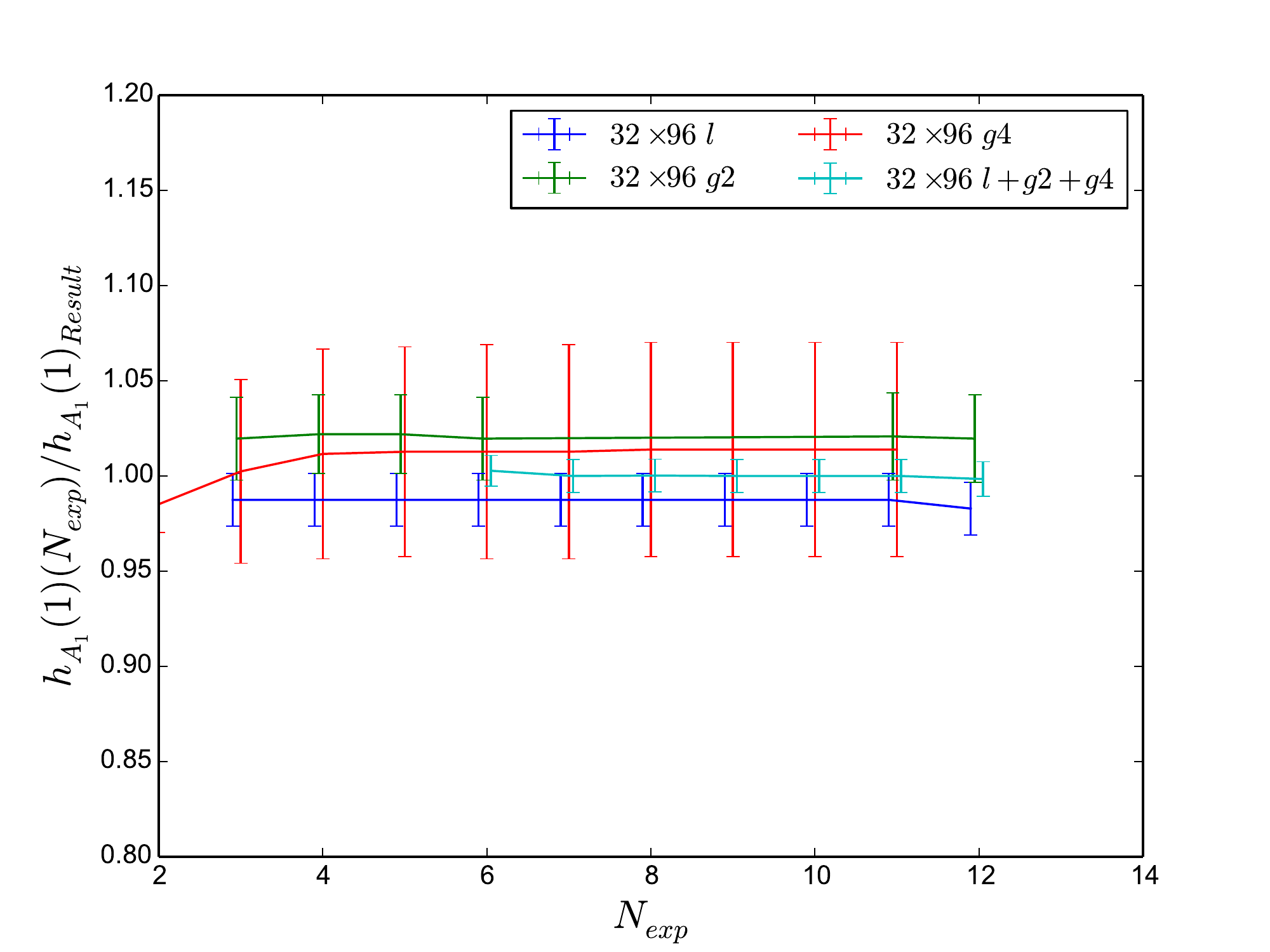}\includegraphics[width=0.4\textwidth]{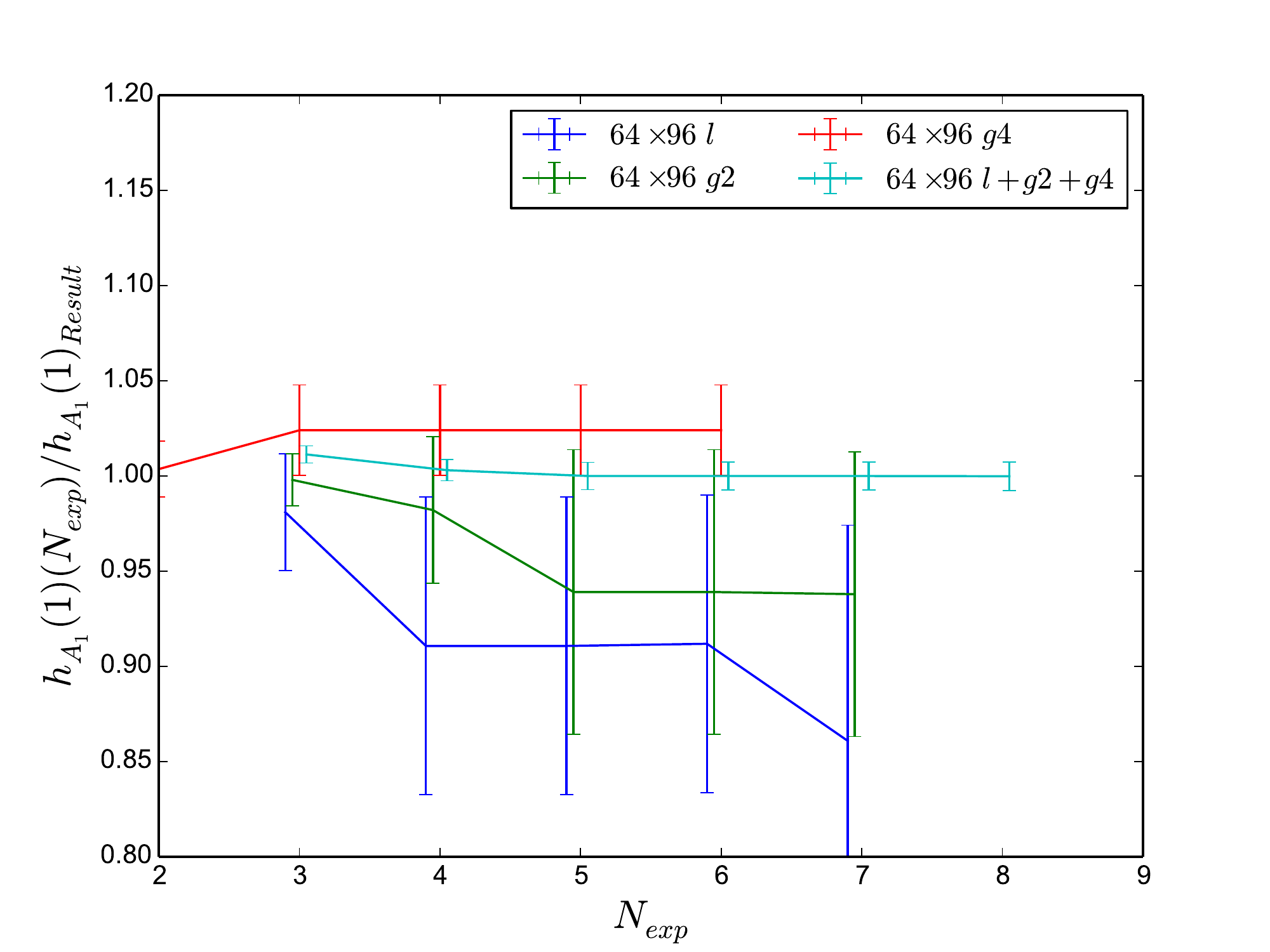}
\caption{Plots of $N_\text{exp}$ fit behaviour on all 8 ensembles (see
  Tab.~\ref{tab:params}). In each plot 4 sets of data points are
  shown: the full fit including all $3\times 3$ source-sink
  combinations, and, for comparison, separate ``diagonal'' fits where
  only one type of source-sink smearing is used.  (The notation is
  defined in Sec.~\ref{sec:params}.)  A significant improvement is
  seen in the full fit.  All diagonal fits show good agreement for
  $N_\text{exp} \ge 4$, but with the increased precision, sometimes
  5 or 6 exponentials are needed to get a good $3\times 3$ matrix fit.}
\label{fig:Nexpplots}
\end{figure}

Fig.~\ref{fig:Nexpplots} shows comparison of the fit results for
$h_{A_1}\!(1)$ when varying numbers of exponentials; the points are
normalized by the value of taken $h_{A_1}\!(1)$ as our result for that
ensemble.  Plots are shown for all 8 ensembles as listed in
Tab.~\ref{tab:params}.  In each plot, we show the full fit results
to the $3\times 3$ matrix of source/sink combinations (local $l$, or
Gaussian with 2 radii, $g2$ and $g4$), as well as ``diagonal'' fits where
only one source/sink is used.  The statistical improvement of using all
the data is apparent.  The flatness of the curves and the constancy of
the error bars shows that, for large enough $N_{\mathrm{exp}}$, the Bayesian
fits are insensitive to adding further exponential terms, i.e.\ effects
of excited states are accounted for.  Our final results typically come
from the $N_{\mathrm{exp}}=5$ fits to the full $3\times 3$ matrix of
correlators; however, on ensembles 3 and 7, we had to include another
exponential.

\section{Chiral continuum fit function}
\label{chipt}

The full expression for the form factor derived in staggered chiral perturbation theory is given by \cite{SCHIPT}
\begin{align}
h_{A_1}\!(1) = 1 + \frac{X(\Lambda_\chi)}{m_c^2} + \frac{g_\pi^2}{48\pi^2 f^2} \Big[\frac{1}{16} \sum_{\delta}(2\bar{F}_{\pi_\delta} +\bar{F}_{K_\delta}) - \frac{1}{2}\bar{F}_{\pi_I} + \frac{1}{6}\bar{F}_{\eta_I}\nonumber\\
+a^2\delta^\prime_V\Big(   \frac{m_{S_V}^2 - m_{\pi_V}^2}{(m_{\eta_V}^2 - m_{\pi_V}^2)(m_{\pi_V}^2 - m_{\eta^\prime_V}^2)}\bar{F}_{\pi_V} +  \frac{m_{\eta_V}^2 - m_{S_V}^2}{(m_{\eta_V}^2 - m_{\eta^\prime_V}^2)(m_{\eta_V}^2 - m_{\pi_V}^2)}\bar{F}_{\eta_V} \nonumber\\
 +  \frac{m_{S_V}^2 - m_{\eta^\prime_V}^2}{(m_{\eta_V}^2 - m_{\eta^\prime_V}^2)(m_{\eta^\prime_V}^2 - m_{\pi_V}^2)}\bar{F}_{\eta^\prime_V}\Big) + (V\rightarrow A)\Big]
\end{align}
where $\bar{F}_X = F[m_X,-\Delta_{m_c}/m_X]$ and
\begin{align}
F[m,x] =& \frac{m^2}{x} \Big\{ x^3 \ln \frac{m^2}{\Lambda_\chi^2} + \frac{1}{3}x^3 -4x+2\pi \nonumber\\
&-\sqrt{x^2-1}(x^2+2)\nonumber \\
&\times \big(\ln\big[ 1-2x(x-\sqrt{x^2-1})\big] -i\pi \big)\Big\} \,.
\label{FDEF}
\end{align}
The masses of the $\eta$ and $\eta^\prime$ are given in \cite{staggeredmasses} as
\begin{align}
m_{\eta_V}^2 &= \frac{1}{2}\left( m_{\pi_V}^2 + m_{S_V}^2 + \frac{3}{4}a^2 \delta^\prime_V - Z \right)\nonumber\\
m_{\eta^\prime_V}^2 &= \frac{1}{2}\left( m_{\pi_V}^2 + m_{S_V}^2 + \frac{3}{4}a^2 \delta^\prime_V + Z \right)\nonumber\\
Z &= \sqrt{(m_{S_V}^2-m_{\pi_V}^2)^2 - \frac{a^2\delta^\prime_V}{2}(m_{S_V}^2-m_{\pi_V}^2) + \frac{9(a^2\delta^\prime_V)^2}{16}}\nonumber\\
&= (m_{S_V}^2-m_{\pi_V}^2) - \frac{a^2 \delta^\prime_V}{4} + \mathcal{O}((a^2\delta^\prime_V)^2)\nonumber\\
m_{\eta_I}^2&=m_{\pi_I}^2/3 + 2m_{S_I}^2/3 \,.
\end{align}
We take the $s\bar{s}$ pseudoscalar taste splittings equal to the pion taste splittings. This is a good approximation in the case of HISQ \cite{Bazavov:2012xda}. We can then write (to order $\mathcal{O}((a^2\delta^\prime_V)^2)$ )
\begin{align}
m_{\eta^\prime_V}^2 - m_{\pi_V}^2 =& m_{S_G}^2 - m_{\pi_G}^2 + a^2 \delta^\prime_V/4\nonumber\\
m_{\eta_V}^2 - m_{\pi_V}^2 =& a^2 \delta^\prime_V/2\nonumber\\
m_{S_V}^2 - m_{\pi_V}^2 =& m_{S_G}^2 - m_{\pi_G}^2 
\end{align}
from which we find
\begin{align}
\frac{m_{S_V}^2 - m_{\eta^\prime_V}^2}{(m_{\eta_V}^2 - m_{\eta^\prime_V}^2)(m_{\eta^\prime_V}^2 - m_{\pi_V}^2)} &= \frac{a^2\delta^\prime_V/4}{(m_{S_G}^2 - m_{\pi_G}^2)^2-(a^2\delta^\prime_V/4)^2}\nonumber\\
\frac{m_{\eta_V}^2 - m_{S_V}^2}{(m_{\eta_V}^2 - m_{\eta^\prime_V}^2)(m_{\eta_V}^2 - m_{\pi_V}^2)} &= \frac{a^2\delta^\prime_V/2 - (m_{S_G}^2 - m_{\pi_G}^2)}{(a^2\delta^\prime_V/4-(m_{S_G}^2 - m_{\pi_G}^2))a^2\delta^\prime_V/2}\nonumber\\
 \frac{m_{S_V}^2 - m_{\pi_V}^2}{(m_{\eta_V}^2 - m_{\pi_V}^2)(m_{\pi_V}^2 - m_{\eta^\prime_V}^2)}&=\frac{ - (m_{S_G}^2 - m_{\pi_G}^2)}{((m_{S_G}^2 - m_{\pi_G}^2)-a^2\delta^\prime_V/4)a^2\delta^\prime_V/2} \,.
\end{align}
The expression for $h_{A_1}\!(1)$ then reduces to
\begin{align}
h_{A_1}\!(1) = 1 + \frac{X(\Lambda_\chi)}{m_c^2} + &\frac{g_\pi^2}{48\pi^2 f^2} \Bigg[\frac{1}{16} \sum_{\delta}2\bar{F}_{\pi_\delta} - \frac{1}{2}\bar{F}_{\pi_I} 
+\left( 2-\frac{a^2\delta^\prime_V}{2(m_{S_G}^2 - m_{\pi_G}^2)}\right)\bar{F}_{\eta_V}+\left( 2-\frac{a^2\delta^\prime_A}{2(m_{S_G}^2 - m_{\pi_G}^2)}\right)\bar{F}_{\eta_A}\nonumber\\
-&\left( 2+\frac{a^2\delta^\prime_V}{2(m_{S_G}^2 - m_{\pi_G}^2)}\right)\bar{F}_{\pi_V}-\left( 2+\frac{a^2\delta^\prime_A}{2(m_{S_G}^2 - m_{\pi_G}^2)}\right)\bar{F}_{\pi_A}\Bigg] + \mathcal{O}((a^2\delta^\prime_V)^2)
\label{FITFUNCTIONFINAL}
\end{align}
where we have ignored terms expected to produce normal discretization errors
and pion mass dependence, as these are included elsewhere in the
fit. Following \cite{PIMASSES} we take $\delta A' \approx \delta V' \approx
-\delta t$, which we implement by including $\delta A' =\delta V' = -\delta t
\times 1.0(5)$ as priors. We use the pion masses computed in
\cite{PIMASSES} together with the taste splittings for the pion, $\delta t$, given in \cite{Bazavov:2012xda}.

\end{widetext}

Finite volume effects can be accounted for in heavy meson chiral perturbation theory \cite{Arndt:2004bg} including taste-splitting effects in the staggered pions \cite{SCHIPT}. The functions  $\bar{F}_X$ in (\ref{FITFUNCTIONFINAL})  receive a correction term corresponding to the difference between infinite volume loop integrals and finite volume discrete sums.  Taste-splitting effects in the pions at non-zero lattice spacing moderate the size of the finite volume corrections because some of the pions in the loops have heavier masses than the Goldstone pion.  Consequently, some of the finite volume effect appears as a lattice-spacing effect, which is dealt with by our chiral-continuum fit.

\begin{figure}
    \includegraphics[width=0.45\textwidth]{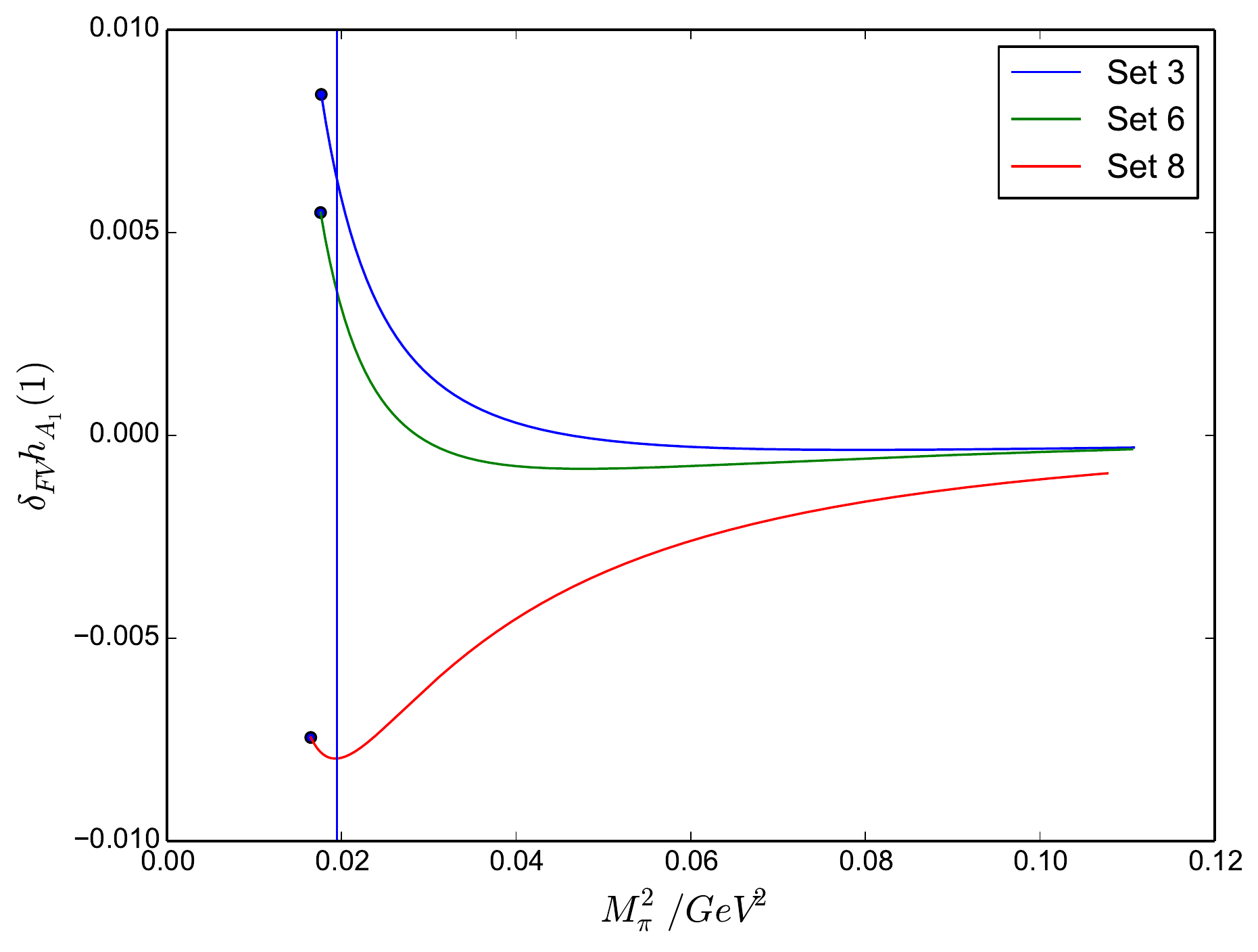}
\caption{Pion mass dependence of the finite volume corrections to $h_{A_1}(1)$, as determined from staggered chiral perturbation theory \cite{SCHIPT}, with parameters corresponding to the physical-mass lattices used here. The curves for the heavier-mass lattices used here show much smaller finite volume corrections, of $O(0.1\%)$.  The vertical blue line is the physical pion mass and the solid point at the end of each curve is at the measured value of the pion mass on each lattice.}
\label{fig:FVeffects}
\end{figure}

We incorporated the finite volume corrections into our fit by subtracting from our data $\delta_{FV} h_{A_1}(1)$, found by adding $\delta \bar{F}_X$ to each $\bar{F}_X$ appearing in (\ref{FITFUNCTIONFINAL}). In Fig.~\ref{fig:FVeffects} we show $\delta_{FV} h_{A_1}(1)$ as a function of pion mass for the parameters appropriate for the physical pion mass lattices, Sets 3, 6, and 8 (see Table~\ref{tab:params}).  For the other lattices, $|\delta_{FV} h_{A_1}(1)| \approx O(0.1\%)$ over the $M_\pi$ range where we have data and is not significant.

In Table \ref{tab:fit_variations} we give fit results for plausible variations on our chosen fit function as a demonstration of stability under such nontrivial choices. Neglecting different powers of $a^2$  we see that our result is only sensitive to leading $\mathcal{O}(a^2)$ errors. The $M_\pi^2$ dependence we included does not affect the central value if removed, nor do changes in the assumed correlations between NRQCD systematics between ensembles. Removing taste splitting terms in the chiral perturbation theory result down to the continuum formula results in only a small change to the central value. Adding $\alpha_s\Lambda_{\subrm{QCD}}/M_B$, which we have excluded from our fit due to Luke's theorem, results in a slight increase in the central value as well as the expected increase in error. Our result is also only mildly sensitive to different choices of $\Lambda_{\subrm{QCD}}$ which we vary by $\pm 50\%$. Taken collectively we note that no tested variations result in more than a $0.25\sigma$ change to the central value.

\begin{table}
\caption{\label{tab:fit_variations}Fit results for $h_{A_1}\!(1)$ for different chiral-continuum fit functions }
\begin{tabular}{ l c c c} 
\hline\hline
  Fit function        &           $h_{A_1}\!(1)$ &           $h^s_{A_1}\!(1)$ \\            
\hline
Eq.\ (\ref{FITFUNCTIONFINAL}) &   0.895(26) &   0.883(31)\\
excluding hairpin terms &   0.895(26) &   0.883(31)\\
continuum $\chi PT$ formula &   0.897(25) &   0.882(31)\\
$\Lambda_{\subrm{QCD}} = 750$ MeV &  0.900(35) &  0.882(38)\\
$\Lambda_{\subrm{QCD}} = 250$ MeV &   0.897(24) &  0.890(23)\\
 excluding polynomial $\mathcal{O}(a^6)$ terms &   0.895(26) &   0.883(31)\\
 excluding polynomial $\mathcal{O}(a^4)$ terms &    0.895(26) &   0.883(31)\\
 excluding polynomial $\mathcal{O}(a^2)$ terms &    0.898(26) &   0.891(25)\\
 excluding polynomial $M_\pi^2$ dependence &   0.895(27) &   0.883(31)\\
 excluding $(\Lambda/M_B)^2$ uncertainty &   0.895(25) &   0.883(31)\\
totally correlated $(\Lambda_{\subrm{QCD}}/M_B)^2$ errors  &   0.895(27) &   0.883(31)\\
 \hline\hline
\end{tabular}
\end{table}

\section{Fits to Experimental Data}
\label{app:exptfits}

The fully differential decay rate is given by \cite{Richman:1995wm,Abdesselam:2017kjf}
\begin{align}
  &\frac{d\Gamma( \bar B \to D^{*}  \ell \, \bar \nu_\ell)}{dw \, d\cos\theta_v \, d\cos\theta_{\ell} \, d\chi} = \frac{3G_F^2
    \left|\bar\eta_{EW} V_{cb}\right|^2}{1024\pi^4}
  \frac{M_{D^*}^2}{M_B} q^2\, \sqrt{w^2-1}   
  \nonumber\\
&~~\times \Big[ (1-\cos\theta_{\ell})^2\sin^2\theta_vH_+^2 + (1+\cos\theta_{\ell})^2\sin^2\theta_vH_-^2 \nonumber \\
& ~~+ 4\sin^2\theta_{\ell}\cos^2\theta_vH_0^2 - 2\sin^2\theta_{\ell}\sin^2\theta_v\cos2\chi H_+H_-\nonumber \\
&~~ -4\sin\theta_{\ell}(1-\cos\theta_{\ell})\sin\theta_v\cos\theta_v\cos\chi H_+H_0\nonumber \\
    & ~~+ 4\sin\theta_{\ell}(1+\cos\theta_{\ell})\sin\theta_v\cos\theta_v\cos\chi H_-H_0 \Big]
  \label{eq:dGdwdcvdcldchi}
\end{align}
where $H_+(w)$, $H_-(w)$, and $H_0(w)$ are helicity amplitudes. In
principle these amplitudes could be determined from lattice QCD, but
presently these must be parametrized and fit to experiment, with a
lattice calculation of the zero recoil form factor providing the
normalization.  Integrating (\ref{eq:dGdwdcvdcldchi}) over the angular
variables gives Eq.~(\ref{eq:dGdwF}), with
\begin{align}
  \chi(w)|\mathcal{F}(w)|^2 = \frac{r(1-2wr+r^2)}{12 M_B^2(1-r)^2}
    \sum_{i=\pm,0} |H_i|^2 \,.
\end{align}
with $r = M_{D^*}/M_B$.  Although not necessary in this work, it is
conventional to factor out the kinematic function
\begin{align}
  \chi(w) = \frac{(w+1)^2}{12}
  \left[1 + \frac{4w}{w+1}\frac{1-2wr + r^2}{(1-r)^2}\right] \,.
\end{align}
Note that $\chi(1)=1$ here, although different normalizations appear
in the literature.

The CLN parametrization expresses the helicity amplitudes as follows
\cite{Caprini:1997mu,Neubert:1993mb}. The reduced helicity amplitudes $\tilde{H}_i$ are defined by
\begin{equation}
  H_i(w) = (M_B - M_{D^*}) (1 + w) \sqrt{\frac{M_B \,M_{D^*}}{q^2}} \,h_{A_1}\!(w) \,
  \tilde{H}_i(w) \,.
\end{equation}
Then
\begin{align}
  \tilde{H}_\pm(w) & = \frac{\sqrt{1 - 2wr + r^2}}{1-r}\left[1 \mp
    \sqrt{\frac{w-1}{w+1}}\,R_1(w)\right] \nonumber \\
  \tilde{H}_0(w) & = 1 + \frac{w-1}{1-r}[1 - R_2(w)] \,
\end{align}
where $r = M_{D^*}/M_B$.  The $h_{A_1}\!(w)$, $R_1(w)$, $R_2(w)$
then expanded in $z$ or $w-1$, as given in (\ref{eq:CLNseries}).

In the BGL parametrization \cite{Boyd:1997kz} (and in the simplified
BCL parametrization we employ) the helicity amplitudes are written
in terms of the $f$, $F_1$, and $g$ form factors as follows
\begin{align}
  H_\pm(w) & = f(z) \mp M_B \,M_{D^*} \sqrt{w^2-1}\, g(z) \nonumber \\
  H_0(w) & = \frac{F_1(z)}{\sqrt{q^2}} \,.
\end{align}
These form factors are then expressed as in (\ref{eq:zexpansion}).

\end{document}